\documentclass[acmtog]{acmart}

\usepackage{ae}
\usepackage{aecompl}
\usepackage[utf8]{inputenc}
\usepackage{tikz}
\usetikzlibrary{arrows}
\tikzstyle{int}=[draw, fill=blue!20, minimum size=2em]
\tikzstyle{init} = [pin edge={to-,thin,black}]
\usepackage{multirow}
\usepackage{array}
\usepackage{booktabs} 
\usepackage{subcaption} 
\usepackage{float} 
\usepackage{wrapfig}
\usepackage{lipsum}
\usepackage[utf8]{inputenc}
\usepackage{amssymb}
\usepackage{pgfplots}

\definecolor{darkgreen}{rgb}{0,0.6,0}

\def\voroplusplus{{Voro\nolinebreak[4]\hspace{-.05em}\raisebox{.4ex}{\tiny\bf ++}}}

\newcommand{\ddarrow}{\ensuremath{{\downarrow\downarrow}}}

\newcommand{\seedset}{\ensuremath{\mathcal{S}}}

\newcommand{\surfaceSeeds}{\ensuremath{\seedset^\updownarrow}}
\newcommand{\interiorSeeds}{\ensuremath{\seedset^\ddarrow}}
\newcommand{\surfaceTime}{\ensuremath{T^\updownarrow}}
\newcommand{\interiorTime}{\ensuremath{T^\ddarrow}}

\newcommand{\surf}{\ensuremath{\mathcal{M}}}

\newcommand{\dsurf}{\ensuremath{\mathcal{T}}}

\newcommand{\vol}{\ensuremath{\mathcal{O}}} 
\newcommand{\sharpF}{\ensuremath{\mathcal{F}}}
\newcommand{\vcTheta}{\ensuremath{\theta^\sharp}}
\newcommand{\vcThetaS}{\ensuremath{\theta^\flat}}
\newcommand{\rmax}{\ensuremath{sz}}

\newcommand{\ballunion}{\ensuremath{\mathcal{U}}}
\newcommand{\ballset}{\ensuremath{\mathcal{B}}}

\newcommand{\AlphaK}{\ensuremath{\mathcal{W}}}

\newcommand{\kdtree}{$k$-d tree}
\newcommand{\kdtrees}{$k$-d trees}

\newcommand{\smoothCoverage}{C1}
\newcommand{\smoothOverlaps}{C2}
\newcommand{\Lipschitz}{C3}
\newcommand{\DeepCoverage}{C4}

\graphicspath{{lowres/}}

\makeatletter
\renewcommand\@formatdoi[1]{\ignorespaces}
\makeatother

\usepackage[ruled]{algorithm2e} 
\SetAlFnt{\small}
\SetAlCapFnt{\small}
\SetAlCapNameFnt{\small}
\SetAlCapHSkip{0pt}

\acmJournal{TOG}

\setcopyright{rightsretained}

\begin{document}

\title{VoroCrust: Voronoi Meshing Without Clipping}
\author{Ahmed Abdelkader}
\affiliation{%
  \institution{University of Maryland, College Park}}
\author{Chandrajit L. Bajaj}
\affiliation{%
  \institution{University of Texas, Austin}}
\author{Mohamed S. Ebeida$^\ast$}
\affiliation{%
  \institution{Sandia National Laboratories}}
\author{Ahmed H. Mahmoud}
\affiliation{%
  \institution{University of California, Davis}}
\author{Scott A. Mitchell}
\affiliation{%
  \institution{Sandia National Laboratories}}
\author{John D. Owens}
\affiliation{%
  \institution{University of California, Davis}}
\author{Ahmad A. Rushdi}
\affiliation{%
  \institution{Sandia National Laboratories}}

\authorsaddresses{}
\renewcommand{\shortauthors}{Abdelkader, Bajaj, Ebeida, Mahmoud, Mitchell, Owens, Rushdi}
\begin{abstract}
Polyhedral meshes are increasingly becoming an attractive option with particular advantages over traditional meshes for certain applications. What has been missing is a robust polyhedral meshing algorithm that can handle broad classes of domains exhibiting arbitrarily curved boundaries and sharp features. In addition, the power of primal-dual mesh pairs, exemplified by Voronoi-Delaunay meshes, has been recognized as an important ingredient in numerous formulations. The VoroCrust algorithm is the first provably-correct algorithm for conforming polyhedral Voronoi meshing for non-convex and non-manifold domains with guarantees on the quality of both surface and volume elements. A robust refinement process estimates a suitable sizing field that enables the careful placement of Voronoi seeds across the surface circumventing the need for clipping and avoiding its many drawbacks. The algorithm has the flexibility of filling the interior by either structured or random samples, while preserving all sharp features in the output mesh. We demonstrate the capabilities of the algorithm on a variety of models and compare against state-of-the-art polyhedral meshing methods based on clipped Voronoi cells establishing the clear advantage of VoroCrust output.
\end{abstract}

\thanks{
$^\ast$ Correspondence address: \href{mailto:msebeid@sandia.gov}{msebeid@sandia.gov}.

Author names are listed in alphabetical order.

This material is based upon work supported by the U.S. Department of Energy, Office of Science, Office of Advanced Scientific Computing Research (ASCR), Applied Mathematics Program, and the Laboratory Directed Research and Development program (LDRD) at Sandia National Laboratories.
Sandia National Laboratories is a multi-mission laboratory managed and operated by National Technology and Engineering Solutions of Sandia, LLC., a wholly owned subsidiary of Honeywell International, Inc., for the U.S. Department of Energy's National Nuclear Security Administration under contract DE-NA0003525.
A. Abdelkader acknowledges the support of the \grantsponsor{GS100000001}{National Science Foundation}{http://dx.doi.org/10.13039/100000001} under grant number~\grantnum{GS100000001}{CCF-1618866}. The research of C. Bajaj was supported in part by the \grantsponsor{GS100000002}{National Institute of Health}{http://dx.doi.org/10.13039/100000002} under grant number~\grantnum{GS100000002}{R01GM117594}. A. Mahmoud and J. Owens acknowledge the support of the \grantsponsor{GS100000003}{National Science Foundation}{http://dx.doi.org/10.13039/100000001} under grant number~\grantnum{GS100000003}{CCF-1637442}.
This paper describes objective technical results and analysis. Any subjective views or opinions that might be expressed in the paper do not necessarily represent the views of the U.S. Department of Energy, or the United States Government, or any of the funding agencies.}


\usetikzlibrary{spy,shapes.misc}
\begin{teaserfigure}
	\centering
\setlength{\tabcolsep}{4pt}
\resizebox{\linewidth}{!}{
	\begin{tabular}{cc}	 
	\begin{tikzpicture}
	[spy using outlines={rounded rectangle, magnification=7,width=10.0cm, height=5.0cm, connect spies}]
	\node {\includegraphics[width=1.0\linewidth]{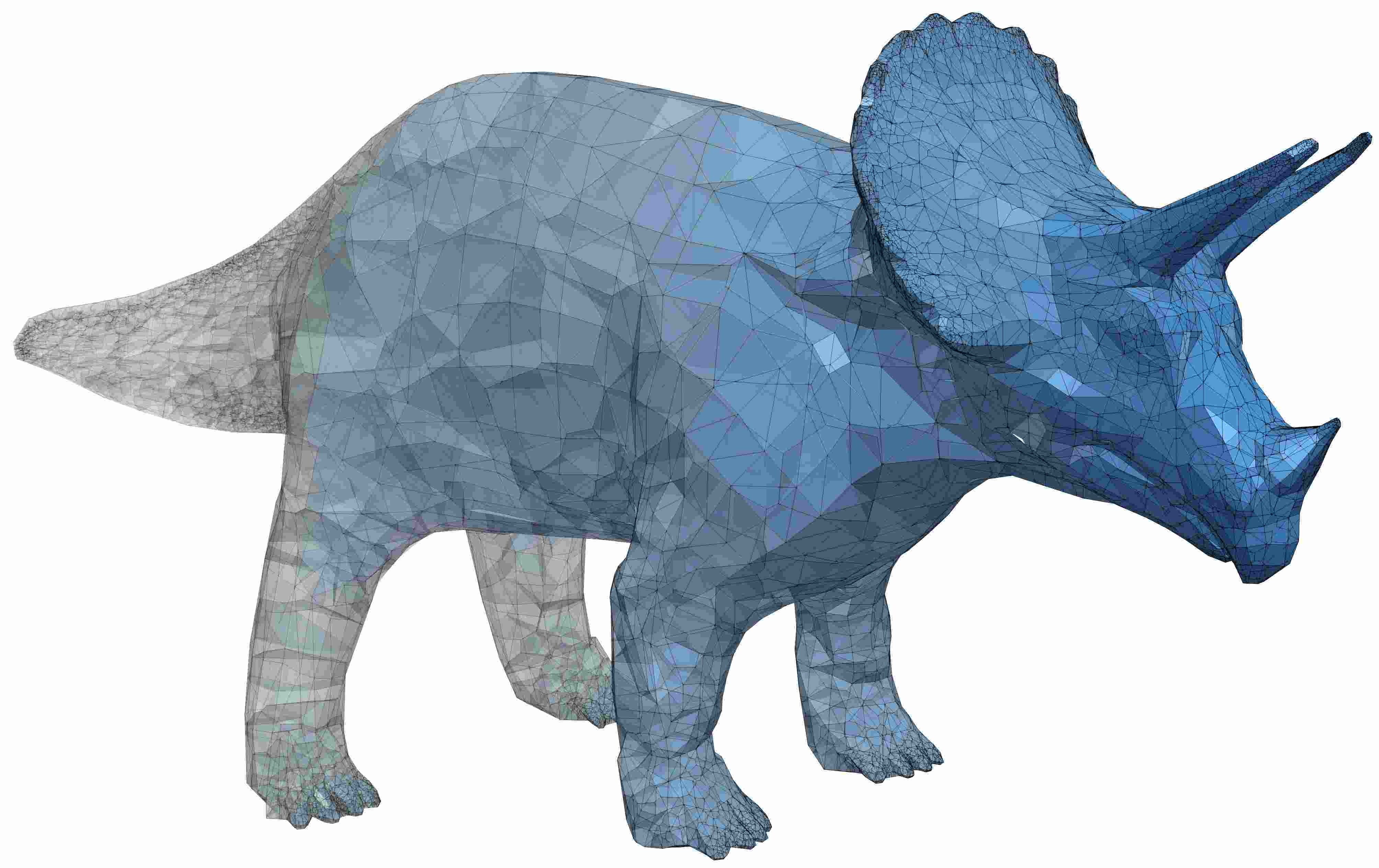}};
	\spy on (3.3,5.20) in node at (-3.6,-5);	
	\node at (7.75, -4.75) {\includegraphics[width=0.6\linewidth]{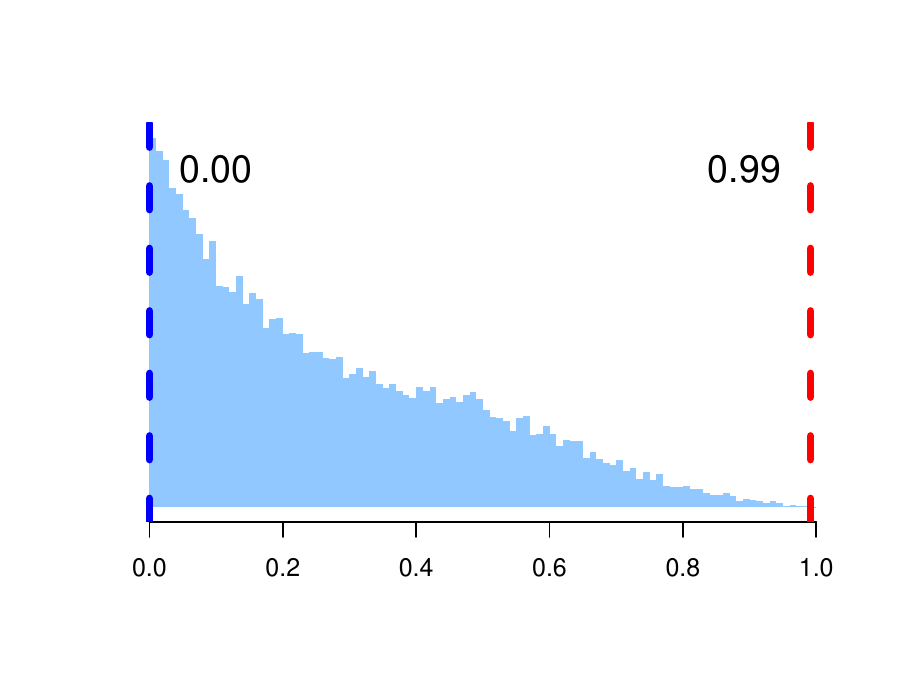}};
	\end{tikzpicture}
 	&
 	\begin{tikzpicture}
	[spy using outlines={rounded rectangle, magnification=7,width=10.0cm, height=5.0cm, connect spies}]
	\node {\includegraphics[width=1.0\linewidth]{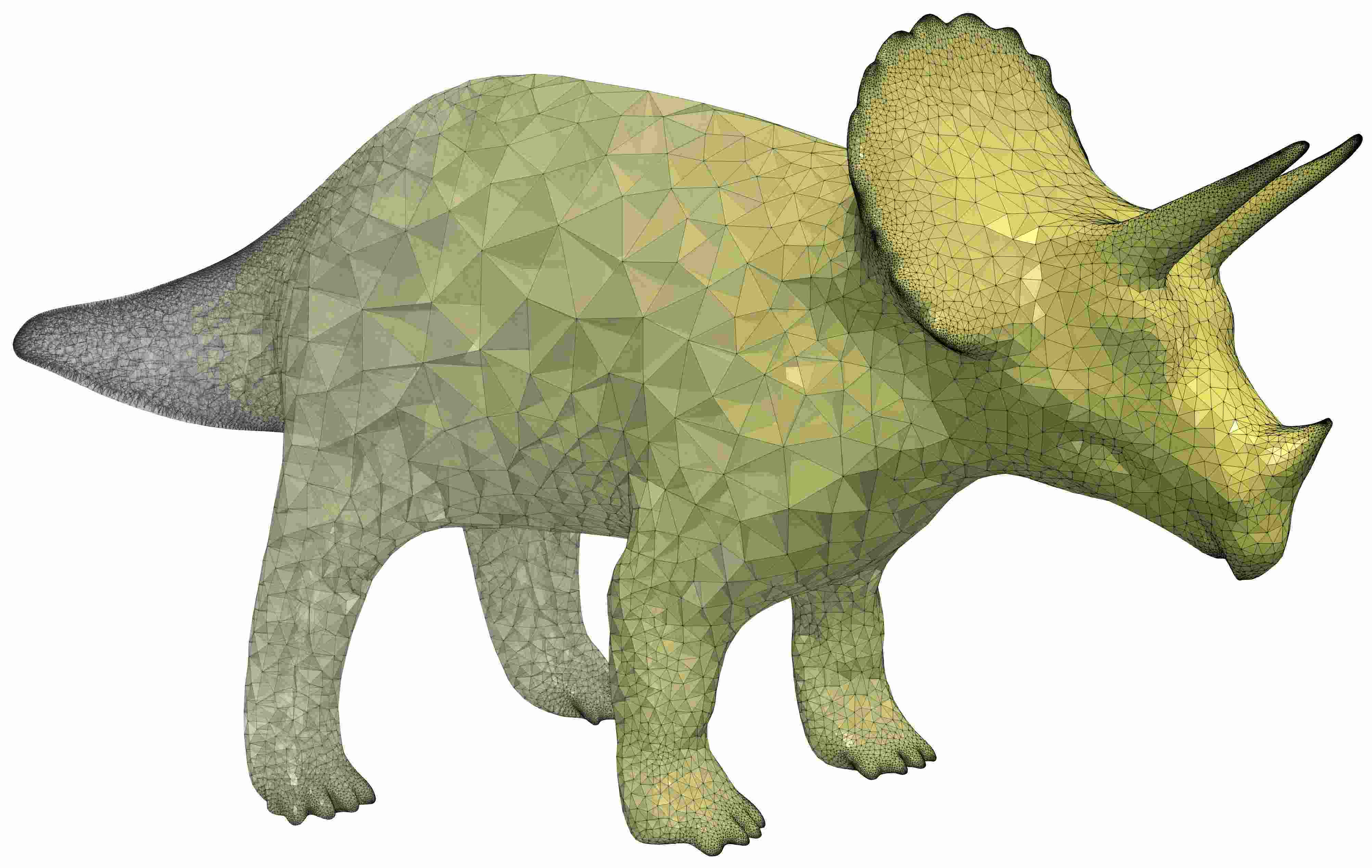}};
	\spy on (3.3,5.20) in node at (-3.6,-5);	
    \node at (7.75, -4.75) {\includegraphics[width=0.6\linewidth]{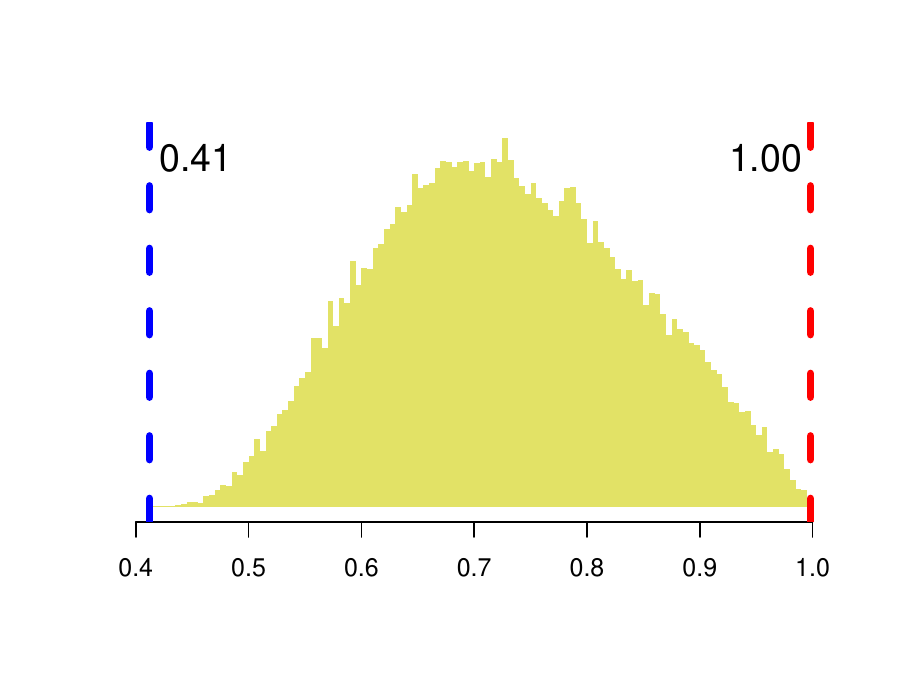}};
	\end{tikzpicture} 		
	\end{tabular}}
	\vspace{-16pt}
	\caption{State-of-the-art methods for conforming Voronoi meshing clip Voronoi cells at the bounding surface. The Restricted Voronoi Diagram~\cite{yan2010efficient} (left) is sensitive to the input tessellation and produces surface elements of very low quality, per the shortest-to-longest edge ratio distribution shown in the inset. In contrast, VoroCrust (right) generates an unclipped Voronoi mesh conforming to a high-quality surface mesh.}
	\label{fig:teaser}
\end{teaserfigure}

\maketitle
.
\newpage
\section{Introduction}
\label{sec:intro}
The computational modeling of physical phenomena requires robust numerical algorithms and compatible high-quality domain discretizations.
Finite element methods traditionally use simplicial meshes, where well-known angle conditions prohibit skinny elements~\cite{ShewchukQuality}.
The limited degrees of freedom of linear tetrahedral elements often lead to excessive refinement when modeling complex geometries or domains undergoing large deformations.
This motivated generalizations to general polyhedral elements, which enjoy larger degrees of freedom and have recently been in increasing demand in computer graphics~\cite{Martin:2008}, physically-based simulations~\cite{Bishop:2014}, applied mathematics~\cite{Sukumar:2014}, computational mechanics~\cite{Paulino:2014} and computational physics~\cite{MFD:2014}.

While the generation of tetrahedral meshes based on Delaunay refinement~\cite{cheng2012delaunay} or variational optimization~\cite{Alliez:2005} is well established, research on polyhedral mesh generation is less mature.
To further ensure the fidelity of the discrete model, the fundamental properties of continuum equations have to be preserved~\cite{Desbrun:2008}.
A well-principled framework is enabled through the combined use of primal meshes and their orthogonal duals~\cite{Mullen:2011:HOT}.
The power of orthogonal duals, exemplified by Voronoi-Delaunay meshes, has recently been demonstrated on a range of applications in computer graphics~\cite{Goes:2014:WTG} and computational physics~\cite{Engwirda:2018}.
It is therefore imperative to develop new algorithms for primal-dual polyhedral meshing.

In this paper, we present the design and implementation of VoroCrust:
the first algorithm for meshing non-convex, non-smooth, and even non-manifold domains by conforming polyhedral Voronoi meshes.
The implicit output mesh, compactly encoded by a set of Voronoi seeds, comes with an orthogonal dual defined by the corresponding Delaunay tetrahedralization.
This makes VoroCrust one of the first robust and efficient algorithms for primal-dual polyhedral meshing.
The crux of the algorithm is a robust refinement process that estimates a suitable sizing function to guide the placement of Voronoi seeds.
This enables VoroCrust to protect all sharp features, and mesh the surface and interior into quality elements.
We demonstrate the performance of the algorithm through a variety of challenging models, see Figure~\ref{fig:vc_features}, and compare against state-of-the-art polyhedral meshing methods based on clipped Voronoi cells; see Figures~\ref{fig:teaser} and~\ref{fig:RVD_nonconvex}. 

\subsection{Background}
\label{sec:background}

Conventional mesh elements, as in tetrahedral and hexahedral meshes, often require excessive refinement when modeling complex geometries or domains undergoing large deformations, e.g., cutting, merging, fracturing, or adaptive refinement~\cite{Wojtan:2009,Wicke:2010,Clausen:2013,Chen:2014}. A key advantage of general polyhedral elements is their superior ability to adjust to deformation~\cite{Martin:2008,Gain:2014} and topological changes~\cite{cut_survey_2015}, while being less biased to principal directions compared to regular tessellations~\cite{Talischi:2013}. In addition, polyhedral elements typically have more neighbors, even at corners and boundaries, enabling better approximation of gradients and possibly higher accuracy using the same number of conventional elements~\cite{cdadapco:polyhedral}.

Unfortunately, robust polyhedral meshing algorithms are still lacking. State-of-the-art approaches often rely on \textit{clipping}, i.e., truncating cells of an initial mesh to fit the domain boundaries~\cite{yan2010efficient}. Such an initial mesh can be obtained as a Voronoi mesh, e.g., with seeds randomly generated inside the domain~\cite{Ebeida2012} or optimized by centroidal Voronoi tessellations (CVT)~\cite{yan2010efficient}, possibly taking anisotropy into account~\cite{Budninskiy:2016:OVT}. Alternatively, an initial Voronoi mesh can be obtained by dualizing a conforming tetrahedral mesh~\cite{Garimella:2014}. Although no clipping is needed if the tetrahedralization is \textit{well-centered}, generating such meshes is very challenging and only heuristic solutions are known~\cite{doi:10.1137/090748214}. A weaker \textit{Gabriel property} ensures all tetrahedra have circumcenters inside the domain and can be guaranteed for polyhedral domains with bounded minimum angles~\cite{Si2010}; however, the dual Voronoi cells still need to be clipped.

\usetikzlibrary{spy,shapes.misc}

\begin{figure}[htb]
\centering
\setlength{\tabcolsep}{4pt}
\resizebox{\linewidth}{!}{
	\begin{tabular}{cc}	 
	\begin{tikzpicture}
	[spy using outlines={rounded rectangle, magnification=3,width=4.0cm, height=3.0cm, connect spies}]
	\node {\includegraphics[width=1.0\linewidth,trim={0 4cm 0 4cm},clip]{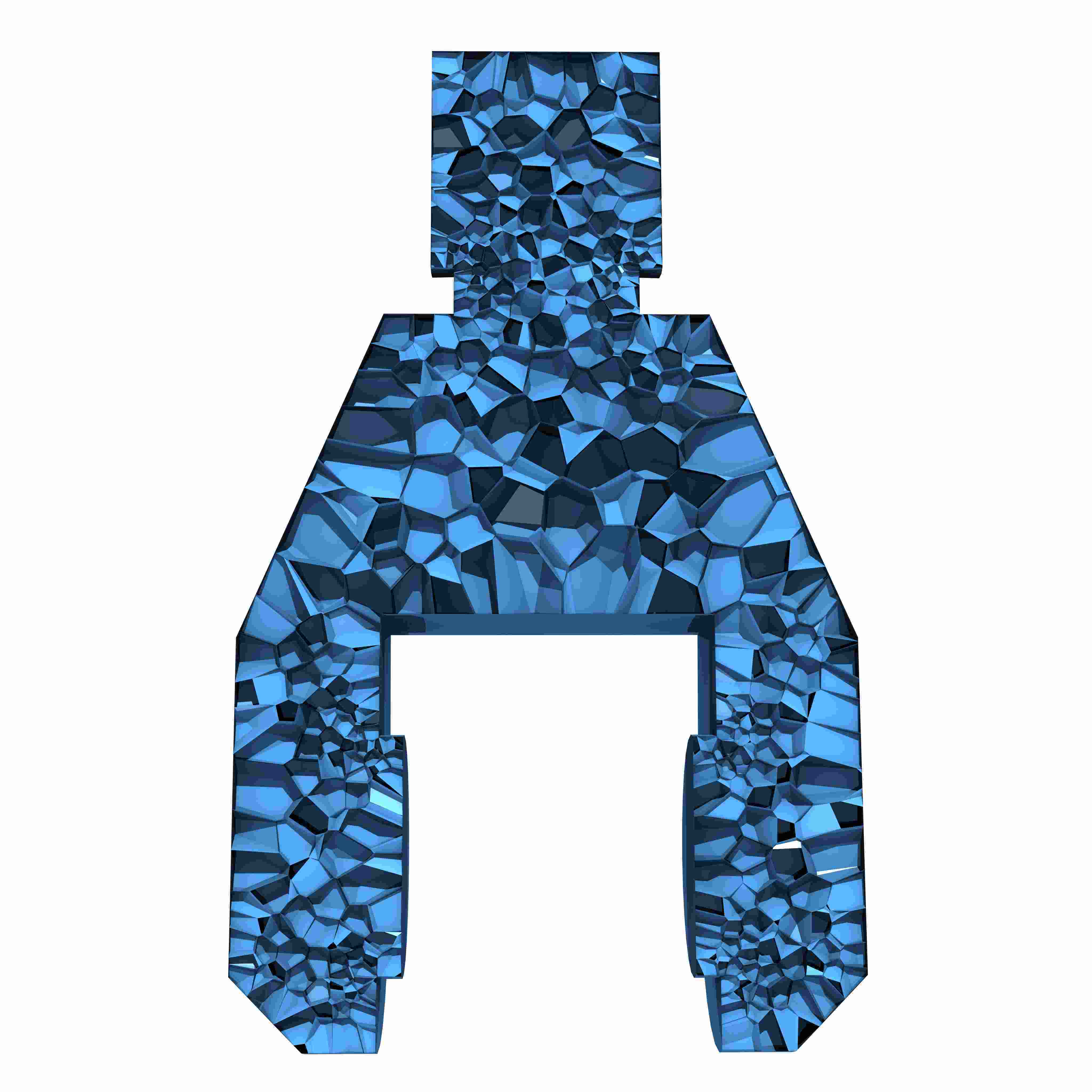}};
	\spy on (-1.3,-0.6) in node at (2.5,-2.7);
	\spy on (-0.8,2) in node at (2.0,2.7);
	\end{tikzpicture}	
 	&
 	\begin{tikzpicture}
	[spy using outlines={rounded rectangle, magnification=3,width=4.0cm, height=3.0cm, connect spies}]
	\node {\includegraphics[width=1.0\linewidth,trim={0 4cm 0 4cm},clip]{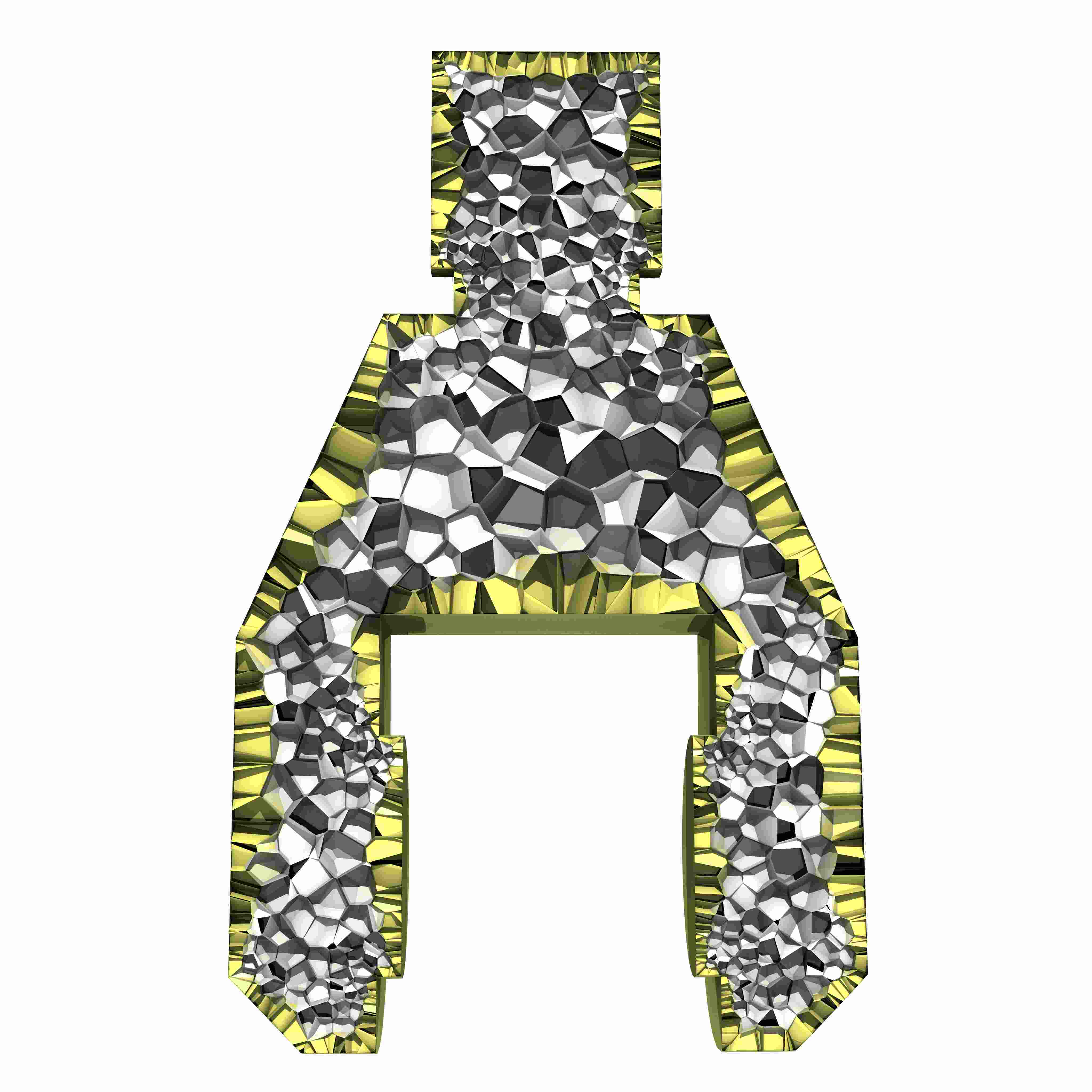}};
	\spy on (-1.3,-0.6) in node at (2.5,-2.7);
	\spy on (-0.8,2) in node at (2.0,2.7);
	\end{tikzpicture} 		
	\end{tabular}}
\caption{State-of-the-art clipping~\cite{yan2010efficient} may create non-convex cells (left); anywhere from $3\%$ up to $96\%$. In contrast, VoroCrust always produces true Voronoi cells conforming to the boundary (right).}
\label{fig:RVD_nonconvex}
\end{figure}


While clipping can be implemented robustly~\cite{yan2010efficient}, it fails to produce true Voronoi cells, sacrificing key geometric properties~\cite{Ebeida2012}. Specifically, clipping at sharp features may yield cells that are not convex as shown in Figure~\ref{fig:RVD_nonconvex}. This violates the requirements of several important applications, e.g., barycentric interpolation~\cite{Warren:2007} and polyhedral finite elements~\cite{Martin:2007}. Even more general formulations, like Virtual Element Methods~\cite{VEM_principles}, require star-shaped cells which clipping cannot guarantee. Unlike prior work, VoroCrust robustly meshes complex domains into Voronoi cells that naturally conform to the boundary circumventing the need for clipping. The combined use of such Voronoi meshes and their orthogonal duals, defined by the corresponding Delaunay tetrahedralizations~\cite{aurenhammer2013voronoi}, has been recognized as an important framework for computational modeling~\cite{Desbrun:2008,Mullen:2011:HOT}.

\subsection{Related Work}
\label{sec:related}
We further motivate Voronoi meshing through a detailed review of primal-dual meshing and its practical relevance. Then, we proceed to review related work on Voronoi-based modeling and meshing piecewise-smooth complexes, which together provide the theoretical underpinnings of the VoroCrust algorithm. \newline

\noindent \textbf{Orthogonal Primal-Dual Meshing.} Orthogonal primal-dual mesh pairs are unstructured staggered meshes~\cite{MAC:1965} with desirable conservation properties~\cite{PEROT200058}, enabling discretizations that closely mimic the continuum equations being modeled~\cite{Hyman:2006,Desbrun:2008}. The power of orthogonal duals~\cite{Mullen:2011:HOT} was recognized in early works on structural design~\cite{Rankine:1864,Maxwell:1870} and numerical methods~\cite{Macneal:1953}, and has recently been demonstrated on a range of applications in computer graphics~\cite{Goes:2014:WTG}, self-supporting structures~\cite{Akbarzadeh:2015}, mesh parameterization~\cite{Mercat:2001}, and computational physics~\cite{Engwirda:2018}. In particular, Voronoi-Delaunay meshes are the default geometric realization of many formulations in numerical methods~\cite{Nicolaides:1997}, fluid animation~\cite{Elcott:2007:SCS}, fracture modeling~\cite{SukuBolander:2009}, and computational cell biology~\cite{Novak:2007}.

Despite many attempts to design a robust Voronoi meshing algorithm, a general solution to the problem remained elusive. In particular, a number of widely used numerical simulators for flow and transport models, e.g., TOUGH2~\cite{Pruess:1991} and PFLOTRAN~\cite{PFLOTRAN}, compute gradients along nodal lines connecting neighboring cells, and hence require that these dual edges are orthogonal to the common primal facets~\cite{Pruess:2004}. Several heuristic approaches to the generation of Voronoi meshes for such simulators were developed~\cite{MeshVoro,IGMESH,Voro2Mesh,Kim2015,Klemetsdal2017}. The situation is further complicated for multi-material domains, where the difficulty of generating conforming meshes necessitates dealing with mixed elements straddling the interface between multiple materials~\cite{Garimella:2011,Kikinzon:2017,Dawes:2017}. In contrast, VoroCrust is a well-principled algorithm for conforming Voronoi meshing that can handle a large class of domains having as boundary either a manifold or non-manifold surface with arbitrarily sharp features. \newline

\begin{figure}
\centering
\begin{subfigure}[b]{0.45\columnwidth}\centering
  \includegraphics[width=0.95\columnwidth]{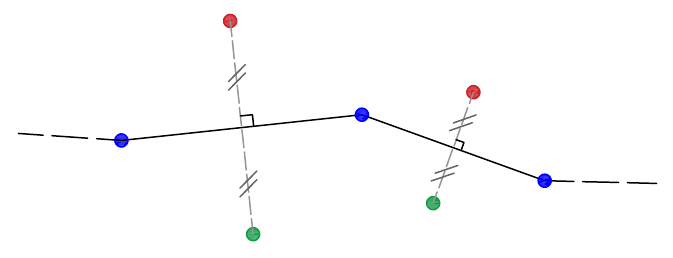}
  \caption{Naive mirroring of seeds.}
\end{subfigure}
\begin{subfigure}[b]{0.5\columnwidth}\centering
  \includegraphics[width=0.95\columnwidth]{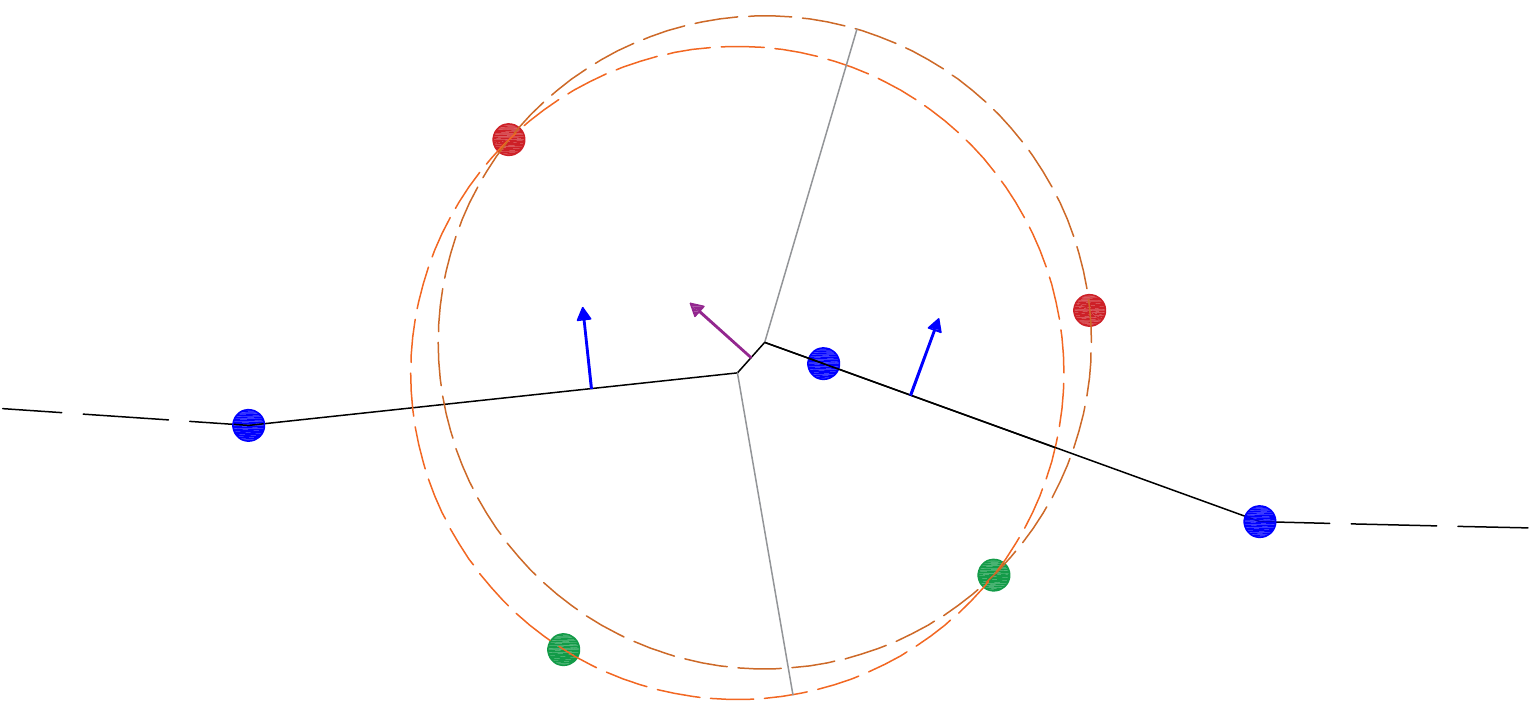}
  \caption{Naive mirroring reconstruction.}
\end{subfigure}
\begin{subfigure}[b]{0.45\columnwidth}\centering
  \includegraphics[width=0.95\columnwidth]{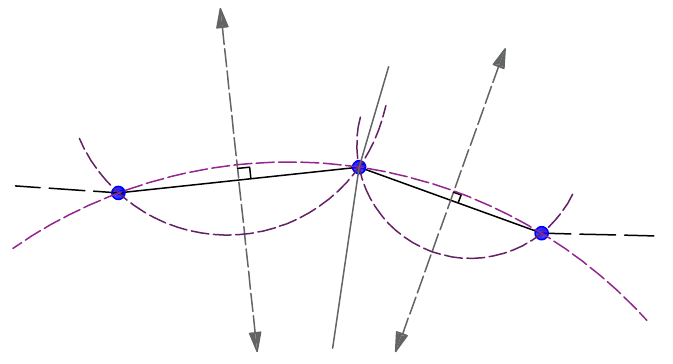}
  \caption{PowerCrust reconstruction.}
\end{subfigure}
\begin{subfigure}[b]{0.5\columnwidth}\centering
  \includegraphics[width=0.95\columnwidth]{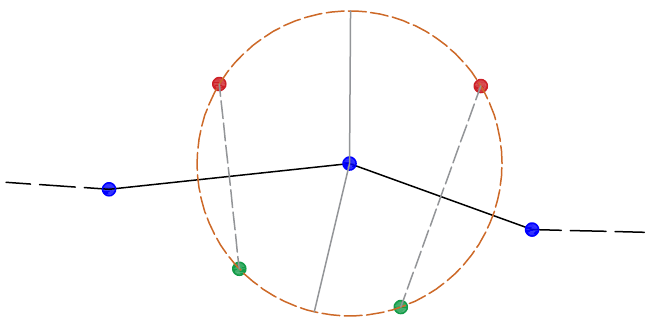}
  \caption{VoroCrust reconstruction.}
\end{subfigure}
   \caption{Voronoi-based reconstruction interpolates boundary samples (blue) using the Voronoi facets generated by seeds on different sides of the boundary, e.g., inside (green) and outside (red). Naive mirroring (a) results in large normal deviations (b) due to Voronoi facets between non-paired seeds. PowerCrust reduces normal deviations by placing weighted seeds on the medial axis away from the boundary (c). VoroCrust eliminates misaligned facets (d) using unweighted seeds.}
    \label{fig:mirroring}
    \vspace{-12pt}
\end{figure}

\begin{wrapfigure}[7]{r}{2\columnsep}
    \vspace{-1\intextsep}
    \hspace*{-1\columnsep}
    \includegraphics[width=0.3\columnwidth]{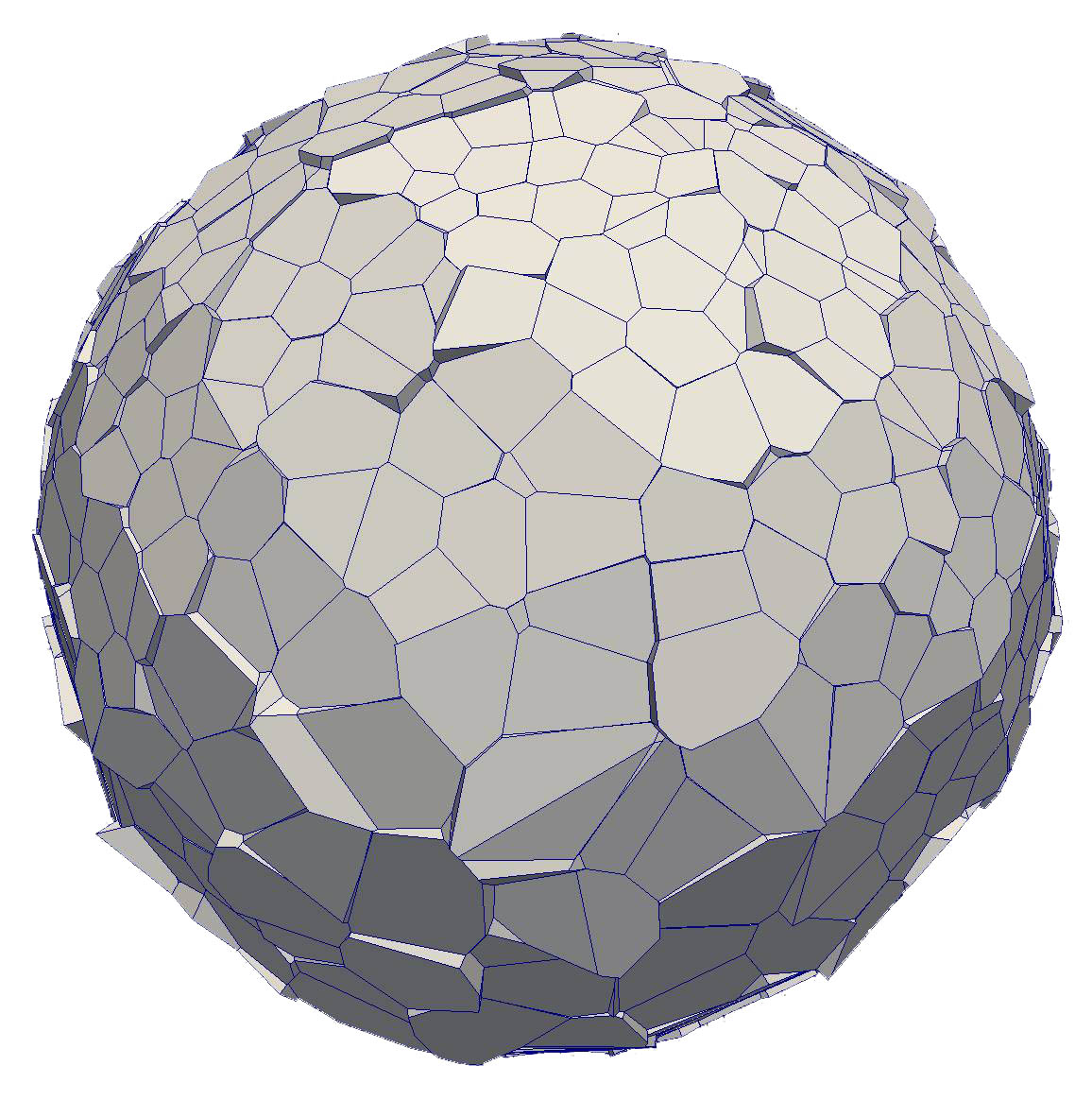}
\end{wrapfigure}

\noindent \textbf{Voronoi-based Modeling.} An intuitive approach to surface approximation is to place pairs of Voronoi seeds \emph{mirrored} across the surface such that their shared Voronoi facets approximate the surface; see Figure~\ref{fig:mirroring}(a). However, a naive implementation of this idea results in a rough surface with spurious misaligned facets; see the inset and Figure~\ref{fig:mirroring}(b). One such mirroring approach relies on an input sizing parameter to segment images into convex polygons assuming no four lines meet in a point~\cite{7298931}.

\begin{figure}[htb]
  \begin{minipage}{1\linewidth}
    \centering
    \includegraphics[width=0.3\columnwidth,trim={0 0 0 3.12cm},clip]{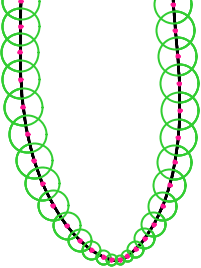}
    \quad
    \includegraphics[width=0.3\columnwidth,trim={0 0 0 3.12cm},clip]{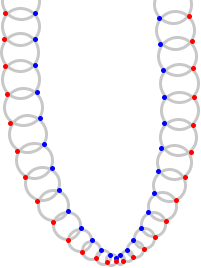}
    \quad
    \includegraphics[width=0.3\columnwidth,trim={0 0 0 3.12cm},clip]{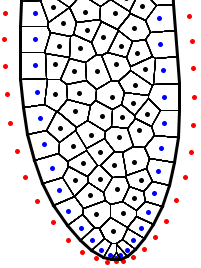}
  \end{minipage}
    \caption{VoroCrust summary: (left) Cover the boundary by a union of balls, (middle) place pairs of Voronoi seeds where balls intersect to capture and isolate the boundary, and finally (right) seed the interior.}
    \label{fig:steps}
  \vspace{-6pt}
\end{figure}

Nonetheless, a more principled mirroring approach provided the first provably-correct surface reconstruction algorithm~\cite{amenta1999surface}. Given an $\epsilon$-sample from an unknown smooth surface, the PowerCrust~algorithm \cite{amenta2001127} places weighted Voronoi seeds at a subset of the vertices in the Voronoi diagram of the input samples. While PowerCrust successfully avoids misaligned facets, the placement of seeds as described is restricted to lie close to the medial axis resulting in very skinny Voronoi cells extending perpendicularly to the surface; see Figure~\ref{fig:mirroring}(c). For the purposes of conforming Voronoi meshing, it is necessary to avoid such skinny cells. In contrast, VoroCrust is able to capture the surface using pairs of unweighted seeds placed close to the surface, enabling further decomposition of the interior using additional seeds; see Figure~\ref{fig:mirroring}(d). A visual summary of the VoroCrust algorithm is provided in Figure~\ref{fig:steps}.

\begin{figure*}
\centering
\begin{subfigure}[b]{0.13\linewidth}\centering
  \includegraphics[width=1\textwidth]{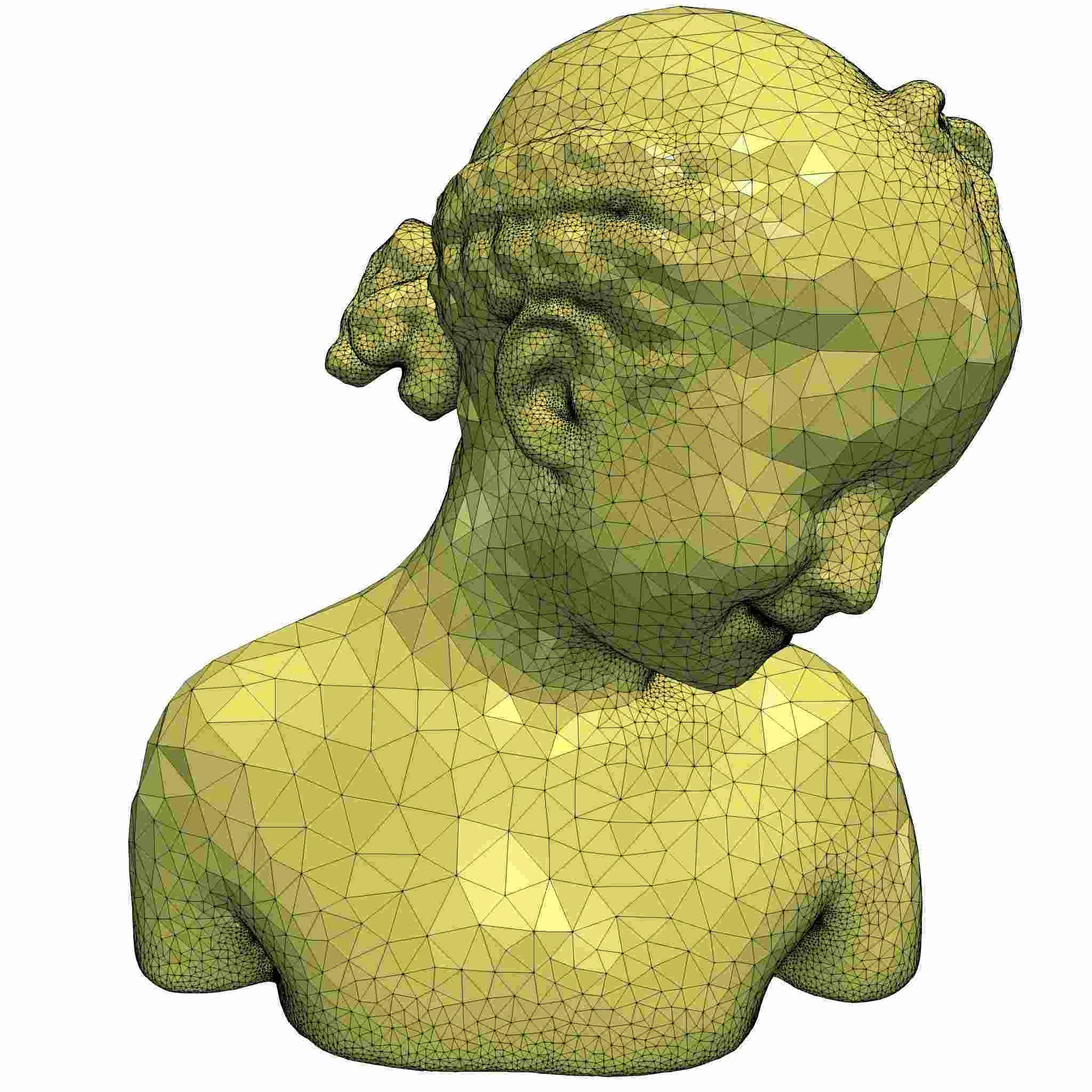}
  \caption{}
\end{subfigure}
\begin{subfigure}[b]{0.13\linewidth}\centering
  \includegraphics[width=1\textwidth]{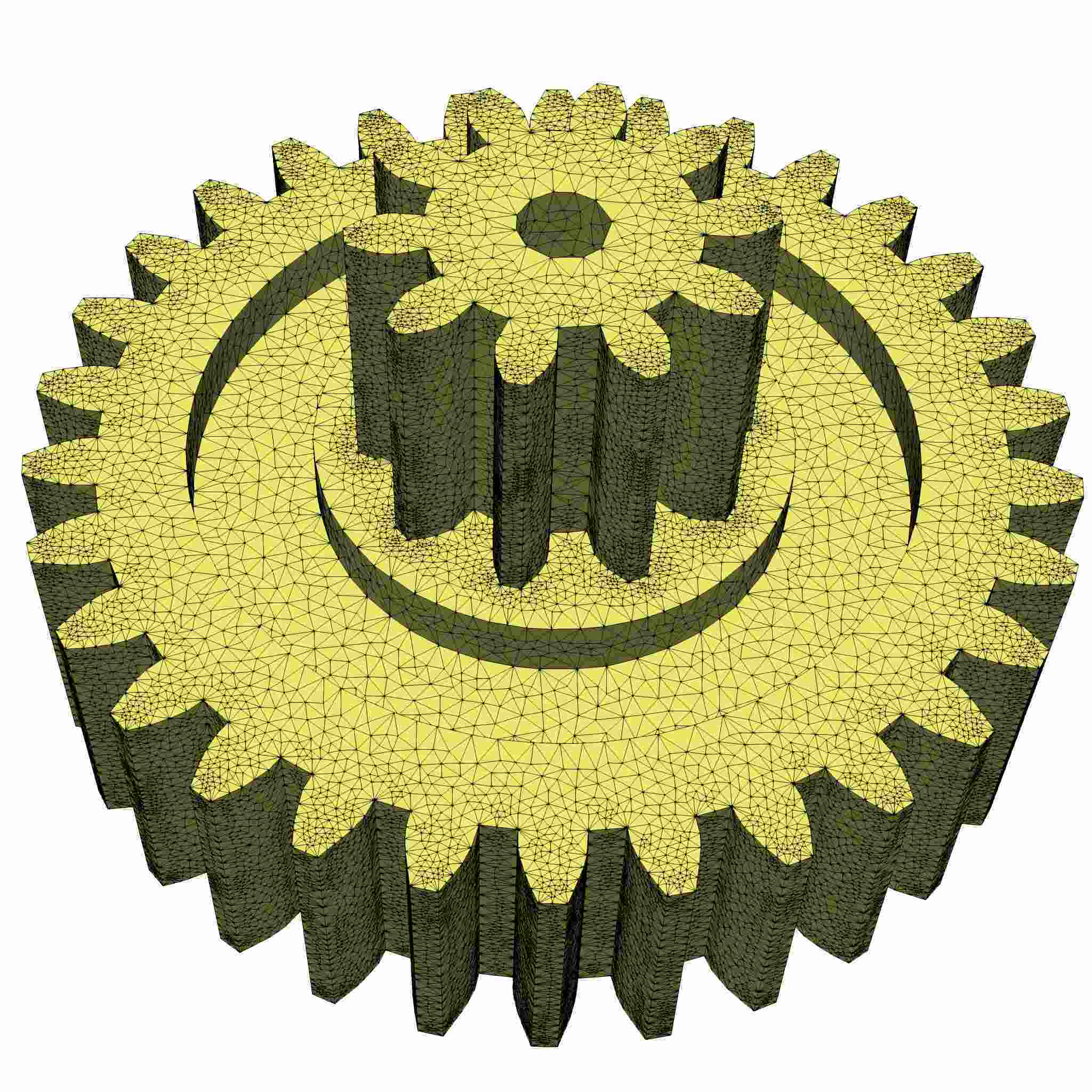}
  \caption{}
\end{subfigure}
\begin{subfigure}[b]{0.13\linewidth}\centering
  \includegraphics[width=1\textwidth]{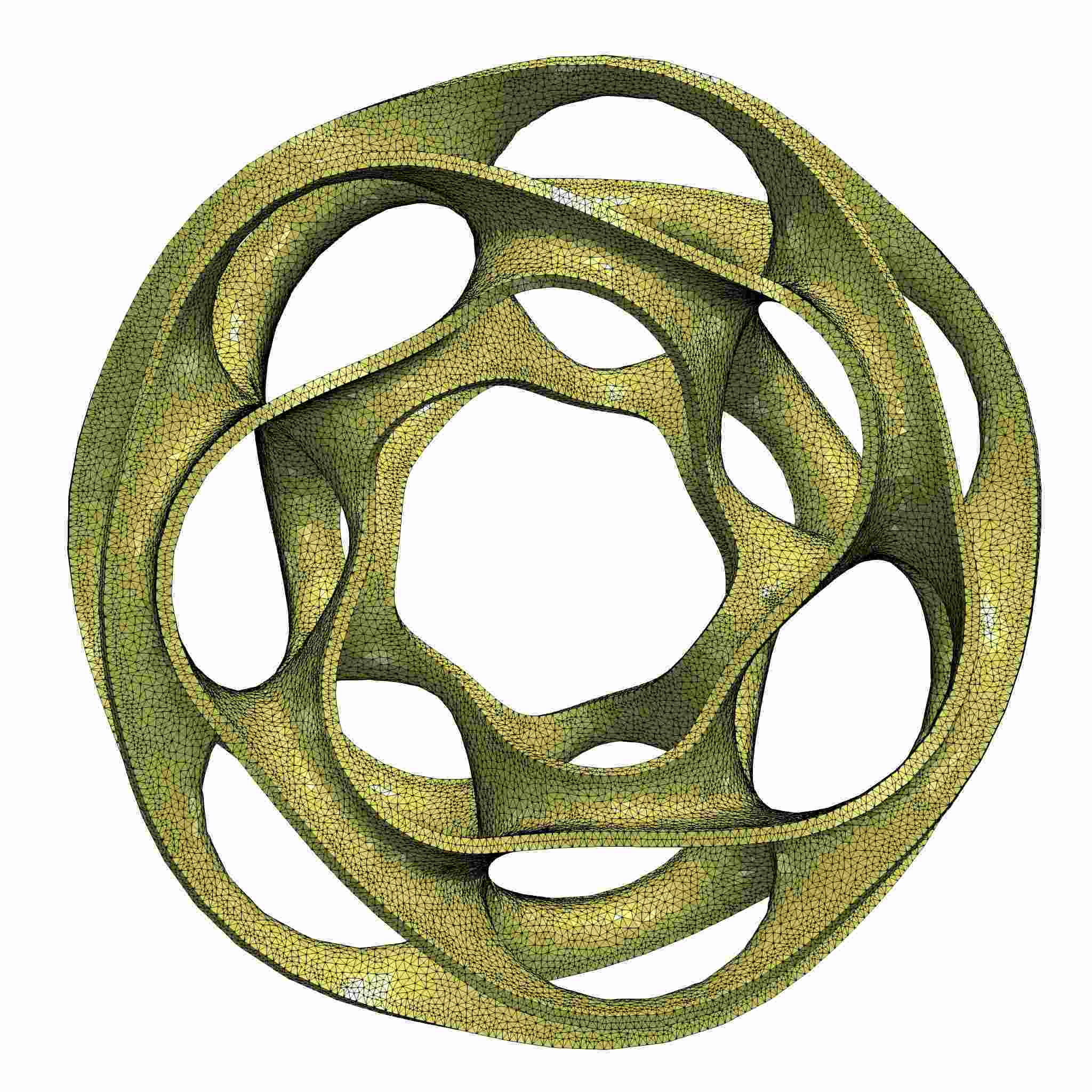}
  \caption{}
\end{subfigure}
\begin{subfigure}[b]{0.13\linewidth}\centering
  \includegraphics[width=1\textwidth]{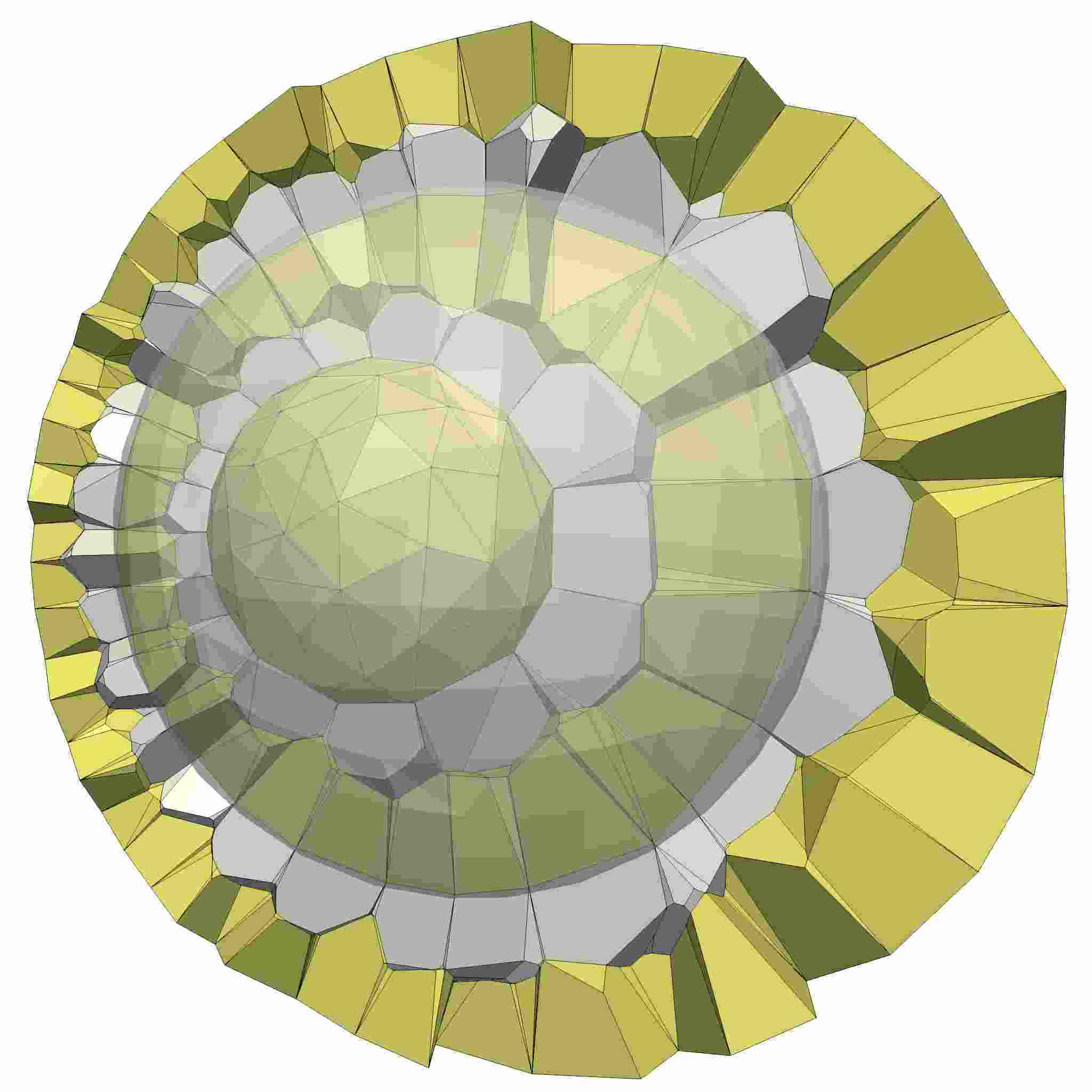}
  \caption{}
\end{subfigure}
\begin{subfigure}[b]{0.13\linewidth}\centering
  \includegraphics[width=1\textwidth]{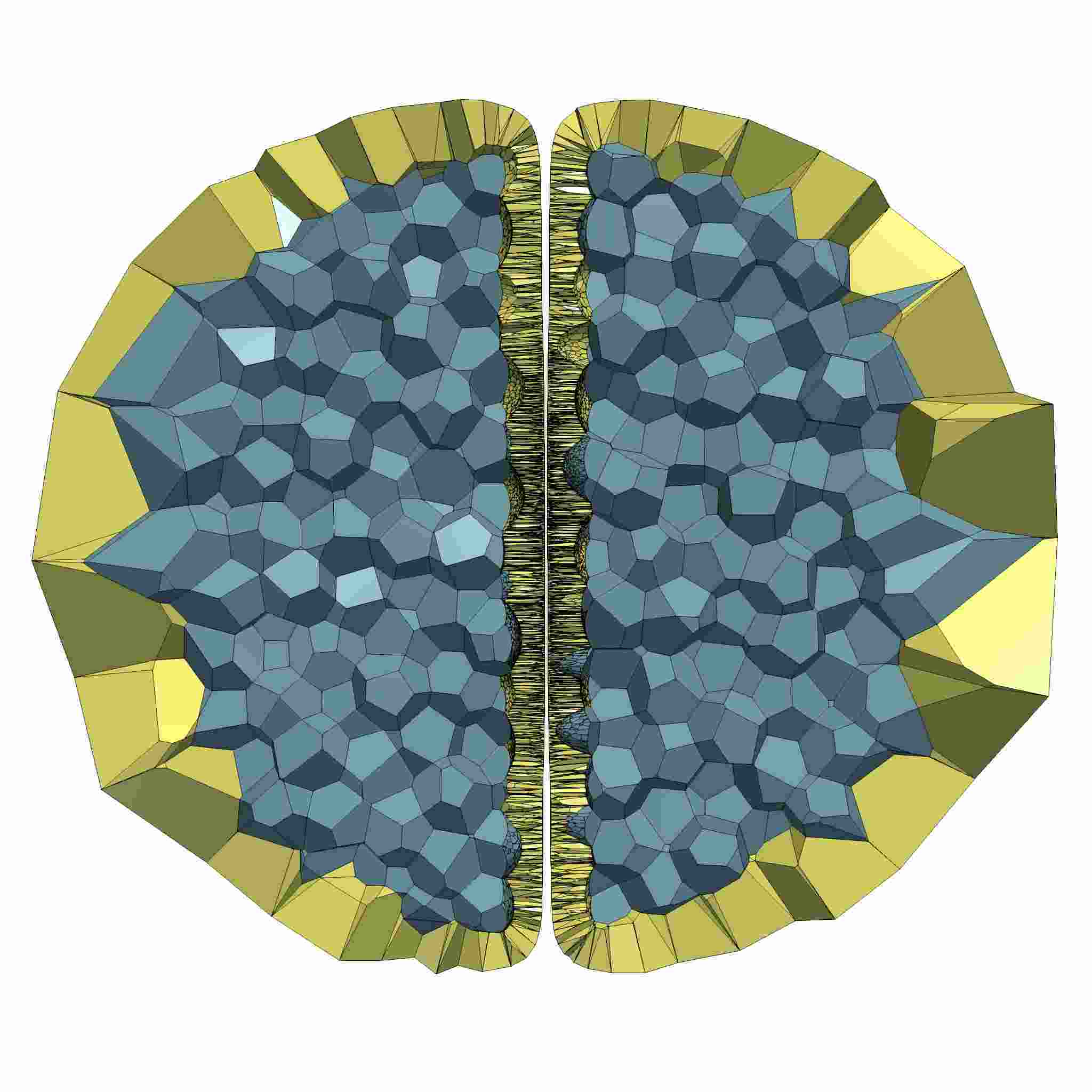}
  \caption{}
\end{subfigure}
\begin{subfigure}[b]{0.13\linewidth}\centering
  \includegraphics[width=1\textwidth]{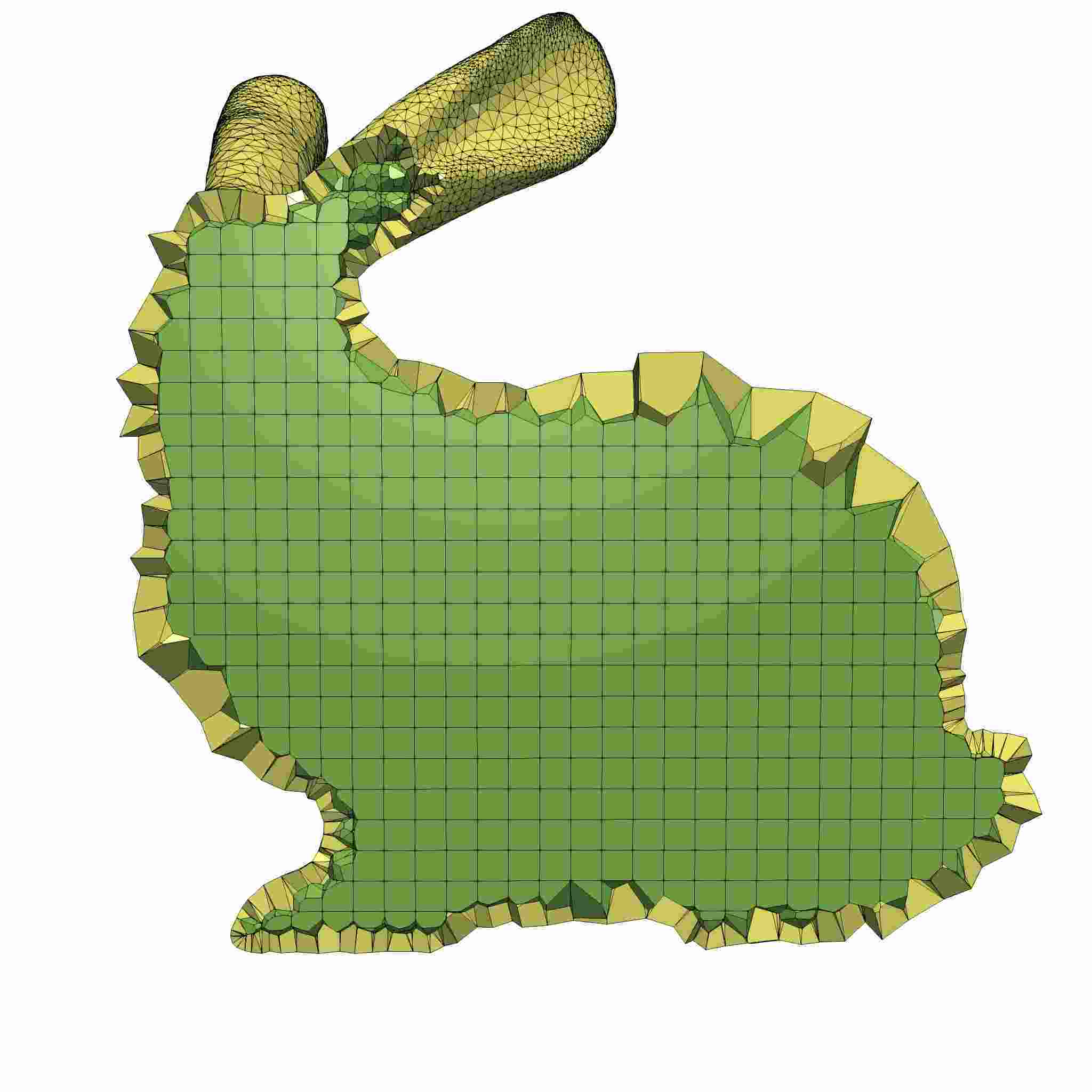}
  \caption{}
\end{subfigure}
\begin{subfigure}[b]{0.13\linewidth}\centering
  \includegraphics[width=1\textwidth]{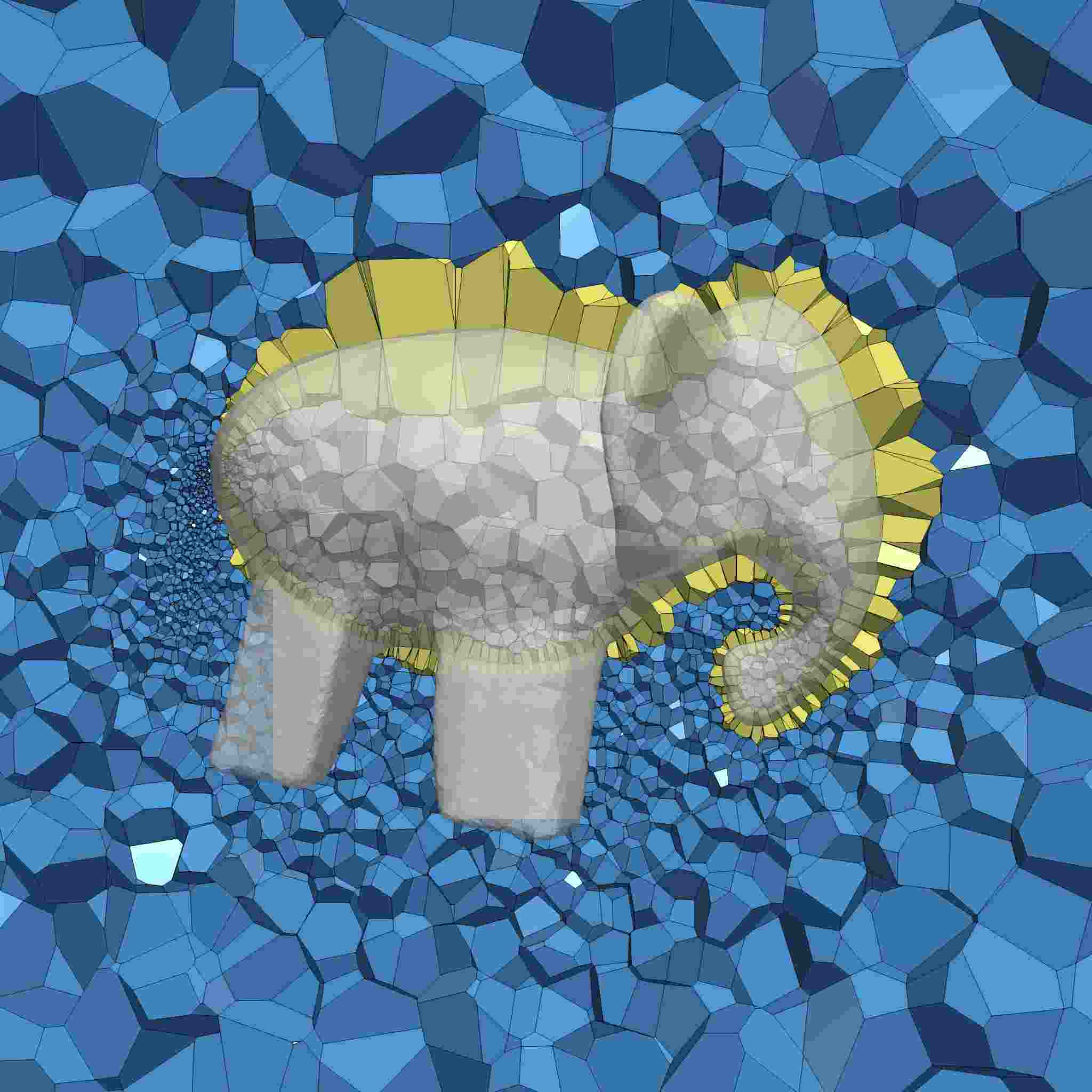}
  \caption{}
\end{subfigure}
\vspace{-10pt}
\caption{VoroCrust can handle inputs having both smooth (a) and sharp (b) features as well as complex topology (c), multi-layers interfacing different types of materials (d), and multiple components (e). The enclosed volume is decomposed into convex unclipped Voronoi cells which can be optimized by CVT (e), controlled to exhibit dominant lattices structures (f), or generated by randomly-sampled seeds (g).}
\label{fig:vc_features}
\vspace{-10pt}
\end{figure*}


An abstract version of the VoroCrust algorithm for smooth manifold surfaces was recently analyzed by Abdelkader et al.~\shortcite{VC_SoCG}. Assuming access to the local feature size with an $\epsilon$-sampling given as input, strong theoretical guarantees on the output quality were established~\cite{VC_SoCG}. In contrast, this paper describes a practical realization of the VoroCrust algorithm for domains with non-manifold boundaries exhibiting arbitrarily sharp features and narrow regions. The VoroCrust refinement produces a union of balls that protects all sharp features while satisfying similar properties to the one analyzed by Abdelkader et al.~\shortcite{VC_SoCG}. Hence, we retain all the approximation and quality guarantees they established, except in the vicinity of sharp features where quality bounds necessarily deteriorate to conform to those features. Furthermore, certain ball configurations yield undesirable sliver artifacts in the output surface and their elimination was left as future work~\cite{VC_SoCG}. The proposed VoroCrust algorithm provably eliminates all such slivers.

The simpler related problem of generating a Voronoi mesh that conforms exactly to restricted classes of piecewise-linear complexes was studied earlier by Abdelkader et al.~\shortcite{shattering2017}. The approach they adopted uses simple rules for the placement of Voronoi seeds to reproduce an input piecewise-linear complex as a set of Voronoi faces similar to earlier works on conforming Delaunay meshing~\cite{MMG2000,Cohen-Steiner:2002,Rand_Collars}. In contrast, VoroCrust always retains the topology of the domain but is not restricted to conform exactly to the boundary; it effectively performs remeshing to improve the output quality within the tolerance specified by the input parameters. \newline

\noindent \textbf{Meshing Piecewise-smooth Complexes.} Delaunay refinement (DR) is a very successful algorithm for the generation of quality unstructured tetrahedral meshes~\cite{cheng2012delaunay}. Since the presence of small angles in the input domain may threaten the termination of DR, a lower bound on input angles may be necessary. A series of works extended DR to more general classes of domains starting with polyhedral domains with no input angles less than $90^\circ$~\cite{shewchuk1998tetrahedral}, and then polyhedral domains with arbitrarily small angles~\cite{Cheng2006}. Motivated by scientific applications dealing with realistic physical domains and engineering designs, the class of inputs with curved boundaries is particularly relevant as treated in~\cite{oudot2010,Tournois:2009} and implemented in the CGAL library~\cite{CGAL}; albeit with assumed lower bounds on the smallest angle in the input. The issue of arbitrarily small input angles was finally resolved by Cheng et al.~\shortcite{Cheng2007} for a large class of inputs called piecewise-smooth complexes. Cheng et al.~\shortcite{Cheng2007} achieved that by deriving a feature size that blends the definitions used for smooth and polyhedral domains, ensuring the protection of sharp features. However, their algorithm is largely impractical as it relies on expensive predicates evaluated using the equations of the underlying surface. To obtain a practical variant as implemented in the DelPSC software, Dey and Levin~\shortcite{a2041327} relied on an input threshold to guide refinement, where topological correctness can only be guaranteed if it is sufficiently small. Another issue with using such a threshold is the uniform sizing of the output mesh, since adaptive sizing requires better sensitivity to the underlying surface. In contrast, the proposed VoroCrust refinement leverages the quality of the input mesh to automatically estimate a sizing similar to the one defined by Cheng et al.~\shortcite{Cheng2007,Cheng2010}; this enables VoroCrust to retain the superior guarantees they established while being practical as shown in our results.

\subsection{Contributions}
\label{sec:contributions}
VoroCrust is the first algorithm for conforming polyhedral Voronoi meshing that can handle a large class of domains with both curved boundaries and arbitrarily sharp features. VoroCrust circumvents the need for clipping, which is the current standard for polyhedral Voronoi-based meshing, successfully avoiding its drawbacks. VoroCrust has the flexibility of decomposing the interior into convex Voronoi cells using either structured or randomly generated seeds. The resulting seeds compactly and uniquely encode the Voronoi mesh, which can be explicitly constructed on-the-fly in a local fashion.

In a broader sense, VoroCrust is one of the first robust and efficient algorithms for polyhedral meshing. In particular, the VoroCrust output consisting of true unweighted Voronoi cells decomposes the domain into convex cells and comes with an orthogonal dual Delaunay tetrahedralization. Such convex decompositions and primal-dual mesh pairs are very useful, and sometimes necessary, in many applications.

The crux of the algorithm is a robust and well-principled refinement process that converges to a suitable sizing function enabling the placement of Voronoi seeds to approximate the surface while preserving all sharp features. VoroCrust estimates sizing through refinement~\cite{Rand2008} as applied in modern meshing frameworks~\cite{Tournois:2009}. This paradigm has proven more efficient than the more traditional approach based on medial axis approximations, e.g., Alliez et al.~\shortcite{Alliez:2005}. The advantage of VoroCrust output is demonstrated by an extensive comparison against state-of-the-art polyhedral meshing methods based on clipped Voronoi cells~\cite{yan2010efficient}; see Figures~\ref{fig:teaser} and~\ref{fig:RVD_nonconvex}.

The practicality of the proposed VoroCrust algorithm and the quality of its output further stem from additional design ingredients to speed up various computations while satisfying the requirements on sampling. We demonstrate the performance of the algorithm through a variety of challenging models, see Figure~\ref{fig:vc_features}, and include a comprehensive parameter study to test the algorithm at the limits.

\section{The VoroCrust Algorithm}
\label{sec:algorithm}

Given a representation of a domain $\vol$, the algorithm produces a boundary-conforming Voronoi decomposition. The crux of the algorithm is the generation of a set of weighted surface samples corresponding to a set of balls $\ballset$ whose union $\ballunion = \cup \ballset$ approximates the boundary $\surf = \partial \vol$. Specifically, $\ballunion$ covers $\surf$ and has the same topology. In addition, $\ballunion$ captures the sharp features of $\surf$. To further guarantee the quality of surface approximation, the radii of surface balls vary smoothly and are sufficiently small w.r.t. the local curvature of $\surf$. In other words, the radii of balls in $\ballset$ mimic a local feature size for $\surf$. Finally, certain configurations of balls are perturbed to eliminate undesirable artifacts in the output surface mesh. These requirements are used to design a refinement process that converges to a suitable union of balls. The conforming surface mesh is obtained by essentially dualizing $\ballunion$ to obtain a set of Voronoi seeds $\surfaceSeeds$. Once $\ballunion$ is obtained, the interior is easily meshed by sampling additional seeds $\interiorSeeds$ outside $\ballunion$. The output mesh can then be computed as a subset of the Voronoi diagram of the seeds in $\surfaceSeeds \cup \interiorSeeds$ without any clipping. In the remainder of this section, we elaborate on these steps per the high-level pseudocode in Algorithm~\ref{alg:one} and Figure~\ref{fig:steps}.

\subsection{Input}
\label{sec:input}
VoroCrust can handle a domain $\vol$ having as boundary a piecewise-smooth complex (PSC) $\surf$ that can be either manifold or non-manifold. The boundary PSC $\surf$ possibly contains \emph{sharp features} where the normal to the surface does not vary smoothly. We make no assumption on how small the input angles might be at such sharp features. VoroCrust guarantees the preservation of all sharp features; sharp corners appear exactly as vertices, while sharp creases are approximated by a set of edges. \newline

\noindent \textbf{Input Mesh.} The algorithm takes as input a watertight piecewise-linear complex (PLC) $\dsurf$ approximating the boundary $\surf$. As in \cite{Dey2010}, we assume that $\dsurf$ approximates $\surf$ in terms of both the Hausdorff error and the surface normals; this enables various predicates to be evaluated using the input PLC rather than the equations describing the underlying PSC \cite{Cheng2010}. In particular, we assume that all dihedral angles in the input mesh, except at sharp features, are at least $\pi - \vcThetaS$, where the \emph{smoothness threshold} $\vcThetaS > 0$ is an implicit design parameter. For the current implementation, we assume $\dsurf$ is a triangle mesh with no self-intersection. Well-established methods can be used to obtain such a mesh given a suitable representation of the domain $\vol$ \cite{Tournois:2009,a2041327,TetWild}. \newline

\noindent \textbf{Parameters.} The algorithm also takes the following inputs:
\begin{itemize}
 \item $\rmax$: a sizing field indicating the largest allowed \emph{size} of mesh elements, and defaults to the diameter of $\dsurf$ or $\infty$.
 \item $\vcTheta < \frac{\pi}{2}$: an angle threshold used to identify the \emph{sharp features} in the PLC $\dsurf$ and bound approximation errors.
 \item $L < 1$: a \emph{Lipschitz} parameter that bounds the variation of radii in $\ballset$ and helps speed-up proximity queries.
\end{itemize}

We distinguish the angle parameters $\theta$ by the superscripts inspired from musical notation: $\sharp$ for sharp and $\flat$ for flat.

\setlength{\textfloatsep}{11pt}
\begin{algorithm}[t]
\SetAlgoNoLine
\KwIn{PLC $\dsurf$ approximating the domain $\vol$, sizing field $\rmax$, \\
      {\color{white} eeeeew } and parameters $\vcTheta$ and $L$ {\footnotesize (Section 2.1)}}
$\sharpF \gets$ the set of sharp features w.r.t. $\vcTheta$ {\footnotesize (Section 2.2)} \\
$\ballset \gets$ a set of balls protecting all features in $\sharpF$ {\footnotesize (Section 2.3)} \\
\While{$\ballunion = \cup \ballset$ does not cover $\dsurf$}{
    Add balls to recover the protection of $\sharpF$ and cover $\dsurf$ \\
    Shrink balls violating any ball conditions {\footnotesize (Section 2.3)} \\
    {\color{white} Shrink balls} or forming half-covered seeds {\footnotesize (Section 2.4)} \\
}
$\surfaceSeeds \gets$ pairs of seeds from triplets of balls in $\ballset$ {\footnotesize (Section 2.4)} \\
$\interiorSeeds \gets$ seeds sampled from the interior of $\vol \setminus \ballunion$ {\footnotesize (Section 2.5)} \\
\Return $\surfaceSeeds \cup \interiorSeeds$
\caption{High-level VoroCrust algorithm}
\label{alg:one}
\end{algorithm}

\subsection{Preprocessing}
\label{sec:preprocessing}
Before refinement, VoroCrust indexes the elements of the input PLC $\dsurf$ and enforces the smoothness condition per the parameter $\vcThetaS$. Then, the algorithm constructs a number of data structures for proximity queries against $\dsurf$ and $\ballset$. \newline

\noindent \textbf{Feature Detection.} We define a \emph{sharp edge} as an edge of $\dsurf$ subtending a dihedral angle less than $\pi - \vcTheta$, or any non-manifold edge incident to exactly one or more than two facets. These sharp edges partition the set of facets incident to any fixed vertex into \emph{sectors}. We define a \emph{sharp corner} as a vertex of $\dsurf$ incident to more than two sharp edges, or two sharp edges whose supporting lines make an angle less than $\pi - \vcTheta$, or two facets in the same sector whose normals differ by at least $\vcTheta$. A polyline arising from a chain of connected sharp edges is called a \emph{crease}, and either forms a cycle or connects two sharp corners. The connected components of the boundary containing no sharp features, denoted $\dsurf_S$, are called \emph{surface patches}. The collection of sharp corners, creases and surface patches are collectively referred to as the \emph{strata} of $\dsurf$.

The algorithm uses $\vcTheta$ to test each edge in $\dsurf$, and collects all sharp edges in a set $E$. Then, each vertex is tested using $\vcTheta$ and $E$, and the sharp corners are collected into the set $\sharpF_C$. From $E$ and $\sharpF_C$, connected chains of sharp edges are collected into the set $\sharpF_E$ by flooding through common vertices except for sharp corners. As a byproduct, each crease is given an index and an orientation, applied consistently to all its sharp edges. Similarly, the facets of $\dsurf$ are indexed, oriented and collected into the set of surface patches $\dsurf_S$ by flooding across non-sharp edges. Finally, we set $\sharpF = \sharpF_C \cup \sharpF_E$. \newline

\noindent \textbf{Patch Smoothing.} If the input mesh $\dsurf$ does not satisfy the required bound on dihedral angles in terms of $\vcThetaS$, VoroCrust starts by applying adaptive loop subdivision~\cite{loop1987smooth} to ensure all dihedral angles between neighboring facets in the same surface patch in $\dsurf_S$ are sufficiently large. In our implementation, we run 6 iterations of loop subdivision, applying subdivision adaptively such that facets with all associated dihedral angles larger than $175^\circ$ are not subdivided. Typical values of $\vcThetaS$ resulting from this step range from $10^\circ$ to $15^\circ$. \newline

\noindent \textbf{Proximity Queries.} Upon generating a new sample point $p \in \dsurf$, VoroCrust needs to find the balls in $\ballset$ covering $p$, and estimate its distance to the elements of $\dsurf$ satisfying certain conditions w.r.t. $\vcTheta$. To speed up such queries, the algorithm constructs three \emph{boundary \kdtrees} to index the elements in $\sharpF_C$, $\sharpF_E$ and $\dsurf_S$. The \kdtrees\ for $\sharpF_E$ and $\dsurf_S$ are populated by supersampling the respective elements with a large number of samples proportional to their sizes. Similarly, the balls in $\ballset$ are indexed into three \emph{ball \kdtrees}. When querying the ball \kdtrees\ for balls in the neighborhood of a given point, the $L$-Lipschitzness of ball radii helps to bound the range and overhead of such queries; see the appendix for more details.

\subsection{Ball Refinement}
\label{sec:refinement}
At a high level, the desired union of balls $\ballunion$ has to (1) protect the sharp features of $\dsurf$ as in~\cite{Cheng2010}, and (2) cover $\dsurf$ while matching its topology as in~\cite{VC_SoCG}. VoroCrust achieves this through a set of \emph{ball conditions} imposed on the balls in $\ballset$. Violations of these conditions drive a refinement process which converges to a suitable union of balls. Before describing this process, we introduce a number of definitions and subroutines. \newline

\begin{figure}[H]
 \centering
  \includegraphics[width=0.6\columnwidth]{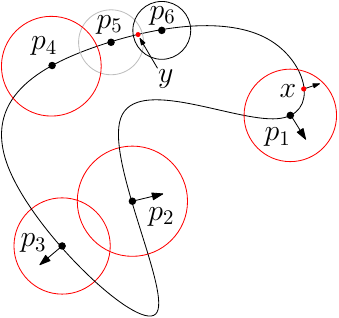}
 \caption{Ball conditions. \smoothCoverage{} is violated at $x$ by $b_{p_1}$. \smoothOverlaps{} is violated by $b_{p_2}$ and $b_{p_3}$. \Lipschitz{} is violated by $b_{p_4}$ and $b_{p_5}$. \DeepCoverage{} is violated at $y$.}
 \label{fig:ball_conditions}
\end{figure}

\noindent \textbf{Smooth Neighborhoods.} As in~\cite{Cheng2010}, we appeal to the curvature of the surface to infer a suitable notion of sizing. Fix a point $x \in \dsurf$ and let $\sigma$ be a face of $\dsurf$ containing $x$. If $\sigma$ is a sharp edge, define $v_{x, \sigma}$ as a unit vector parallel to $\sigma$. If $\sigma$ is a surface patch, define $v_{x, \sigma}$ as a unit vector normal to $\sigma$. $v_{x, \sigma}$ inherits the orientation of the stratum, i.e., the crease or surface patch, containing $\sigma$. A path $\gamma$ lying entirely in a unique stratum $\Sigma$ is called a \emph{smooth path} iff for all $x, y \in \gamma$ we have that $\angle v_{x,\sigma}, v_{y,\tau} \leq \vcTheta$, where $\sigma$ and $\tau$ are the two top-dimensional faces of $\Sigma$ containing $x$ and $y$, respectively. Two points $x, y \in \dsurf$ are called \emph{co-smooth} iff they can be connected by a smooth path. For example, for the curve shown in Figure~\ref{fig:ball_conditions}, if $\vcTheta = \pi/4$, then $p_1$ is not co-smooth with $x$ while $p_5$ is co-smooth with $p_6$. \newline

\noindent \textbf{Ball Conditions.} For a sample point $p \in \dsurf$, let $b_p \in \ballset$ denote the ball centered at $p$ and let $r_p$ denote its radius. The following conditions drive the refinement process and are ensured for $\ballset$ upon termination; see Figure~\ref{fig:ball_conditions}.

\textbf{(\smoothCoverage) Smooth Coverage.} For any $b_p \in \ballset$ and all $x \in b_p \cap \dsurf$, we require that $p$ and $x$ are co-smooth.

\textbf{(\smoothOverlaps) Smooth Overlaps.} For any $b_p, b_q \in \ballset$ s.t. $b_p \cap b_q \neq \emptyset$, we require that $b_p \cup b_q$ contains a smooth path from $p$ to $q$.

\textbf{(\Lipschitz) Local $L$-Lipschitzness.} For any two balls $b_p, b_q \in \ballset$ such that $p, q \in \sharpF_C$, or $p, q \in \sharpF_E$, or $p, q \in \dsurf_S$, we require that $r_p \leq r_q + L \cdot \|p - q\|$.

\textbf{(\DeepCoverage) Deep Coverage.} Fix a constant $\alpha \in (0, 1)$. For all $x \in \dsurf$, we require that $\|x - p\| \leq (1-\alpha) \cdot r_p$ for some ball $b_p \in \ballset$. In addition, we require that $\|p - q\| \geq (1-\alpha) \cdot \max(r_p, r_q)$ for all balls $b_p, b_q \in \ballset$. \newline

\noindent \textbf{Sizing Estimation.} A \emph{sizing} assigns to each new sample $p$ a radius $r_p$. We seek a sizing at most $\rmax$ that satisfies all ball conditions. VoroCrust computes such a sizing by dynamically evolving the assignments $r_p$ for each ball $b_p \in \ballset$ in the course of the refinement process. To speed up convergence, a newly generated ball $b_p$ is initialized with a conservative estimate that is more likely to satisfy all ball conditions. To help avoid \smoothCoverage{} and \smoothOverlaps{} violations, the boundary \kdtrees\ are queried using $p$ to obtain a surrogate point $q^\ast$ for the nearest non-co-smooth point on $\dsurf$. To help avoid \Lipschitz{} violations, the ball \kdtrees\ are queried to find the ball $b_q$ whose center is nearest to $p$. With that, we set $r_p = \min(\rmax(p), 0.49 \cdot \|p - q^\ast\|, r_q + L \cdot \|p - q\|)$. \newline

\noindent \textbf{Termination.} Since VoroCrust uses the PLC $\dsurf$, which only provides a discrete approximation to the PSC $\surf$, and approximates various distance queries, the sizing estimates as defined above may later be found to violate some ball conditions. By similar arguments to those in~\cite{a2041327}, refinement terminates satisfying all ball conditions. The intuition is that for each region on a crease or surface patch, there exists a positive lower bound on ball radii below which neither of the first two conditions can be violated. The refinement process resolves violations by \emph{shrinking} some balls, effectively adjusting all sizing estimates, before recursing to restore protection and coverage. As demonstrated through a variety of challenging models, our algorithm is tuned to avoid excessive refinement; see Section 3. \newline

\noindent \textbf{Sampling Basics.} The refinement process uses Maximal Poisson-Disk Sampling (MPS) \cite{ebeida2012simple,yan2013gap,guo201572} to generate the balls needed to protect the creases and cover the surface patches. The MPS procedure maintains an \emph{active pool}, initialized by all faces on the stratum at hand. To generate a new sample, MPS starts by sampling a face $\sigma$ from the active pool with a probability proportional to its measure, defined as the length for edges and the area for facets. Then, a point $p$ is sampled from $\sigma$ uniformly at random. If $p$ is not covered by the balls in $\ballset$, it is assigned a radius $r_p$ and the ball $b_p$ is added into $\ballset$. Otherwise, $p$ is discarded and a \emph{miss counter} is incremented. Upon counting 100 successive misses, all faces in the active pool are \emph{subdivided} into \emph{subfaces} and the miss counter is reset; edges are split in half and facets are evenly split into four by connecting edge midpoints. Any subface whose points are all deeply covered is discarded, and the remaining subfaces become the new active pool. \newline

\noindent \textbf{Deep Coverage.} For any point $x \in \dsurf$, condition \DeepCoverage{} dictates a stronger form of coverage by the balls in $\ballset$. We say that $x \in \dsurf$ is \emph{$\alpha$-deeply covered} by a ball $b_p \in \ballset$ if $\|p - x\| \leq (1-\alpha) \cdot r_p$; see Figure~\ref{fig:ball_conditions}. We set $\alpha = 1 - \sqrt{3}/2 \approx 0.13$ in our implementation. Equivalently, we require adjacent balls to intersect \emph{deeply}. The reason for that is twofold. First, any point $x$ in the proximity of a crease $\Sigma$ must be closer to the weighted samples on $\Sigma$ than the samples on any other stratum of $\dsurf$~\cite{a2041327}. Second, a sufficient distance between pairs of seeds is needed to bound the aspect ratio of Voronoi cells~\cite{VC_SoCG}. The refinement process ensures \DeepCoverage{} by modifying the coverage test for MPS as follows. First, a new sample is only accepted if it is \emph{not} deeply covered. Second, upon subdividing a face in the active pool, a subface is discarded only if it is completely deeply covered by a single ball with a co-smooth center. Third, the requirements of protecting sharp features prohibit deep overlaps between balls of different types; we elaborate on this further below following the description of our MPS implementation. \newline

\noindent \textbf{Detecting Violations.} Before MPS discards a subface $\sigma$, the algorithm checks for violations of \smoothCoverage{} or \smoothOverlaps, and shrinks encroaching balls as follows. The algorithm starts by finding the nearest sample to $\sigma$ on each stratum using the respective ball \kdtree. Then, the algorithm queries the trees for neighboring balls and checks whether $\sigma$ is deeply covered by any of these balls. For each such ball $b_p$, the algorithm also checks whether $p$ is co-smooth with the points of $\sigma$. If not, the algorithm finds the point $q^\ast \in \sigma$ minimizing the distance to $p$ and shrinks $b_p$ if necessary to ensure $r_p \leq 0.49 \cdot \|p - q^\ast\|$. By ensuring such $b_p$ does not overlap $\sigma$, \smoothCoverage{} violations are avoided. In addition, letting $\tau$ denote the subface containing $p$, any ball $b_q$ with $q \in \sigma$ cannot overlap $b_p$. This effectively avoids \smoothOverlaps{} violations as the algorithm ensures $\max(r_p, r_q) \leq 0.49 \cdot \|p - q\|$ before $\sigma$ and $\tau$ are both discarded.
Finally, whenever the algorithm shrinks a ball, it needs to check for violations of \Lipschitz{} and possibly shrink more balls; the algorithm in~\cite{Tournois:2009} is similar in that regard. However, violations of \Lipschitz{} are not checked during the MPS procedure, which possibly terminates with such violations. As we describe below, enforcing \Lipschitz{} is interleaved with a later step to speed up convergence. \newline

\noindent \textbf{Testing Co-smoothness.} Given two subfaces $\sigma, \tau$ on a stratum $\Sigma$ and a point $p \in \tau$, our implementation uses a more practical test rather than computing smooth paths on $\Sigma$. This test is based on the observation that smooth paths starting at a subface $\sigma$ are confined to small (co)cones of aperture $2\vcTheta$ emanating from the boundary of $\sigma$. In particular, the smooth neighborhood is nearly collinear or coplanar with $\sigma$ if $\Sigma$ is a crease or surface patch, respectively.

The algorithm starts by finding the point $q^\ast \in \sigma$ minimizing the distance to $p$, and sets $v_{pq^\ast} = p - q^\ast$. Then, the co-smoothness test is relaxed to only require that (1) $\angle v_{\sigma, q^\ast}, v_{\tau, p} \leq \vcTheta$ and (2) $\angle v_{\sigma, q^\ast}, v_{pq^\ast} \leq \vcTheta$ if $\Sigma$ is a crease, or $\angle v_{\sigma, q^\ast}, v_{pq^\ast} \leq \frac{\pi}{2} - \vcTheta$ if $\Sigma$ is a surface patch. We argue that this relaxed test suffices for the refinement process to eventually guarantee both \smoothCoverage{} and \smoothOverlaps. Let $\gamma \in \Sigma$ be any path from $p$ to $\sigma$. If $\gamma$ is a smooth path, then the test passes on all subfaces along $\gamma$. Otherwise, the test fails for some subface $\sigma' \in \gamma$. Hence, if no smooth path exists from $p$ to $\sigma$, then every such path $\gamma$ encounters a subface $\sigma'$ for which the test fails before reaching $\sigma$. By applying the relaxed test to every subface $\sigma$ and each ball in a sufficiently large neighborhood around $\sigma$, any remaining violations of \smoothCoverage{} or \smoothOverlaps{} can be detected before MPS terminates. To further validate this claim, we implemented the strict test and verified that both \smoothCoverage{} and \smoothOverlaps{} are always satisfied when MPS terminates. \newline


\begin{figure}
  \begin{minipage}{1\linewidth}
    \centering
    \includegraphics[width=0.3\linewidth]{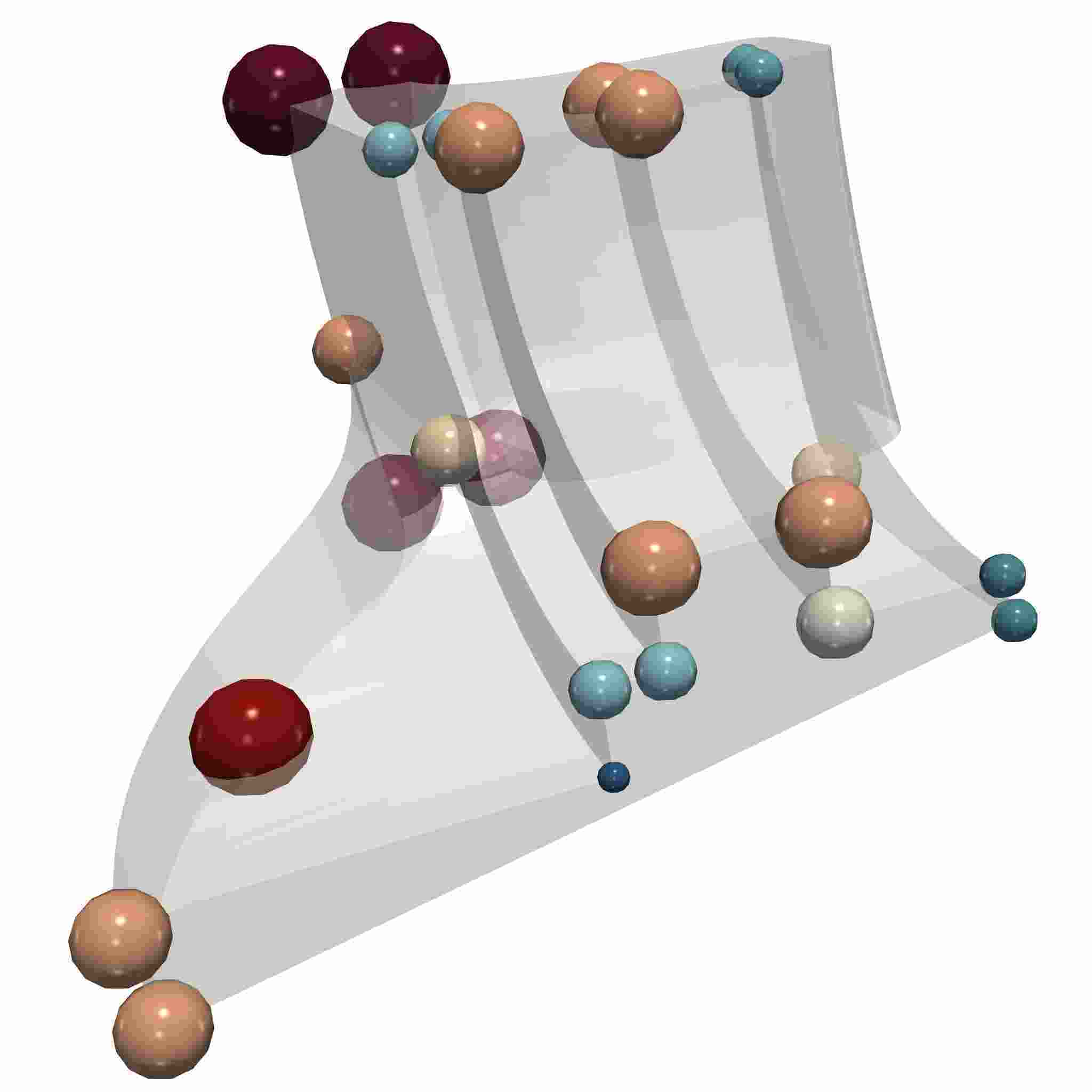}
    \includegraphics[width=0.3\linewidth]{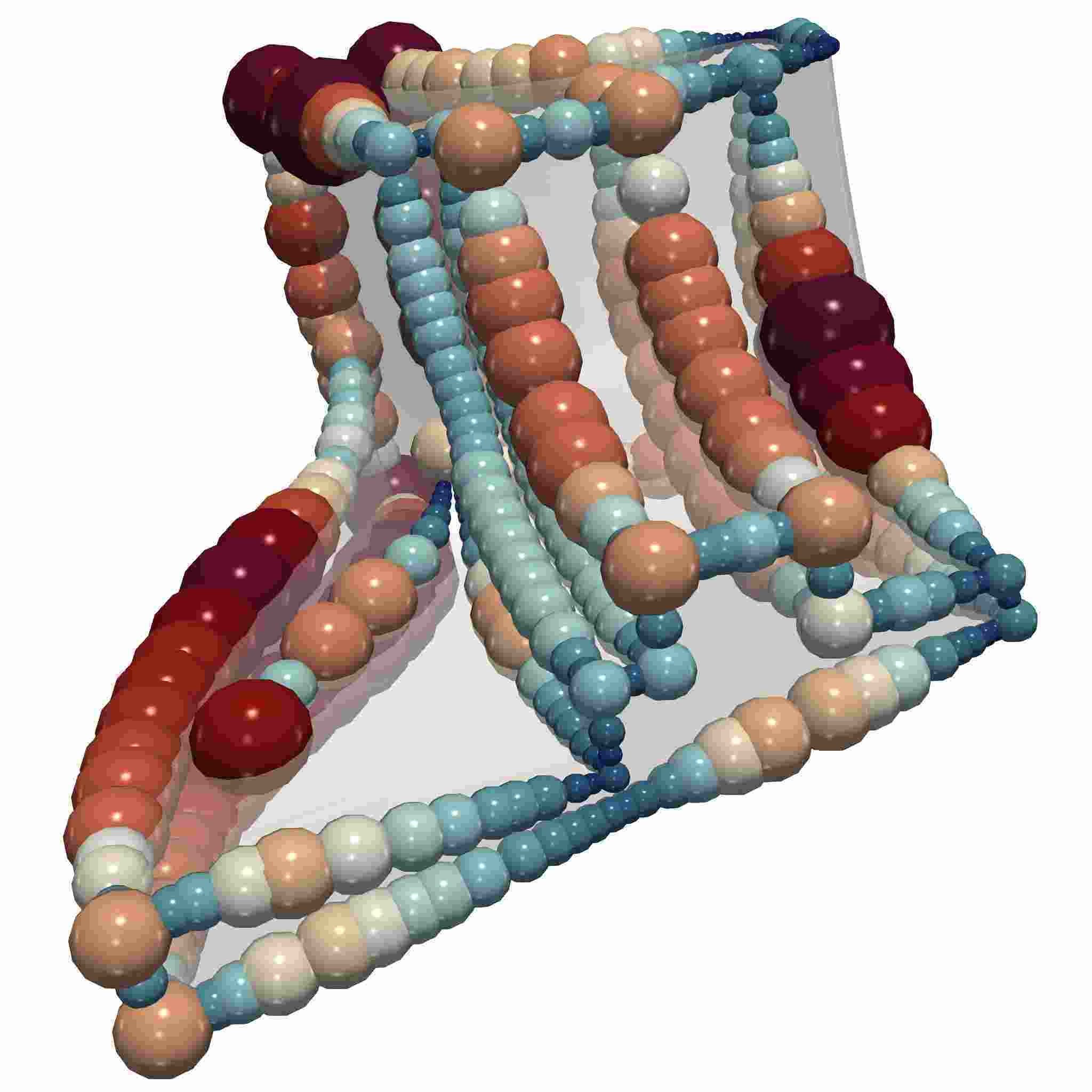}
    \includegraphics[width=0.3\linewidth]{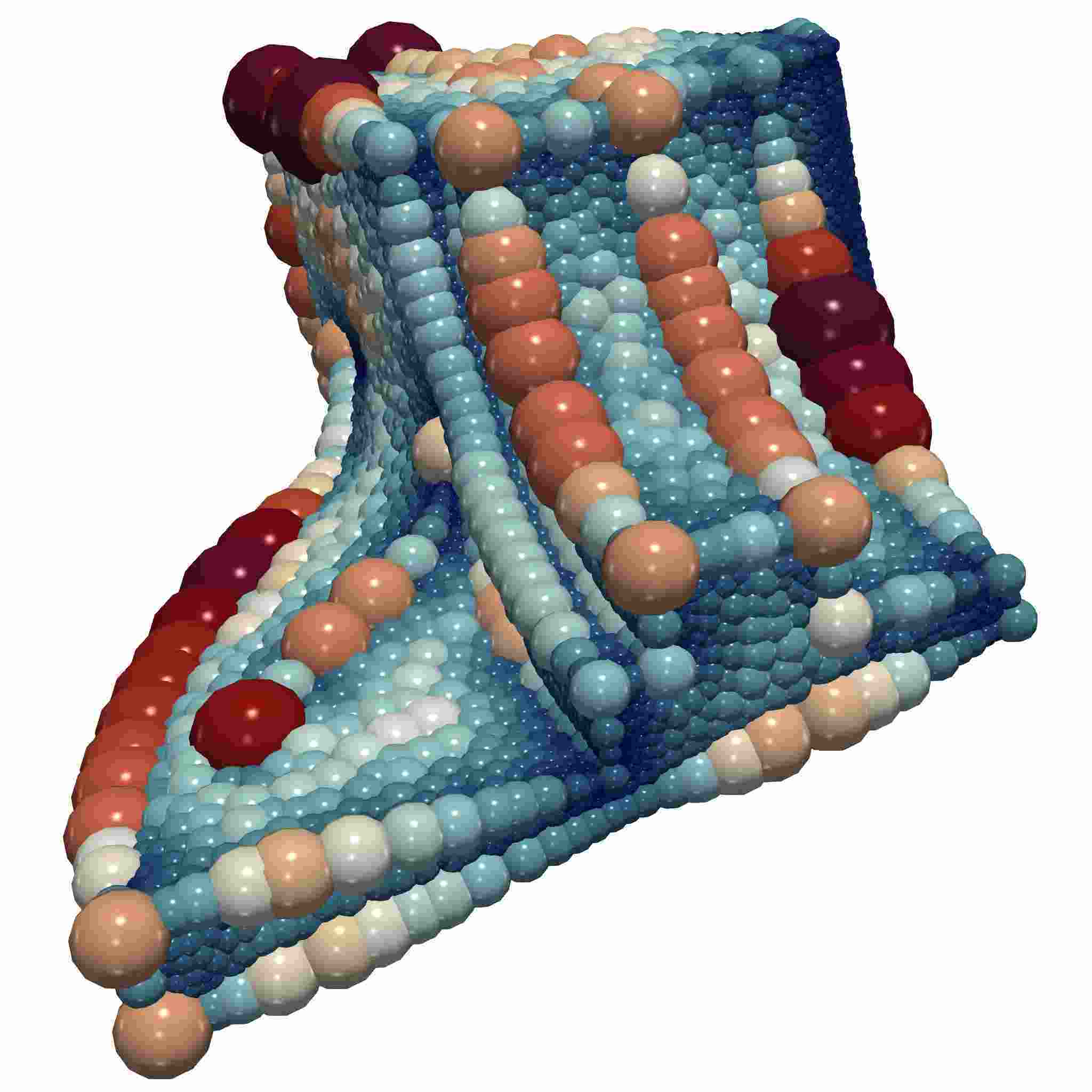} 
  \end{minipage}
  \caption{The three phases of VoroCrust refinement demonstrated on the Fandisk model: protection by corner balls (left) followed by edge balls (center), and finally coverage by surface balls (right).}
  \label{fig:overview_features}
\end{figure}

\noindent \textbf{Protection and Coverage.} The refinement process is realized as a recursive MPS procedure (RMPS) that goes through three phases, ordered by the dimension of the underlying stratum, starting with the protection of sharp corners to the protection of creases and finally the coverage of surface patches; see Figure~\ref{fig:overview_features}. At each phase, if refinement shrinks any of the balls belonging to a previous phase, the algorithm recurses by rerunning RMPS on the affected lower-dimensional strata before proceeding. The process starts by initializing the set of balls with one \emph{corner ball} centered at each sharp corner. As the base case of RMPS, the algorithm enforces \Lipschitz{} among corner balls, shrinking balls as needed. Then, each crease $\Sigma$ is protected by a set of \emph{edge balls} by running RMPS on $\Sigma$. If any corner ball had to be shrunk, RMPS immediately recurses to adjust the corner balls. Whenever RMPS terminates on all creases, the algorithm enforces \Lipschitz{} on all edge balls and reruns RMPS as needed to restore protection. After successfully protecting all sharp corners and creases, the algorithm proceeds to cover each surface patch $\Sigma$ by a set of \emph{surface balls} by running RMPS on $\Sigma$. Similarly, if any corner or edge ball had to be shrunk, RMPS immediately recurses to the respective phase. Finally, the algorithm enforces \Lipschitz{} on surface balls. Before rerunning RMPS as needed to restore protection and coverage, the algorithm perturbs slivers, as we describe in Section~\ref{sec:surface_meshing}; this helps refinement converge in fewer iterations.

We now turn back to the restrictions on overlaps between balls of different type. Whenever a subface encountered by RMPS is completely contained in a corner ball, it is excluded from RMPS in higher phases on neighboring strata. Similarly, whenever a subface is completely contained in an edge ball, it is excluded from RMPS on neighboring surface patches. This is necessary to ensure the protection of sharp features. As a consequence, the deep coverage condition \DeepCoverage{} may be violated in the vicinity of sharp features. This contributes to the deterioration of element quality in these neighborhoods but otherwise does not threaten the termination of the algorithm; see Section~\ref{sec:surface_meshing} and the supplemental materials. \newline

\noindent \textbf{Density Regulation.} Extra care is needed to avoid the well-known clustering phenomenon resulting from the greedy generation of samples. This can be mitigated by biasing the sampling to avoid introducing new sample points near the boundaries of existing balls. In particular, whenever the radius assigned to a new sample $p$ results in the ball $b_p$ violating \DeepCoverage{} by containing an existing sample, $p$ is rejected with a small constant probability; we set this constant to $0.1$ in our implementation. If $p$ is not rejected, $b_p$ is shrunk to ensure it satisfies \DeepCoverage. As demonstrated in Section 3, VoroCrust successfully avoids unnecessarily dense clusters of samples.

\subsection{Surface Meshing}
\label{sec:surface_meshing}
VoroCrust populates the set of \emph{surface seeds} $\surfaceSeeds$ using triplets of overlapping balls in $\ballset$. The bounding spheres of each such triplet intersect in exactly two points on either side of the boundary. The algorithm places one labeled Voronoi seed at each such point as long as it does not lie in the interior of any fourth ball in $\ballset$. Then, the Voronoi facets common to two Voronoi seeds on different sides of the boundary constitute the resulting VoroCrust surface mesh which coincides with the weighted $\alpha$-shape of the samples \AlphaK{} inheriting the topology of $\ballunion$~\cite{AMENTA200125}. The deep coverage condition \DeepCoverage{} guarantees that all samples $p$ appear as vertices in the Voronoi diagram of $\surfaceSeeds$, with at least 4 seeds lying on $\partial b_p$. We point out that VoroCrust effectively remeshes the surface on-the-fly to reduce the complexity of the output within the tolerance specified by the input parameters. The quality of surface elements follows from $L$-Lipschitzness~\cite{VC_SoCG}, with the exception of elements formed by corner or edge balls in the vicinity of sharp features. \newline

\begin{figure}[H]
\centering
\begin{subfigure}[b]{0.48\columnwidth}\centering
  \includegraphics[width=1\columnwidth,trim={5.5cm 1cm 6cm 1cm},clip]{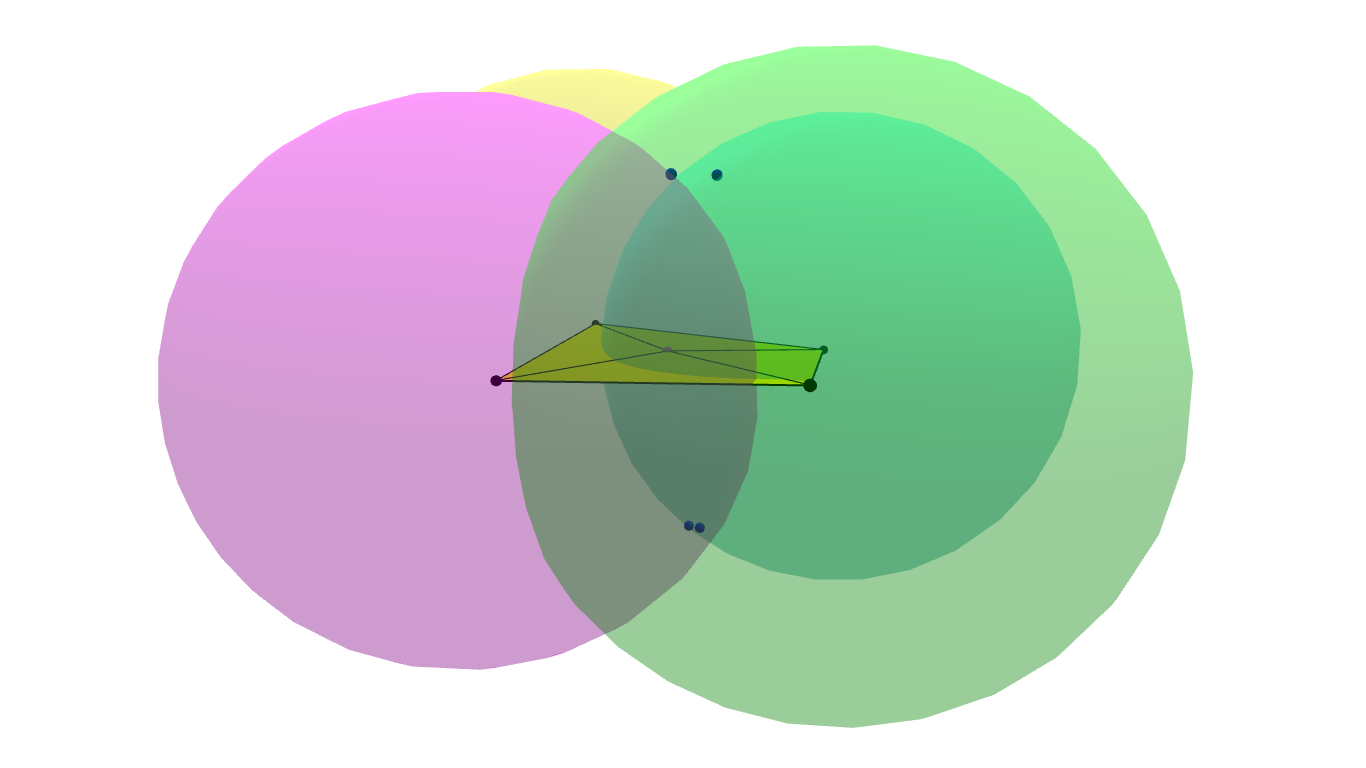}
\end{subfigure}
\begin{subfigure}[b]{0.48\columnwidth}\centering
  \includegraphics[width=1\columnwidth,trim={5.5cm 1cm 6cm 1cm},clip]{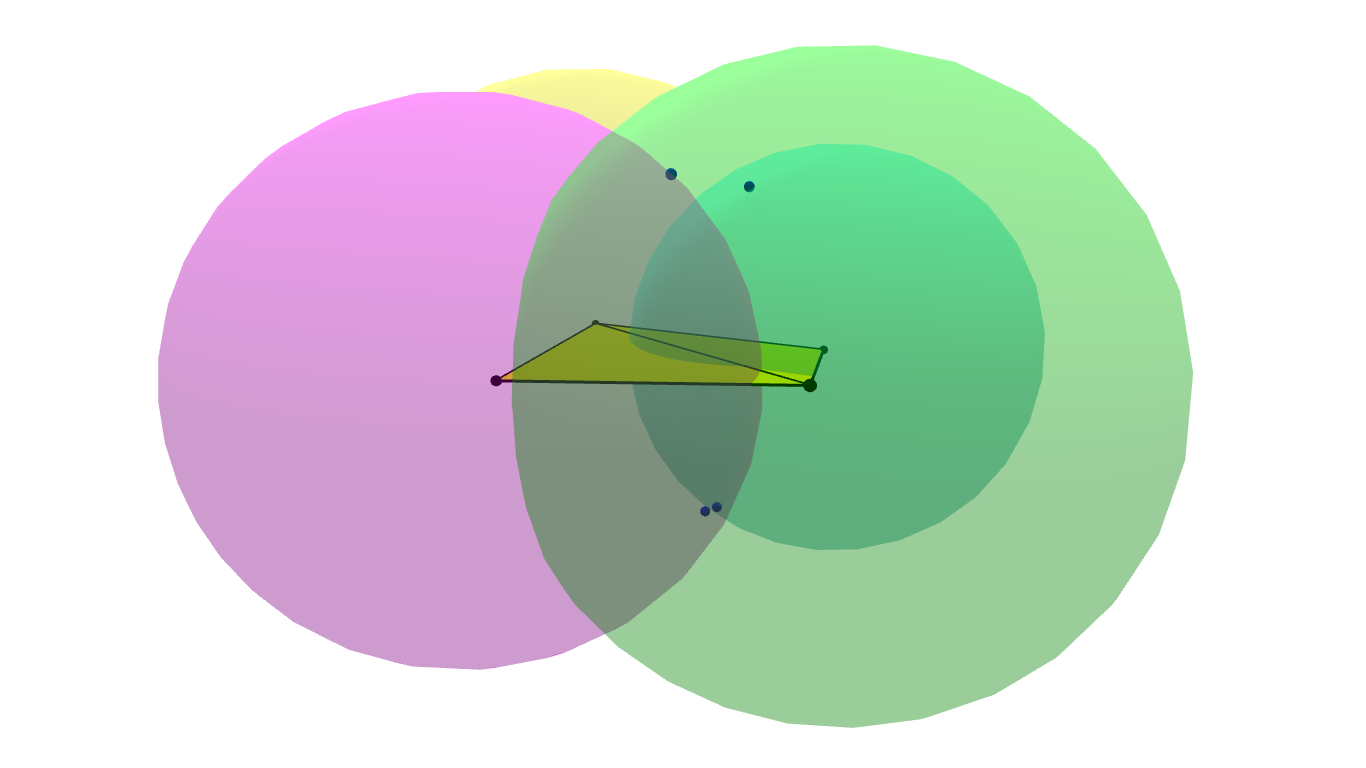}
\end{subfigure}
   \caption{Sliver elimination: (left) A quartet of balls centered at four samples (black) with four half-covered seeds (blue) yielding a Steiner vertex (pink) with four incident facets. (right) Shrinking one ball resolves half-covered seeds eliminating the Steiner vertex to yield only two facets; see the supplemental materials for the numerical values.}
   \label{fig:slivers}
\end{figure}

\noindent \textbf{Sliver Elimination.} VoroCrust applies further refinement to the set of balls $\ballset$ to eliminate undesirable artifacts in the output. When a triplet of overlapping balls yield only one Voronoi seed, we have a \emph{half-covered seed pair}. The four samples yielding the problematic configuration of balls are typically the vertices of a nearly flat tetrahedron appearing as a \emph{regular component} in \AlphaK~\cite{VC_SoCG}; we refer to such regular components as \emph{slivers}. These slivers result in extra \emph{Steiner vertices}, besides the samples, appearing in the Voronoi diagram of the seeds and consequently on the output surface mesh. As these Steiner vertices may not lie on the input surface, their incident Voronoi facets may not be aligned with the surface possibly yielding large deviations in surface normals; see Figure~\ref{fig:slivers}. To eliminate such slivers, the algorithm determines one ball to shrink for each half-covered seed.

For every ball $b_p \in \ballset$, the algorithm queries the ball \kdtrees\ for neighboring balls and collects those overlapping $b_p$ into the set $\ballset_p$. The algorithm iterates over $\ballset_p$ to form triplets of overlapping balls including $b_p$. For each such triplet $t$, the algorithm computes the pair of intersection points on their bounding spheres and tests whether the pair is half-covered by any fourth ball in $\ballset_p$; all candidate fourth balls along with the triplet in $t$ are collected into a secondary set $\ballset_t$. Then, every quartet of balls in $\binom{\ballset_t}{4}$ defining a half-covered seed pair is considered in isolation. For each such quartet, the algorithm determines the ball requiring the least shrinkage to uncover all seeds. Over all quartets in $\binom{\ballset_t}{4}$, the ball requiring the least shrinkage is assigned a smaller radius. For each ball $b$, the algorithm records the smallest radius assigned to $b$ over all quartets it is part of. Once all balls are processed, the algorithm shrinks every ball assigned a smaller radius. Recalling that $L$-Lipschitzness is satisfied for $\ballset$, $|\ballset_p|$ is kept small and the running time of this procedure is linear in $|\ballset|$. The procedure just described eliminates a subset of existing slivers but potentially violates some ball conditions and creates new slivers. The algorithm reruns RMPS to resolve such violations before repeating to eliminate any remaining slivers. \newline

\begin{figure*}[htb]
\centering
\begin{subfigure}[b]{0.32\linewidth}
  \begin{tikzpicture}
    \node {\includegraphics[height=0.8\textwidth,trim={3cm 0 0 0}]{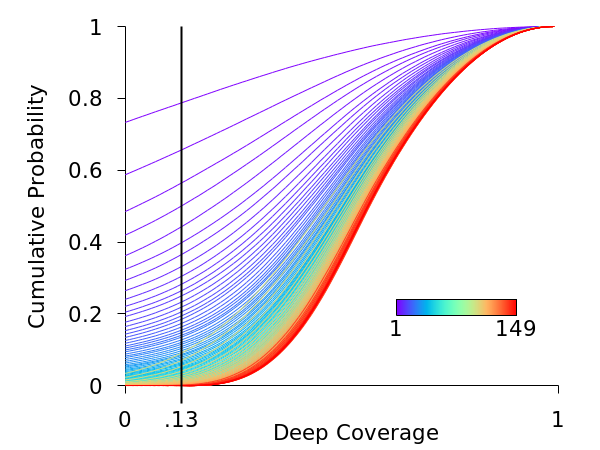}};
    \node at (1.15,-0.55) {{\small $|\ballset| / 100$}};
  \end{tikzpicture}
\end{subfigure}
\begin{subfigure}[b]{0.32\linewidth}\centering
  \begin{tikzpicture}
    \node {\includegraphics[height=0.8\textwidth,trim={2cm 0 2cm 0}]{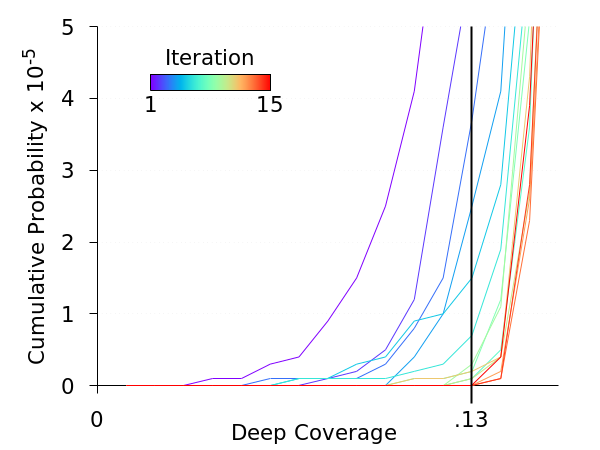}};
  \end{tikzpicture}
\end{subfigure}
\begin{subfigure}[b]{0.32\linewidth}\centering
  \begin{tikzpicture}
    \node {\includegraphics[height=0.8\textwidth, trim={2cm 0 3cm 0}]{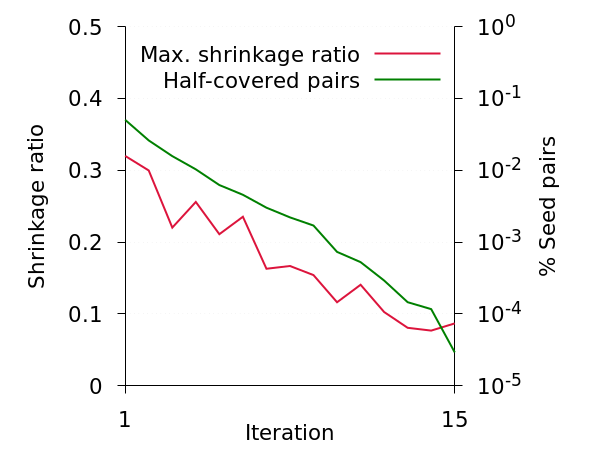}};
  \end{tikzpicture}
\end{subfigure}
    \vspace{-10pt}
   \caption{Empirical analysis of sliver elimination using the Bimba model: (left) evolution of the deep-coverage distribution through the first invocation of RMPS as $\ballset$ grows in increments of 100 balls, (middle) sliver elimination executes 15 iterations where shrinking eventually ceases to violate $\alpha$-deep coverage, (right) the refinement incurred by sliver elimination decreases the maximum shrinkage ratio applied in subsequent iterations. As a result, the number of newly created slivers, measured by the percentage of triplets with half-covered seed pairs, decays rapidly.}
    \label{fig:density_analysis}
\end{figure*}


\noindent \textbf{Termination without Slivers.} Each execution of the above procedure, followed by rerunning RMPS, counts as a single iteration of sliver elimination. The termination of the algorithm requires a finite bound on the number of such iterations, which can be established by bounding the shrinkage that may be applied to any ball through subsequent iterations. The intuition behind this bound is the well-known relationship between increasing the density of sampling and the increased local flatness of the surface approximation. Specifically, shrinkage decreases as the density increases. As it turns out, violations of the deep coverage condition \DeepCoverage{} are the main cause for refinement after shrinking to eliminate slivers. The termination of the algorithm can be guaranteed by accepting a set of balls with no half-covered seeds as long as all boundary points are only $\alpha'$-deeply covered, for some $\alpha' < \alpha$. As we prove in the supplemental materials, the smoothness of the input surface per the parameter $\vcThetaS$ guarantees that shrinkage eventually falls below a threshold that cannot violate $\frac{\alpha}{2}$-deep coverage. \newline

\noindent \textbf{Practical Variant.} Our implementation always reruns RMPS to recover $\alpha$-deep coverage. We argue that this variant terminates with high probability by combining the bounds on shrinkage with the stability of deep coverage as a distribution. In our experiments, VoroCrust always terminates with all slivers eliminated successfully while avoiding excessive refinement; see Section 3. In the unlikely event that sliver elimination fails to terminate in a constant number of iterations, set to 100, we restart in a \textit{safe mode} accepting $\frac{\alpha}{2}$-deep coverage to guarantee termination; we never encountered such cases. \newline

\begin{wrapfigure}[7]{r}{2.5\columnsep}
    \vspace{-1\intextsep}
    \hspace*{-0.75\columnsep}
    \includegraphics[width=0.3\columnwidth,trim={0.75cm 0 3.8cm 0},clip]{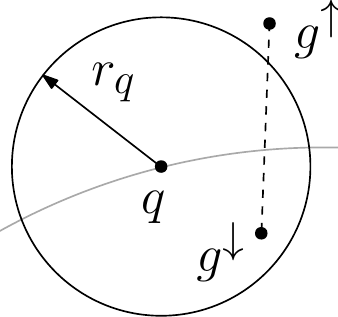}
\end{wrapfigure}

\noindent \textbf{Shrinkage Ratio.} Fix a triplet $t$ and let $g^\uparrow$ and $g^\downarrow$ denote the intersection points of its bounding spheres, such that $t$ has a half-covered seed due to a fourth ball $b_q$. Assume w.l.o.g. that $g^\downarrow \in b_q$ while $g^\uparrow \notin b_q$, i.e., $\|q - g^\downarrow\| < r_q$ while $\|q - g^\uparrow\| \geq r_q$; see the inset. To resolve the half-covered seed, the algorithm shrinks $b_q$ by setting its radius to $\|q - g^\downarrow\|$. Hence, the shrinkage is $r_q - \|q - g^\downarrow\| > 0$. As violations of $\alpha$-deep coverage after shrinking are the main cause for further refinement, we consider shrinkage as a ratio of the original radius which we denote by $\Delta$. The above inequalities imply the following bound:
$\Delta = \frac{r_q - \|q - g^\downarrow\|}{r_q} \leq \frac{\|q - g^\uparrow\| - \|q - g^\downarrow\|}{\|q - g^\downarrow\|} = \frac{\|q - g^\uparrow\|}{\|q - g^\downarrow\|} - 1.$
In particular, as $\frac{\|q - g^\uparrow\|}{\|q - g^\downarrow\|}$ approaches $1$, $\alpha$-deep coverage is less likely to be violated after shrinking. Specifically, if $\Delta \leq \frac{\alpha}{\alpha-2}$, then $\frac{\alpha}{2}$-deep coverage holds. Assuming the input $\dsurf$ is sufficiently smooth per $\vcThetaS$, this observation guarantees the termination of the algorithm if $\frac{\alpha}{2}$-deep coverage is accepted; see the supplemental materials for the proof and further discussion. \newline

\noindent \textbf{Decaying Shrinkage and Violations.} Subsequent invocations of RMPS in the course of sliver elimination increase the density of sampling. A consequence of the ball conditions maintained by RMPS is that the radii of overlapping balls get smaller. In particular, the deviation in normals at the centers of overlapping balls gets smaller, which is equivalent to enforcing the smooth overlap condition \smoothOverlaps{} with a smaller angle threshold. Intuitively, the neighborhood of each sample becomes \emph{nearly flat}. This flatness increases the ratio $\frac{\|q - g^\uparrow\|}{\|q - g^\downarrow\|}$ for all nearby samples $q$, which reduces the shrinkage ratio $\Delta$ and restricts the potential locations of new samples that create new slivers. It follows that the percentage of triplets with half-covered seed pairs decays rapidly; see Figure~\ref{fig:density_analysis}(right). \newline

\noindent \textbf{Deep-coverage Distribution.} Let $f_i$ be a function that maps each $x \in \dsurf$ to $\max\{1 - \frac{\|x - p\|}{r_p} \mid b_p \in \ballset_{i, x}\}$ where $\ballset_{i, x}$ is the subset of balls containing $x$ at iteration $i$. We use the family of functions $\{f_i\}$ to define the deep-coverage distribution as $F_i(\alpha) = \textrm{Pr}[f_i(x) \leq \alpha \mid x \in \dsurf]$ with $\alpha \in [0, 1]$. We estimate $F_i$ by the empirical distribution function over 100 bins using independent random samples of $10^6$ points. Figure~\ref{fig:density_analysis}(left) shows the evolution of the deep-coverage distribution through the first invocation of RMPS until convergence. Every subsequent invocation of RMPS, following shrinking for sliver elimination, converges to a nearly identical distribution. Related aspects of the distributions of MPS samplings were analyzed~\cite{mitchell2012variable}, which are consistent with our experiments\footnote{The total variation distance~\cite{DasGupta08} between the empirical distributions obtained through all subsequent iterations is at most $0.02$.}; see the supplemental materials for further examples and discussion. As seen in Figure~\ref{fig:density_analysis}(middle), shrinking for sliver elimination initially violates $\alpha$-deep coverage, per \DeepCoverage{} requiring a fixed $\alpha \approx 0.13$, but causes no such violations over the last few iterations. The combination of decaying shrinkage and the stability of deep coverage as a distribution bounds the probability of such violations. It follows that subsequent invocations of RMPS are less likely to introduce new balls to recover $\alpha$-deep coverage. As a result, the number of newly created slivers per iteration decays rapidly; see Figure~\ref{fig:density_analysis}(right). Hence, the total number of slivers encountered by the algorithm is bounded in expectation, which implies termination in a finite number of steps with high probability.

\subsection{Volume Meshing}
\label{sec:volume_meshing}
Once the refinement process terminates, the set of balls $\ballset$ is fixed and a conforming surface mesh can be generated. To further decompose the interior into a set of graded Voronoi cells, additional weighted samples $\interiorSeeds$ are generated in the interior of the domain. Similar to $\ballset$, the balls corresponding to interior samples are required to satisfy the $L$-Lipschitzness condition. Standard MPS may be used for sampling the interior. However, to reduce the memory footprint of this step, the spoke-darts algorithm~\cite{spokedarts18} is used instead following a lightweight initialization phase using standard dart-throwing; see the appendix for more details. Alternatively, the interior samples may be chosen as the vertices of a structured lattice. This can be used to output a hex-dominant mesh conforming to the surface; see Figure~\ref{fig:vc_features}(f). The quality of the volume mesh can be further improved by applying CVT optimization to the set of interior seeds; see Figure~\ref{fig:vc_features}(d).

\subsection{Meshing 2D Domains}
\label{sec:2d}
The proposed VoroCrust algorithm can readily be applied to the decomposition of 2D domains into conforming Voronoi meshes. As illustrated in Figure~\ref{fig:steps}, the seed placement strategy can be applied in 2D given a suitable union of balls. The refinement strategy described in this section can easily be applied to generate such a union of balls by regarding the 2D boundary as a set of creases embedded in 3D. In particular, assuming the 2D boundary is available as a set of line segments or a planar straight-line graph (PSLG) as common in 2D meshing, the input segments can be mapped to 3D by adding a third coordinate, e.g., $z = 0$, to all end points. The ball conditions and refinement process for the protection of sharp features, as defined in Section~\ref{sec:refinement}, guarantee a union of balls that approximates the embedded 2D boundary.
\begin{figure}[htb]
  \begin{minipage}{1\linewidth}
    \centering
    \includegraphics[width=0.4\columnwidth]{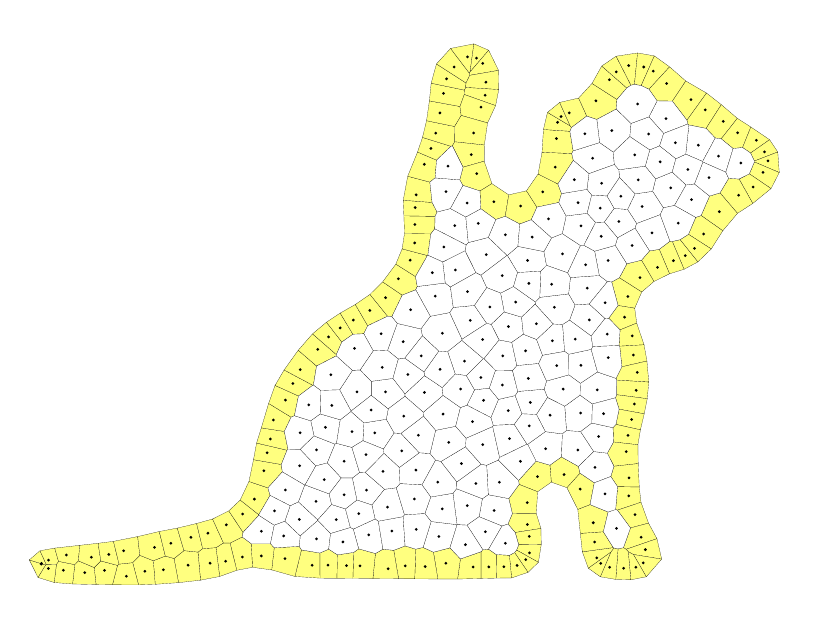}
    \quad
    \includegraphics[width=0.35\columnwidth]{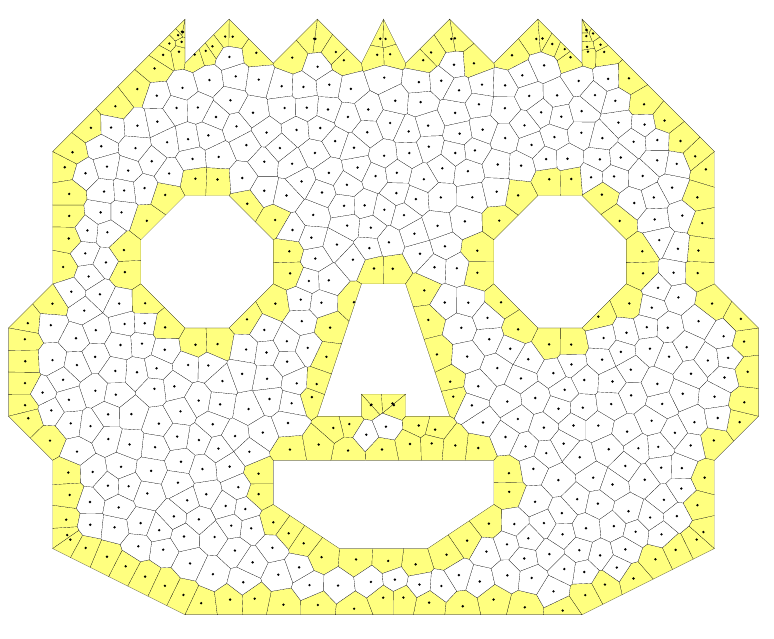}
  \end{minipage}
    \caption{The VoroCrust algorithm readily handles 2D domains.}
    \label{fig:vc_2d}
  \vspace{-6pt}
\end{figure}


Such a union of balls can be used to place Voronoi seeds in 2D as follows. First, all balls are projected onto the 2D plane as circles centered along the boundary. Then, the pairs of intersection points between consecutive circles are computed. Recalling that the edge balls protecting any given crease may only overlap consecutive balls along the same crease, these pairs of intersection points are well-defined. Once the intersection pairs are obtained, the algorithm places Voronoi seeds across the 2D boundary and proceeds to sample additional seeds to mesh the 2D interior. Figure~\ref{fig:vc_2d} shows a number of conforming 2D Voronoi meshes, with uniform sizing in the interior, obtained by a 2D implementation of VoroCrust.
\section{Results}
\label{sec:results}
We demonstrate the capabilities of the VoroCrust algorithm and study the impact of input parameters. Then, we compare against the work of Yan et al.~\shortcite{yan2010efficient} as a representative of state-of-the-art clipping-based methods. All experiments were conducted on a Mac Pro machine with a 3.5 GHz 6-Core Intel Xeon E5 processor and 32 GB of RAM. \newline

\begin{figure}[h]
	\centering
	
\setlength{\tabcolsep}{3pt}
\resizebox{\linewidth}{!}{
	\begin{tabular}{cc}
	\rotatebox{90}{\qquad \qquad \qquad \qquad \begin{Huge}Bimba\end{Huge}}
     \includegraphics[width=1.0\linewidth]{figs/smooth/bimba_vc.jpg} &
     \includegraphics[width=1.0\linewidth]{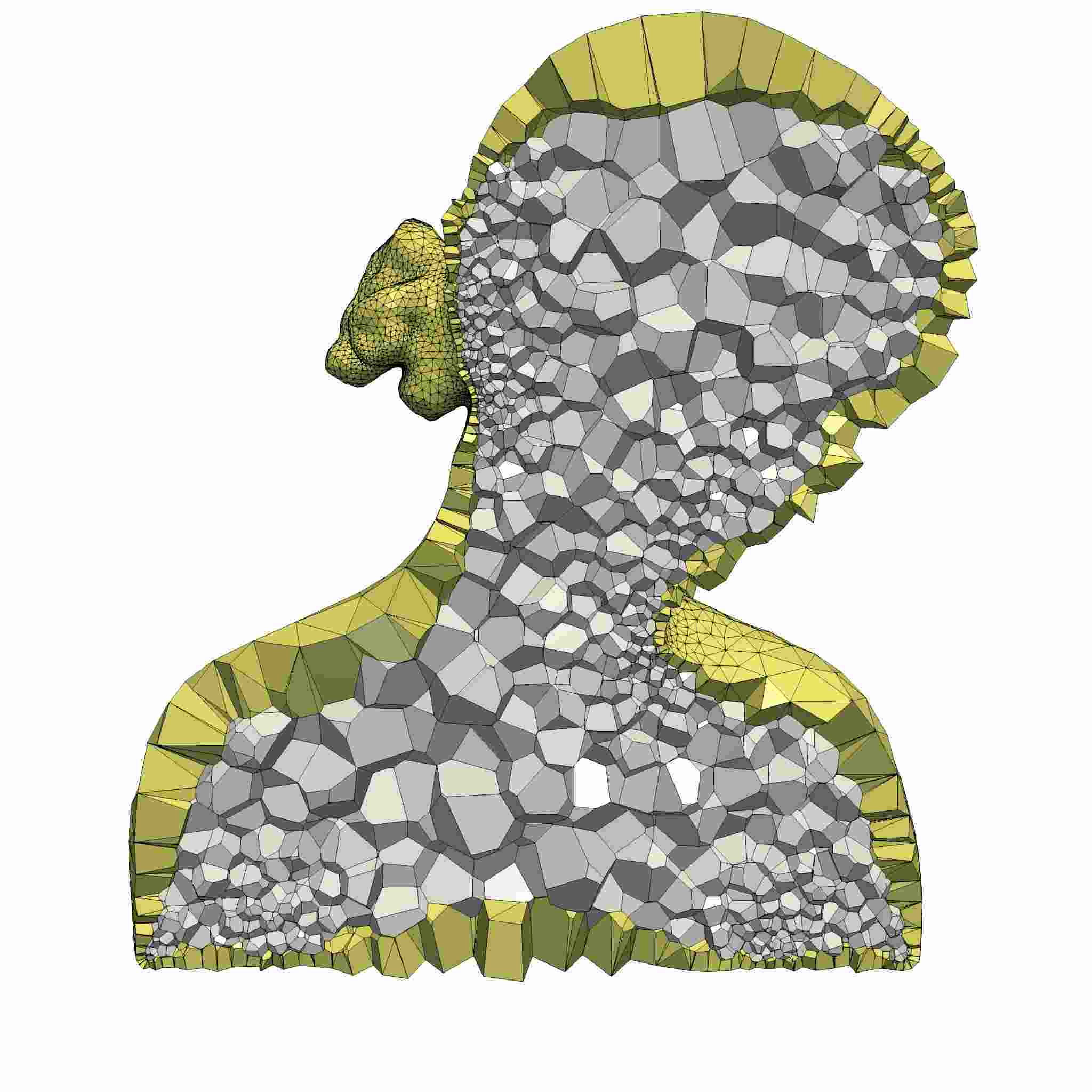}\\	
	\end{tabular}}
\setlength{\tabcolsep}{3pt}
\resizebox{\linewidth}{!}{
	\begin{tabular}{ccccccccc}
  	    $\theta_{<30}\%$& $\theta_{>90}\%$& $Q_{min}$&$\rho_{max}$& $d_{H}(\times 10^{-2})$ & $\surfaceSeeds$ & $\interiorSeeds$ & $\surfaceTime$ & $\interiorTime$ \\
     	$2$& $16$& $0.373$ & $5.345$ & $0.614$ & $68472$ & $17035$ &$935$ & $587$\\
		\bottomrule
	\end{tabular}}

\setlength{\tabcolsep}{3pt}
\resizebox{\linewidth}{!}{
	\begin{tabular}{cc}
	\rotatebox{90}{\qquad \qquad \qquad \qquad \begin{Huge}Loop\end{Huge}}
	\includegraphics[width=1.0\linewidth]{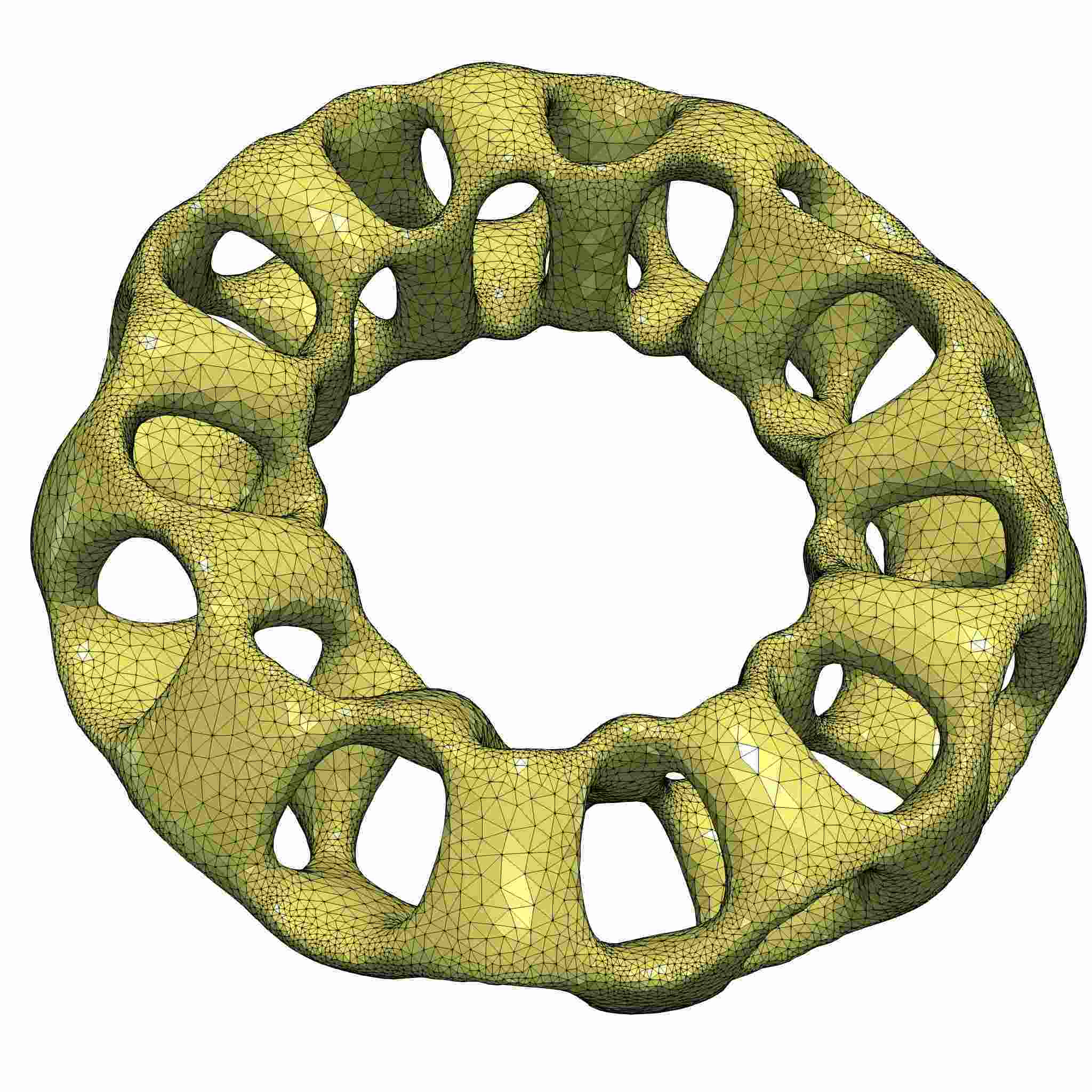} & 
	\includegraphics[width=1.0\linewidth]{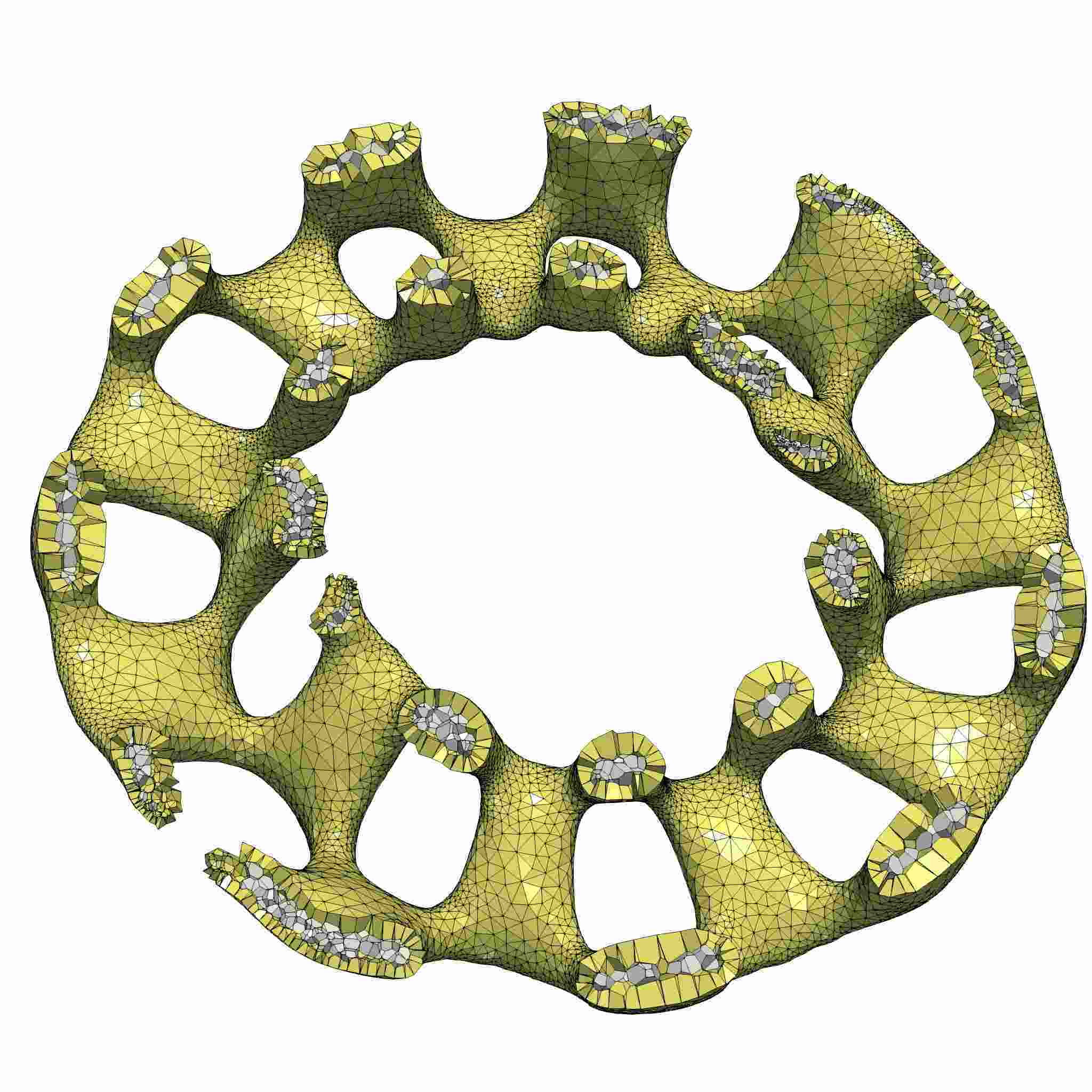}\\	
	\end{tabular}}
\setlength{\tabcolsep}{3pt}
\resizebox{\linewidth}{!}{
	\begin{tabular}{ccccccccc}
  	    $\theta_{<30}\%$& $\theta_{>90}\%$& $Q_{min}$&$\rho_{max}$& $d_{H}(\times 10^{-2})$ & $\surfaceSeeds$ & $\interiorSeeds$ & $\surfaceTime$ & $\interiorTime$ \\
		$2$ & $16$ & $0.383$ & $5.407$ & $0.171$ & $114472$ & $6726$ & $1581$ & $1363$ \\
		\bottomrule
	\end{tabular}}

\setlength{\tabcolsep}{3pt}
\resizebox{\linewidth}{!}{
	\begin{tabular}{cc}
	\rotatebox{90}{\qquad \qquad \qquad \begin{Huge}Close Hemispheres\end{Huge}}
	\includegraphics[width=1.0\linewidth]{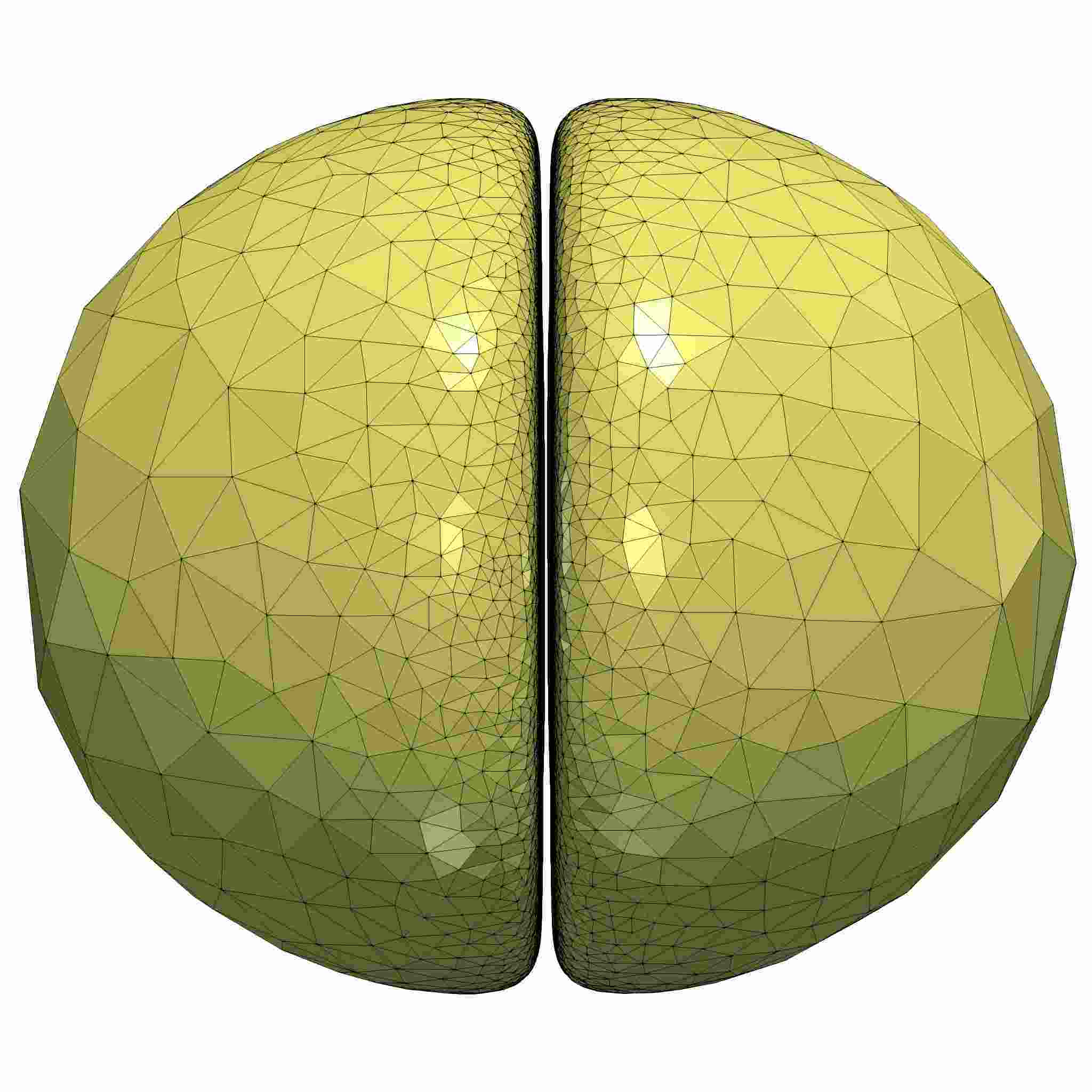} & 
	\includegraphics[width=1.0\linewidth]{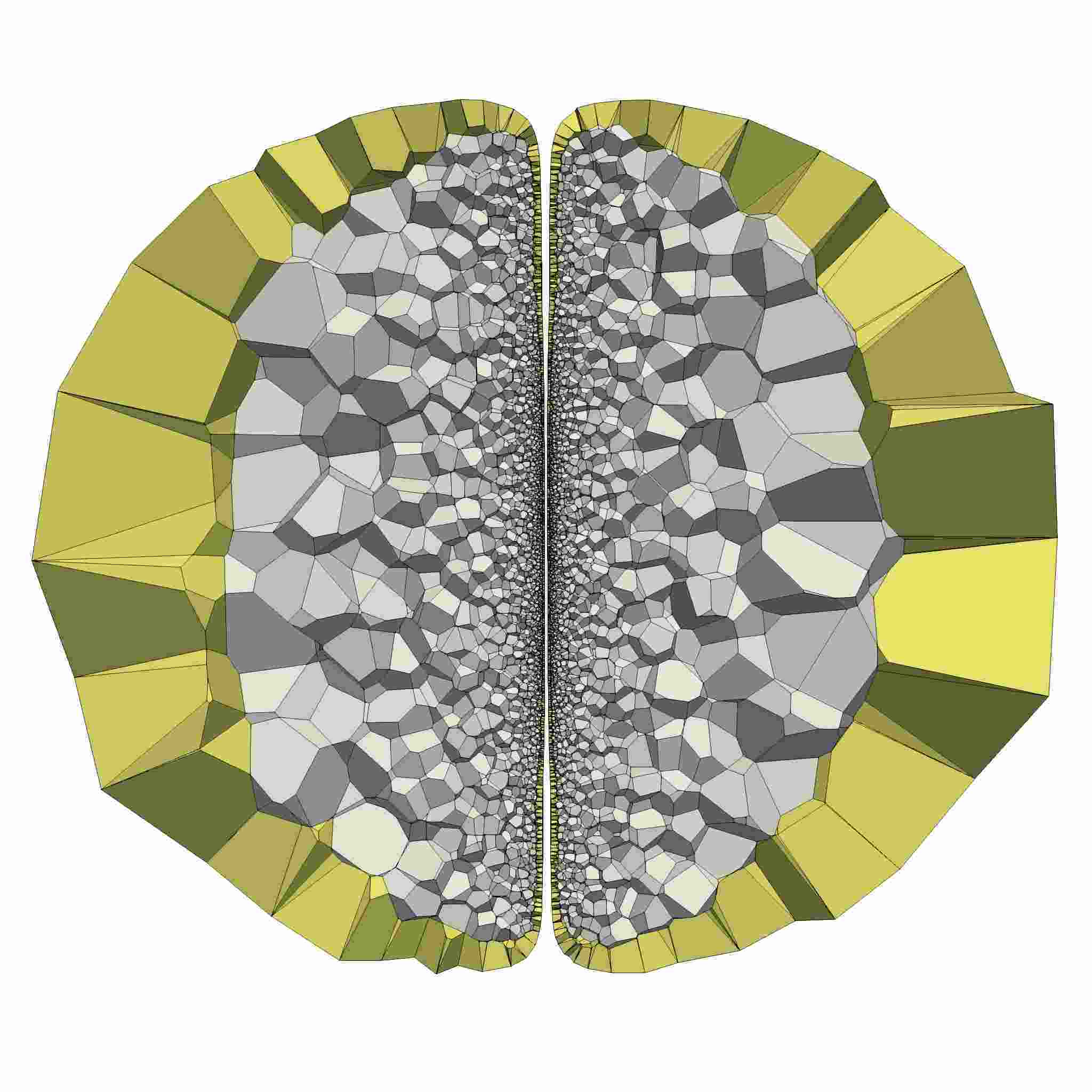}\\	
	\end{tabular}}
\setlength{\tabcolsep}{3pt}
\resizebox{\linewidth}{!}{
	\begin{tabular}{ccccccccc}
  	    $\theta_{<30}\%$& $\theta_{>90}\%$& $Q_{min}$&$\rho_{max}$& $d_{H}(\times 10^{-2})$ & $\surfaceSeeds$ & $\interiorSeeds$ & $\surfaceTime$ & $\interiorTime$ \\
		$0.07$& $15$& $0.4$ & $4.863$ &$0.851$ & $497536$ & $113837$ &$4582$ & $7007$ \\
		\bottomrule
	\end{tabular}}
	 
\caption{Sample results on smooth models exhibiting detailed features with a large range of feature sizes (top), complex topologies with multiple holes and narrow regions (middle), and multiple components nearly in contact (bottom).}
\label{fig:vc_stat}
\vspace{-6pt}
\end{figure}

\begin{figure}
	\centering

\setlength{\tabcolsep}{3pt}
\resizebox{\linewidth}{!}{
	\begin{tabular}{ccc}
	\rotatebox{90}{\qquad \qquad \qquad \qquad \begin{Huge}Joint\end{Huge}}&
     \includegraphics[width=1.0\linewidth]{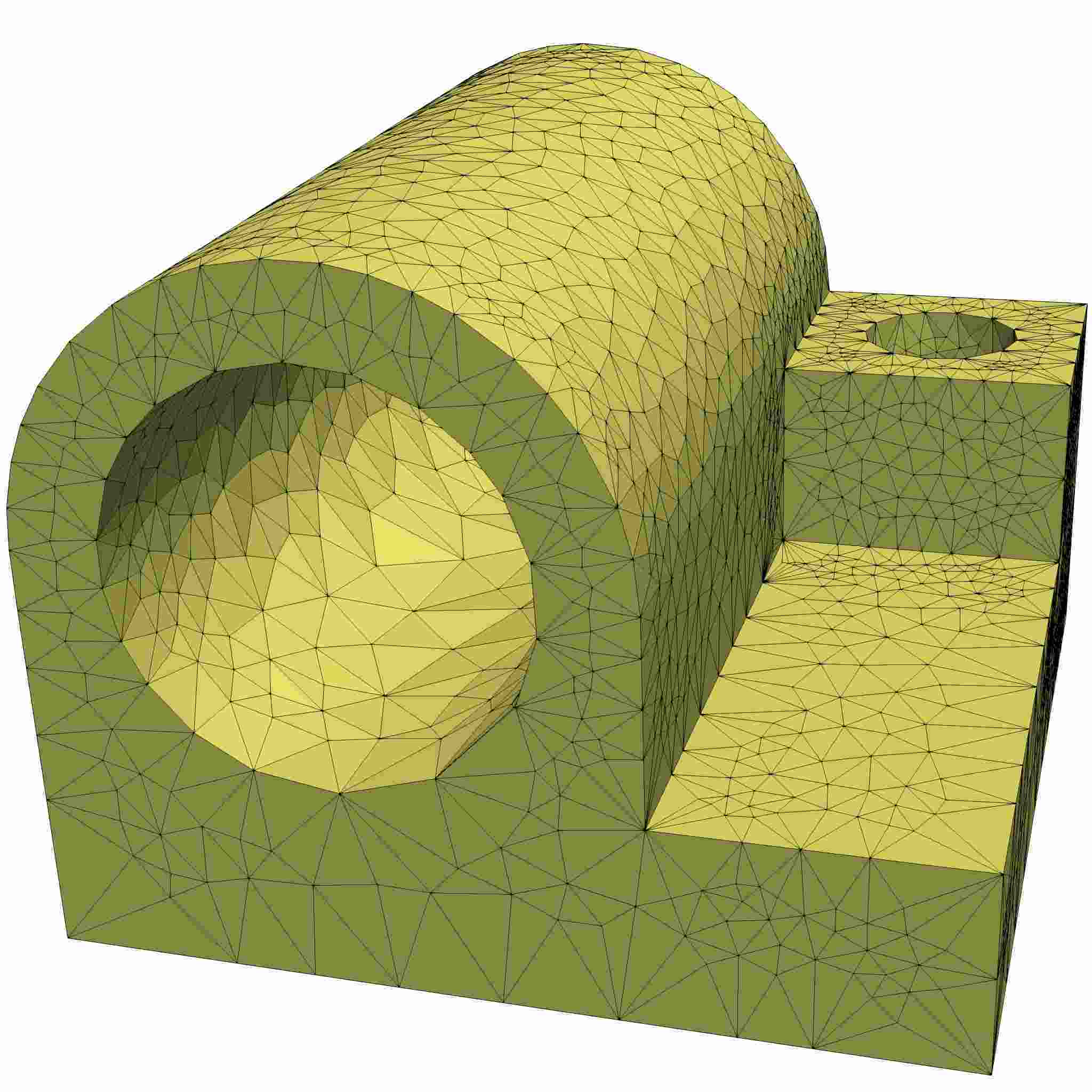} &
     \includegraphics[width=1.0\linewidth]{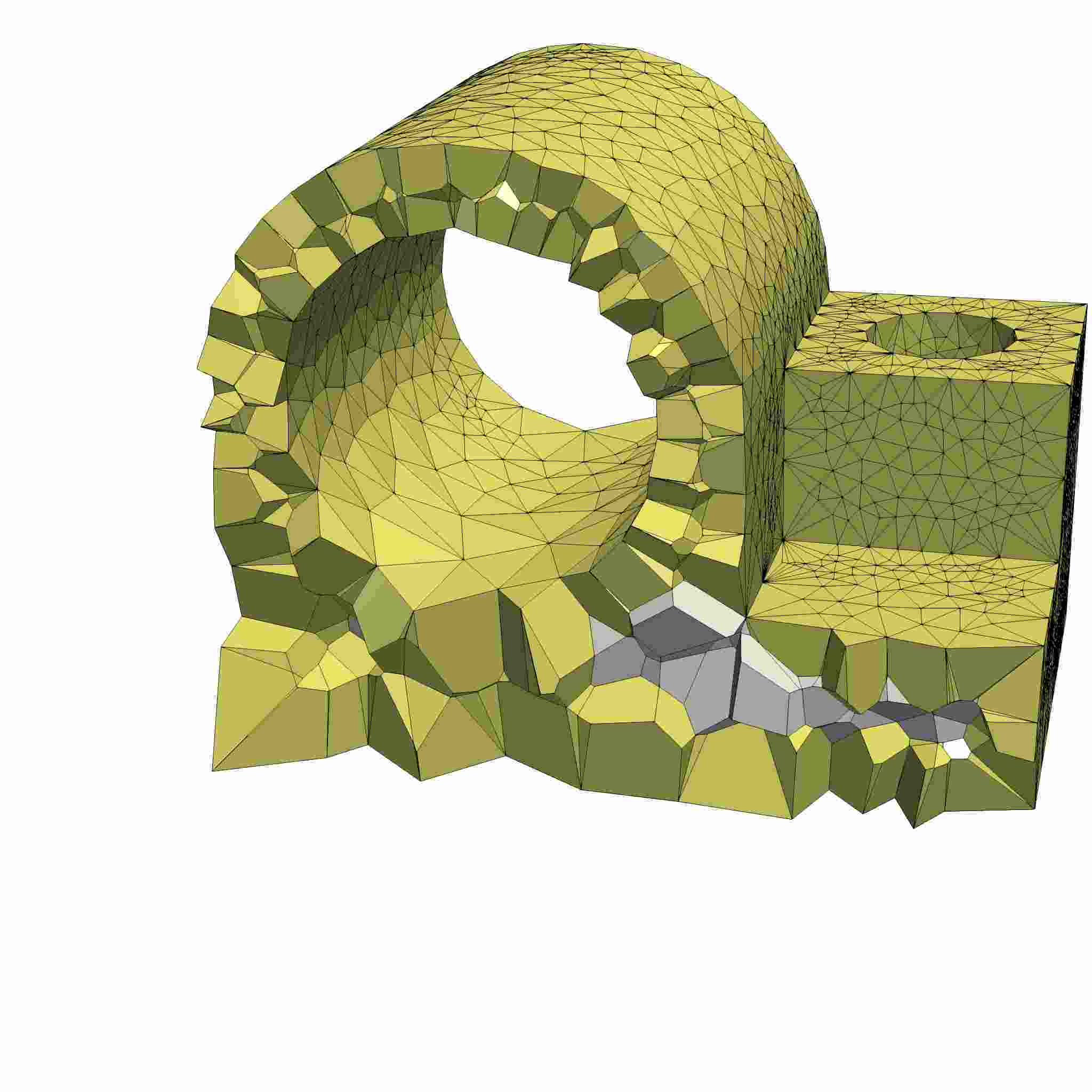}\\	
	\end{tabular}}
\setlength{\tabcolsep}{3pt}
\resizebox{\linewidth}{!}{
	\begin{tabular}{ccccccccc}
  	    $\theta_{<30}\%$& $\theta_{>90}\%$& $Q_{min}$&$\rho_{max}$& $d_{H}(\times 10^{-2})$ & $\surfaceSeeds$ & $\interiorSeeds$ & $\surfaceTime$ & $\interiorTime$ \\
     	$19$ & $23$ & $0.149$ & $12.495$ & $0.569$ & $11480$ & $868$ & $32$ & $34$ \\
		\bottomrule
	\end{tabular}}

\setlength{\tabcolsep}{3pt}
\resizebox{\linewidth}{!}{
	\begin{tabular}{ccc}
	\rotatebox{90}{\qquad \qquad \qquad \qquad \begin{Huge}Heptoroid\end{Huge}}&
     \includegraphics[width=1.0\linewidth]{figs/sharp/heptoroid_vc.jpg} &
    \includegraphics[width=1.0\linewidth]{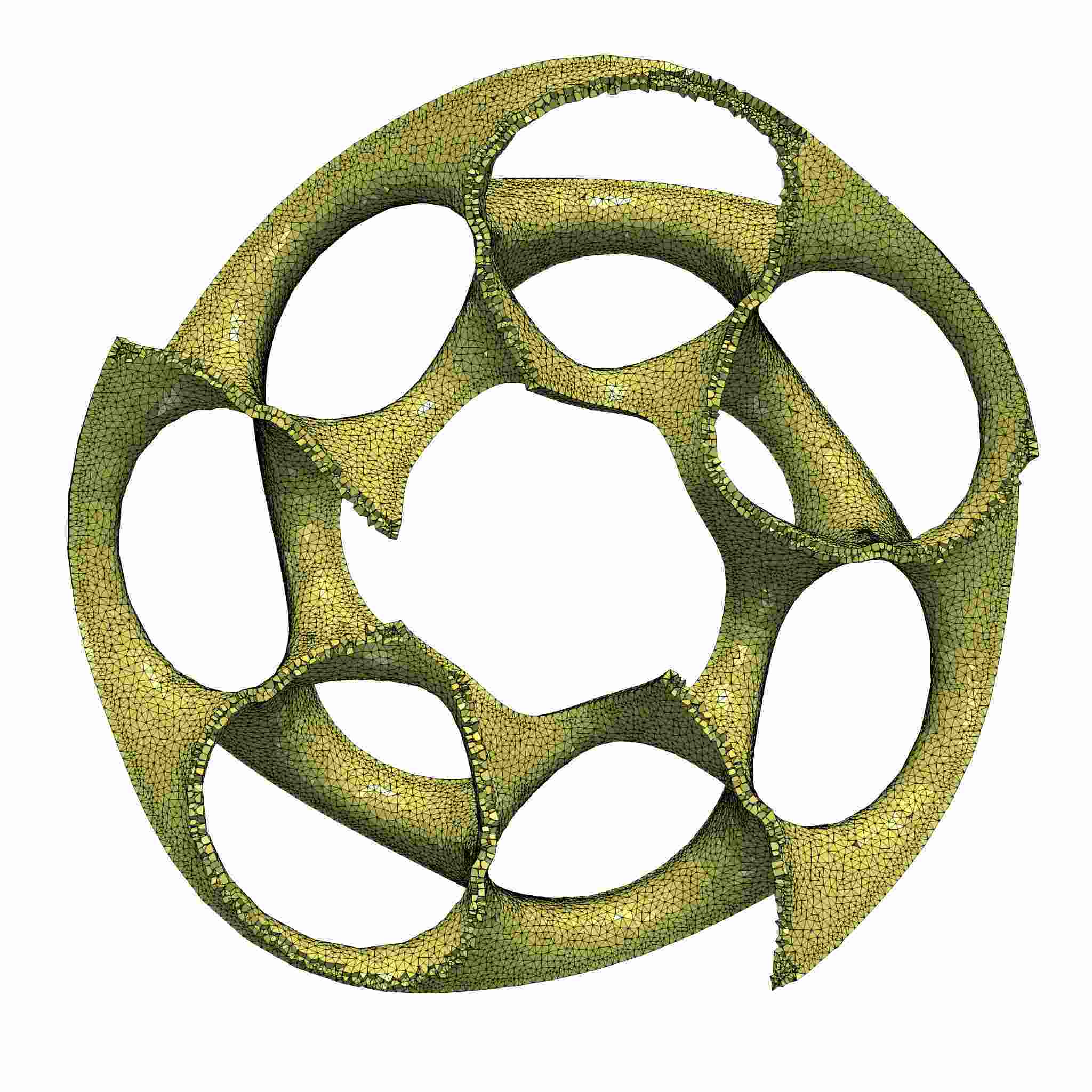}\\	
	\end{tabular}}
\setlength{\tabcolsep}{3pt}
\resizebox{\linewidth}{!}{
	\begin{tabular}{ccccccccc}
  	    $\theta_{<30}\%$& $\theta_{>90}\%$& $Q_{min}$&$\rho_{max}$& $d_{H}(\times 10^{-2})$ & $\surfaceSeeds$ & $\interiorSeeds$ & $\surfaceTime$ & $\interiorTime$ \\
     	$11$& $19$& $0.273$ & $377.029$ & $0.087$ & $258010$ & $0$ & $1464$ & $3432$ \\
		\bottomrule
	\end{tabular}}

\setlength{\tabcolsep}{3pt}
\resizebox{\linewidth}{!}{
	\begin{tabular}{ccc}	
	\rotatebox{90}{\qquad \qquad \qquad \qquad \begin{Huge}Snowflake\end{Huge}}&
     \includegraphics[width=1.0\linewidth]{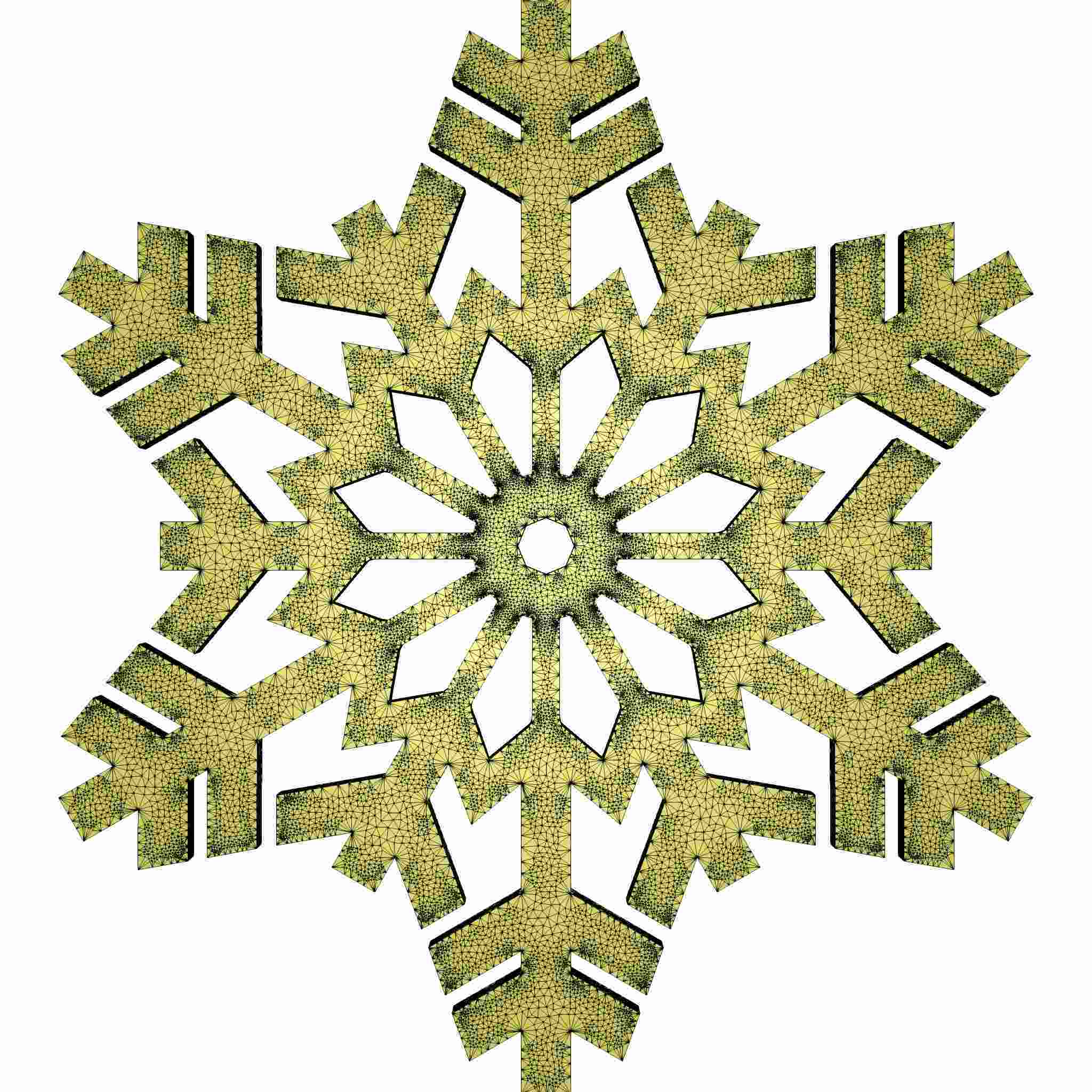} &
     \includegraphics[width=1.0\linewidth]{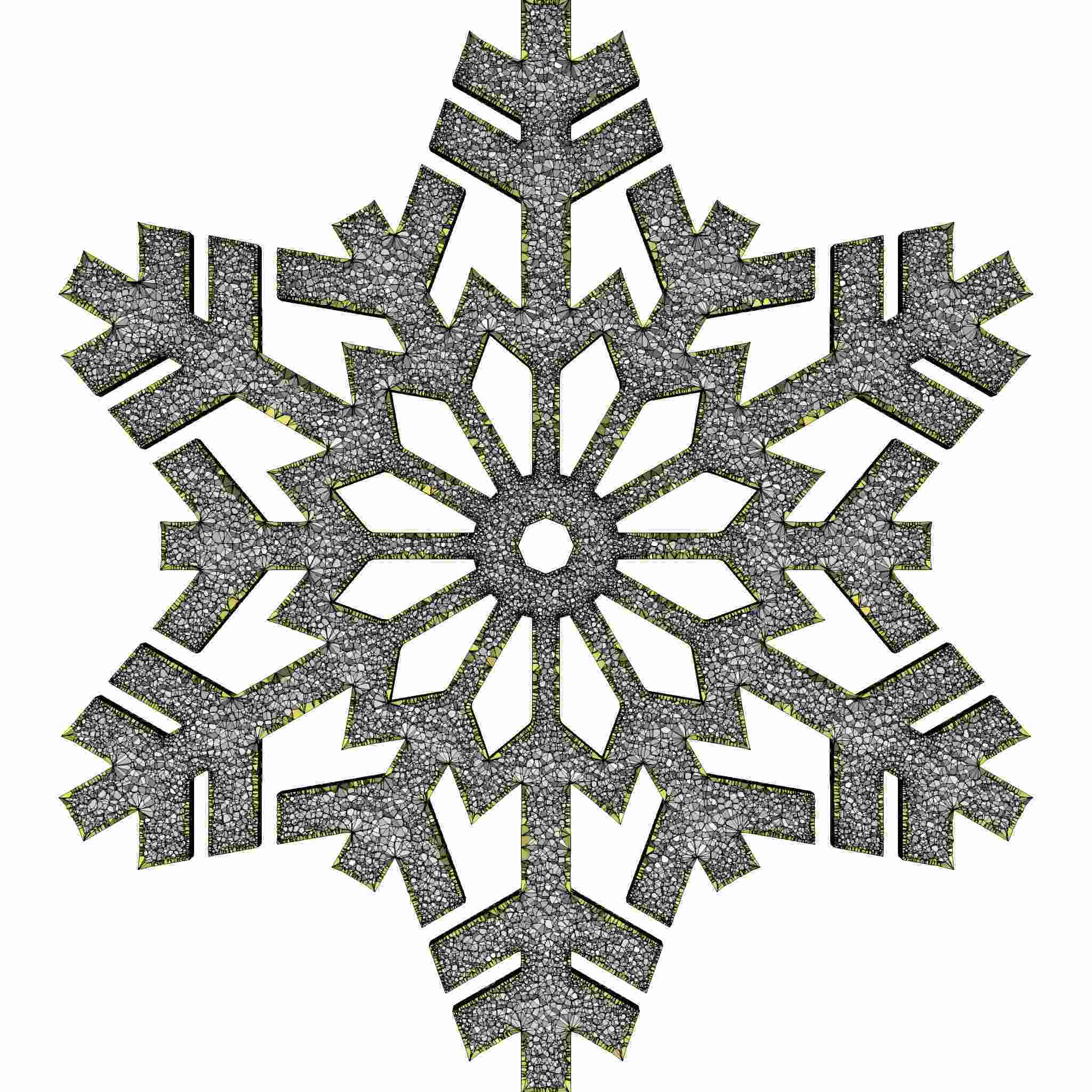}\\	
	\end{tabular}}
\setlength{\tabcolsep}{3pt}
\resizebox{\linewidth}{!}{
	\begin{tabular}{ccccccccc}
  	    $\theta_{<30}\%$& $\theta_{>90}\%$& $Q_{min}$&$\rho_{max}$& $d_{H}(\times 10^{-2})$ & $\surfaceSeeds$ & $\interiorSeeds$ & $\surfaceTime$ & $\interiorTime$ \\
     	$21$ & $25$ & $0.086$ & $63$ & $0.058$ & $85380$ & $57474$ & $2146$ & $9497$ \\

		\bottomrule
	\end{tabular}}
 
\caption{Sample results on models with sharp features including: mechanical models (top), complex topologies and narrow regions (middle and bottom).}
\label{fig:vc_stat_sharp}
\vspace{-6pt}
\end{figure}

\begin{figure}
\centering
\setlength{\tabcolsep}{4pt}
\resizebox{\linewidth}{!}{
	\begin{tabular}{cc}	
	\includegraphics[width=1.0\linewidth]{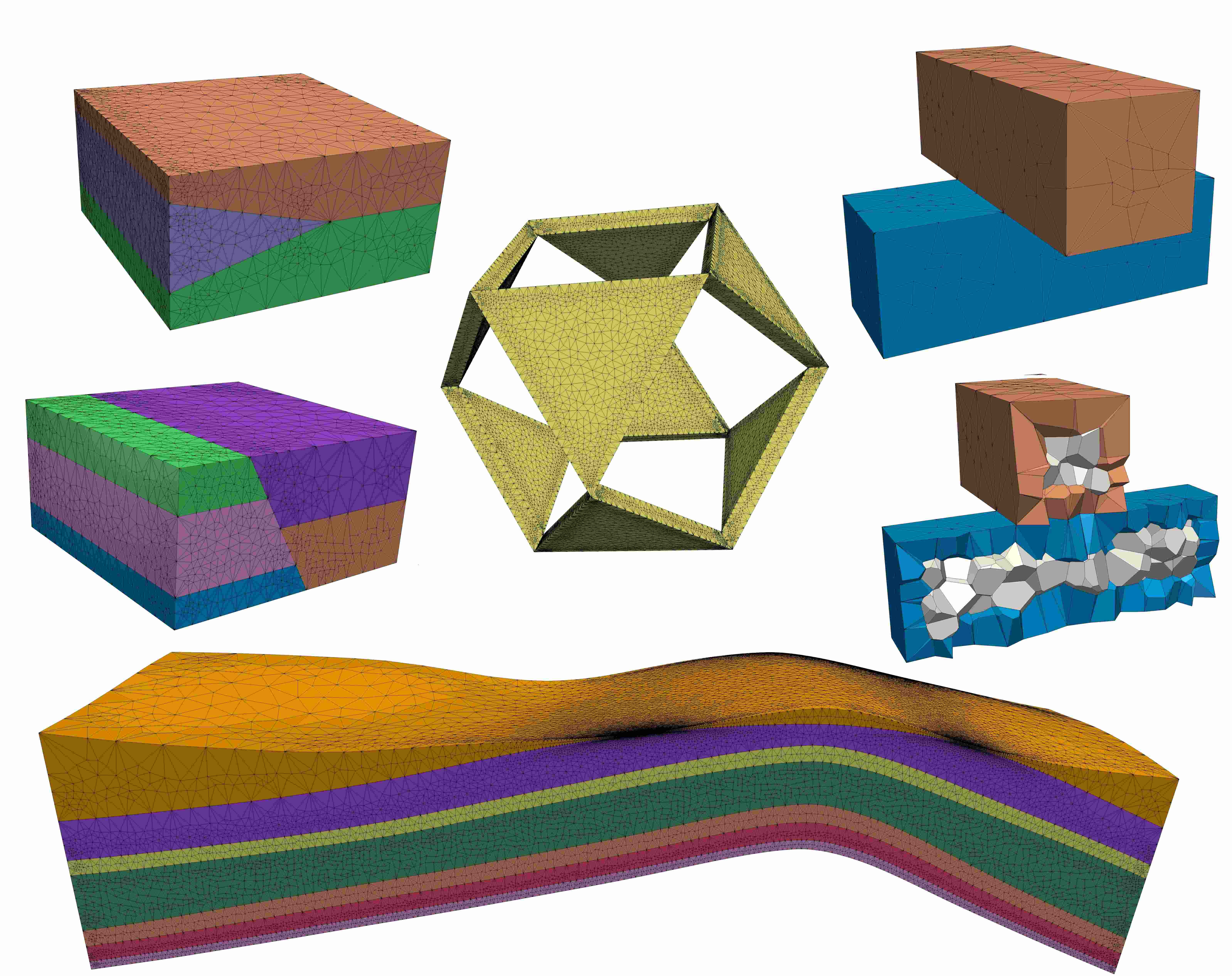}
	\end{tabular}}
\caption{Sample outputs for non-manifold domains consisting of multiple materials depicted in different colors. VoroCrust automatically detects the non-manifold interfaces between the materials (top left) and decomposes each subdomain into Voronoi cells that conform to those interfaces while preserving all sharp features (top right). More challenging cases involve contact at sharp features (top center), or multiple layers tapering into narrow regions towards contact (bottom).}
\label{fig:nonmanifold}
\end{figure}

\noindent \textbf{Robustness and Quality.} We test VoroCrust on a variety of models exhibiting different challenges ranging from smooth models with detailed features and narrow regions as in Figure~\ref{fig:vc_stat}, to sharp features with curvature and holes as in Figure~\ref{fig:vc_stat_sharp}, and even non-manifold boundaries as in Figure~\ref{fig:nonmanifold}. The quality of the surface mesh is measured by the percentage of triangles with angles less than $30^\circ$ or greater than $90^\circ$, as well as the minimum triangle quality\footnote{Triangle quality is defined as $\frac{6S}{\sqrt{3}hP}$, where $S$ is the area, $h$ is the longest edge length, and $P$ is half the perimeter.} $Q_{min}$. The quality of the volume mesh is measured by the maximum aspect ratio\footnote{Aspect ratio is defined as the ratio between the radius of the smallest circumscribing sphere to the radius of the largest inscribed sphere.} $\rho_{max}$, which is often realized by cells incident to the surface. We also report the approximation error in terms of the Hausdorff error $d_H$ (normalized by the diameter of the bounding box). The number of seeds in $\surfaceSeeds$ and $\interiorSeeds$ are reported along with the time in seconds taken to generate each, denoted $\surfaceTime$ and $\interiorTime$, respectively. Meshes were generated from VoroCrust seeds using \voroplusplus~\cite{rycroft2009voro++}.

We encountered no issues with any of the models, which demonstrates the robustness of the algorithm and its implementation. We set $\vcTheta$ to $60^{\circ}$ for smooth models, and choose an appropriate value of $\vcTheta$ for models with sharp features. The value of $L$ was fixed at 0.25 for all inputs. We note that the output surface meshes are of high quality per the minimum triangle quality and angle bounds, while achieving small approximation errors. The demonstrated quality of VoroCrust output, with no skinny elements, is in agreement with the theoretical guarantees established for an abstract version of the algorithm~\cite{VC_SoCG}. Additional results on a variety of models are provided in the supplemental materials. \newline

\noindent \textbf{Parameters.} We start by studying the impact of $L$ on the complexity of the output surface mesh and the running time of the algorithm. Figure~\ref{fig:parameters2} demonstrates this impact on the Joint model. The results of this experiment demonstrate the impact of $L$ on the level of refinement per the number of balls in $\ballset$ generated by the algorithm. In particular, smaller values of $L$ lead to higher refinement. On the other hand, larger values of $L$ slow down the algorithm due to the increased size of ball neighborhoods resulting in processing a larger number of balls for various tasks; see Section~\ref{subsec:ball-neighborhood}. This behavior of the algorithm in terms of $L$ is consistent for different values of $\vcTheta$ as can be seen in Figure~\ref{fig:parameters2}.

\pgfplotsset{every tick label/.append style={font=\huge}}
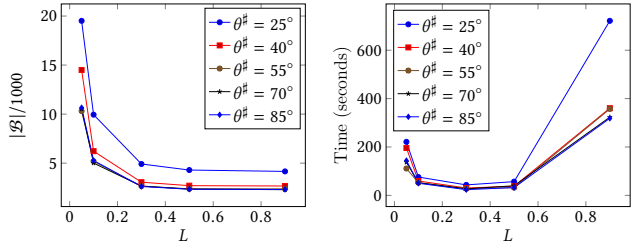
\begin{figure}[h]
\centering
\setlength{\tabcolsep}{4pt}
\resizebox{\linewidth}{!}{
	\begin{tabular}{ccc}
     \begin{tikzpicture}
      \begin{axis}[xlabel={$L$}, ylabel={$|\ballset| / 1000$}, label style={font=\huge}, legend style={font=\huge}]
       \addplot coordinates { 
        (0.05, 19.513)
        (0.10, 9.943)
        (0.30, 4.912)
        (0.50, 4.296)
        (0.90, 4.157)
       };
       \addplot coordinates { 
        (0.05, 14.498)
        (0.10, 6.22)
        (0.30, 3.057)
        (0.50, 2.704)
        (0.90, 2.669)
       };
       \addplot coordinates { 
        (0.05, 10.314)
        (0.10, 5.191)
        (0.30, 2.656)
        (0.50, 2.378)
        (0.90, 2.379)
       };
       \addplot coordinates { 
        (0.05, 10.534)
        (0.10, 5.008)
        (0.30, 2.659)
        (0.50, 2.355)
        (0.90, 2.294)
       };
       \addplot coordinates { 
        (0.05, 10.657)
        (0.10, 5.267)
        (0.30, 2.619)
        (0.50, 2.338)
        (0.90, 2.312)
       };
      \legend{$\vcTheta=25^\circ$,$\vcTheta=40^\circ$,$\vcTheta=55^\circ$,$\vcTheta=70^\circ$,$\vcTheta=85^\circ$}
      \end{axis}
     \end{tikzpicture} & $~~$ &
     \begin{tikzpicture}
      \begin{axis}[xlabel={$L$}, ylabel={Time (seconds)}, legend pos = north west, label style={font=\huge}, legend style={font=\huge}]
       \addplot coordinates { 
        (0.05, 221.286)
        (0.10, 75.8507)
        (0.30, 43.3653)
        (0.50, 56.851)
        (0.90, 721.486)
       };
       \addplot coordinates { 
        (0.05, 196.013)
        (0.10, 59.3934)
        (0.30, 30.6658)
        (0.50, 37.7637)
        (0.90, 360.286)
       };
       \addplot coordinates { 
        (0.05, 111.096)
        (0.10, 52.5207)
        (0.30, 27.8777)
        (0.50, 31.9331)
        (0.90, 357.286)
       };
       \addplot coordinates { 
        (0.05, 136.867)
        (0.10, 52.5421)
        (0.30, 26.9879)
        (0.50, 39.8934)
        (0.90, 323.139)
       };
       \addplot coordinates { 
        (0.05, 144.02)
        (0.10, 50.1347)
        (0.30, 23.6367)
        (0.50, 32.4707)
        (0.90, 318.492)
       };
      \legend{$\vcTheta=25^\circ$,$\vcTheta=40^\circ$,$\vcTheta=55^\circ$,$\vcTheta=70^\circ$,$\vcTheta=85^\circ$}
      \end{axis}
     \end{tikzpicture}
     \vspace{-20pt}
	\end{tabular}} 
\caption{Impact of the parameter $L$ on the Joint model for varying values of $\vcTheta$. While the level of refinement is inversely proportional to $L$, increasing $L$ slows down the algorithm due to larger ball neighborhoods.}
\label{fig:parameters2}
\end{figure}

\begin{figure*}
	\centering
	\setlength{\tabcolsep}{5pt}
	\resizebox{1.0\linewidth}{!}{
		\begin{tabular}{c|*{5}{m{0.18\linewidth}}}
			\rotatebox{90}{\shortstack[c]{$L = 0.05$}}
			& \includegraphics[width=1.0\linewidth]{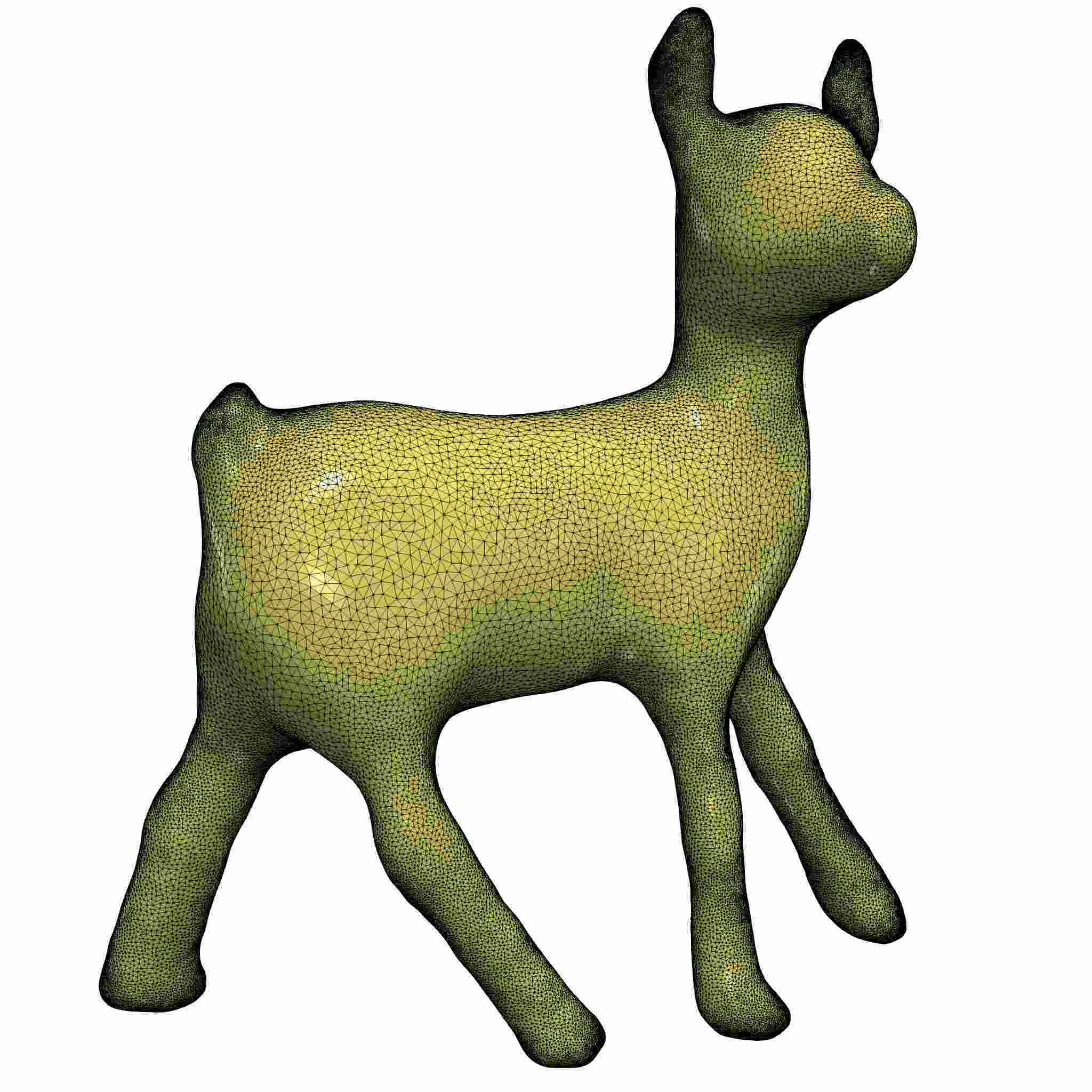}
			& \includegraphics[width=1.0\linewidth]{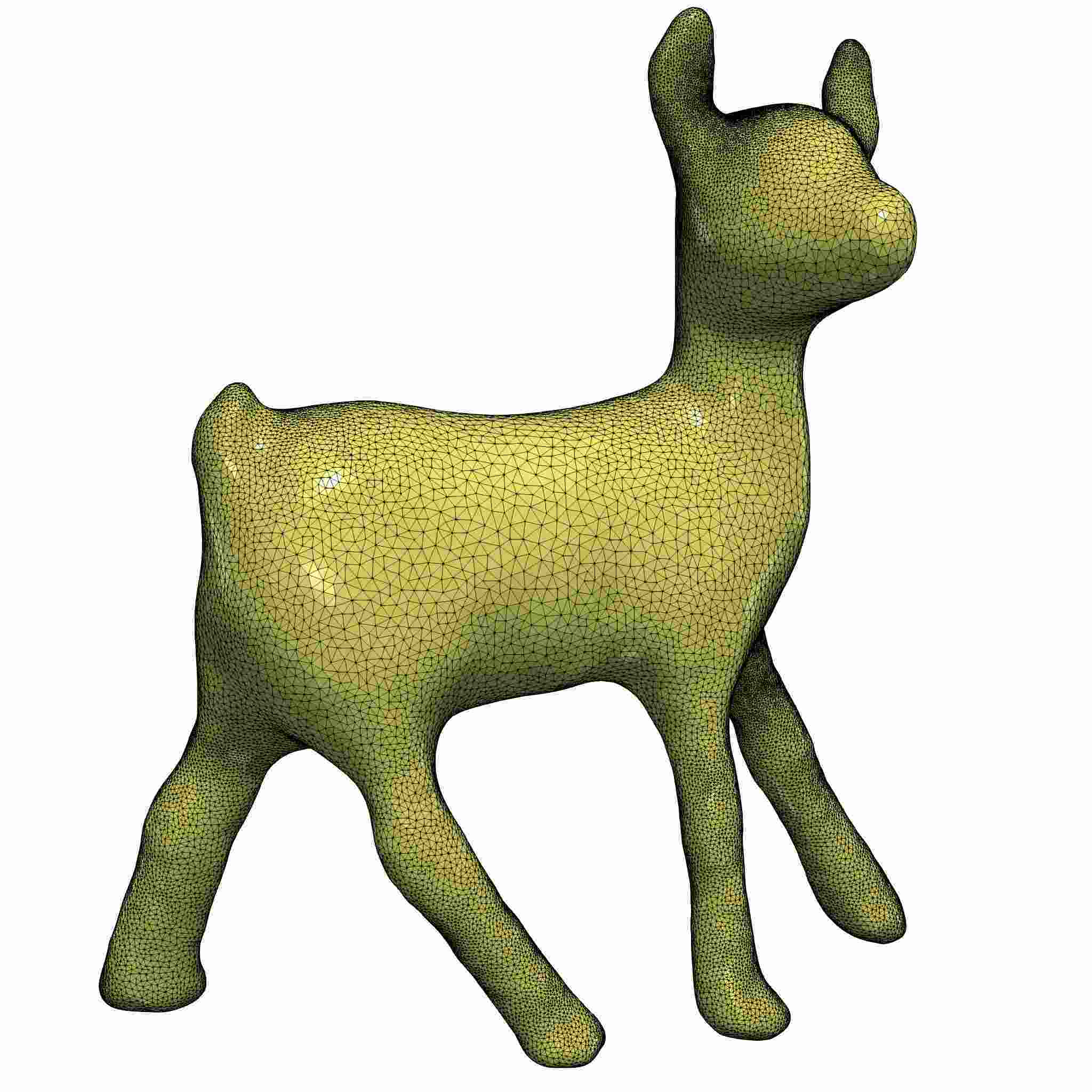}
			& \includegraphics[width=1.0\linewidth]{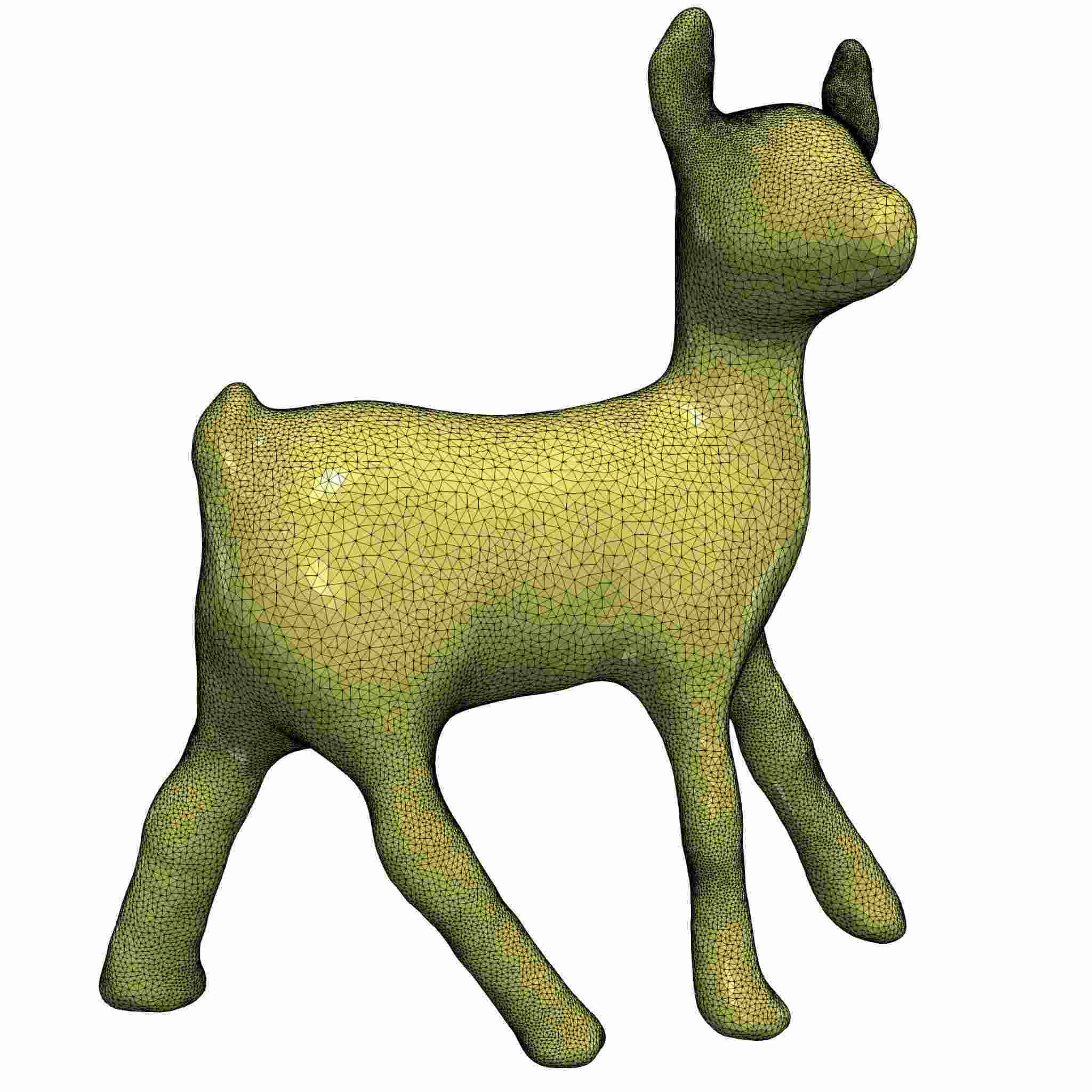}
			& \includegraphics[width=1.0\linewidth]{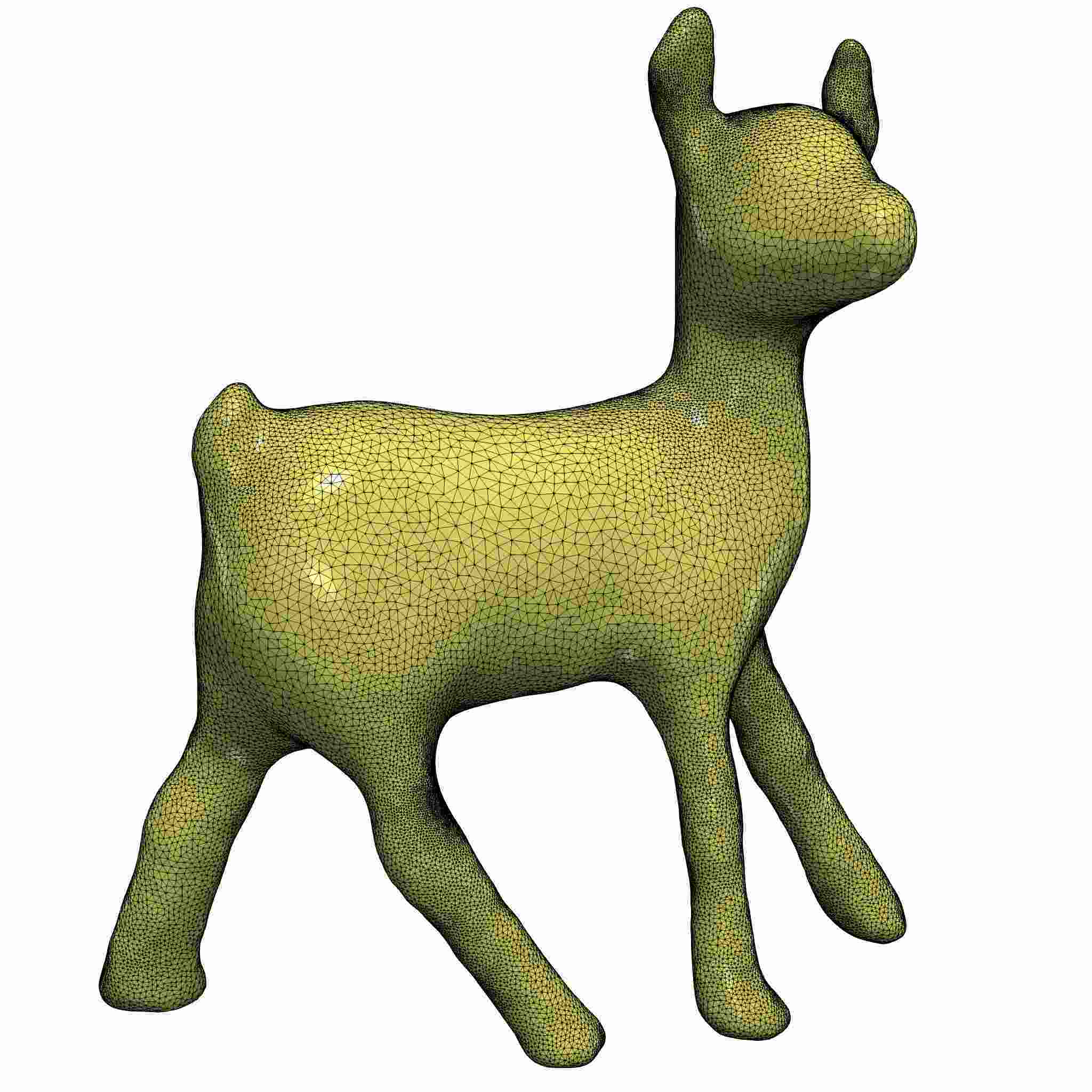}
			& \includegraphics[width=1.0\linewidth]{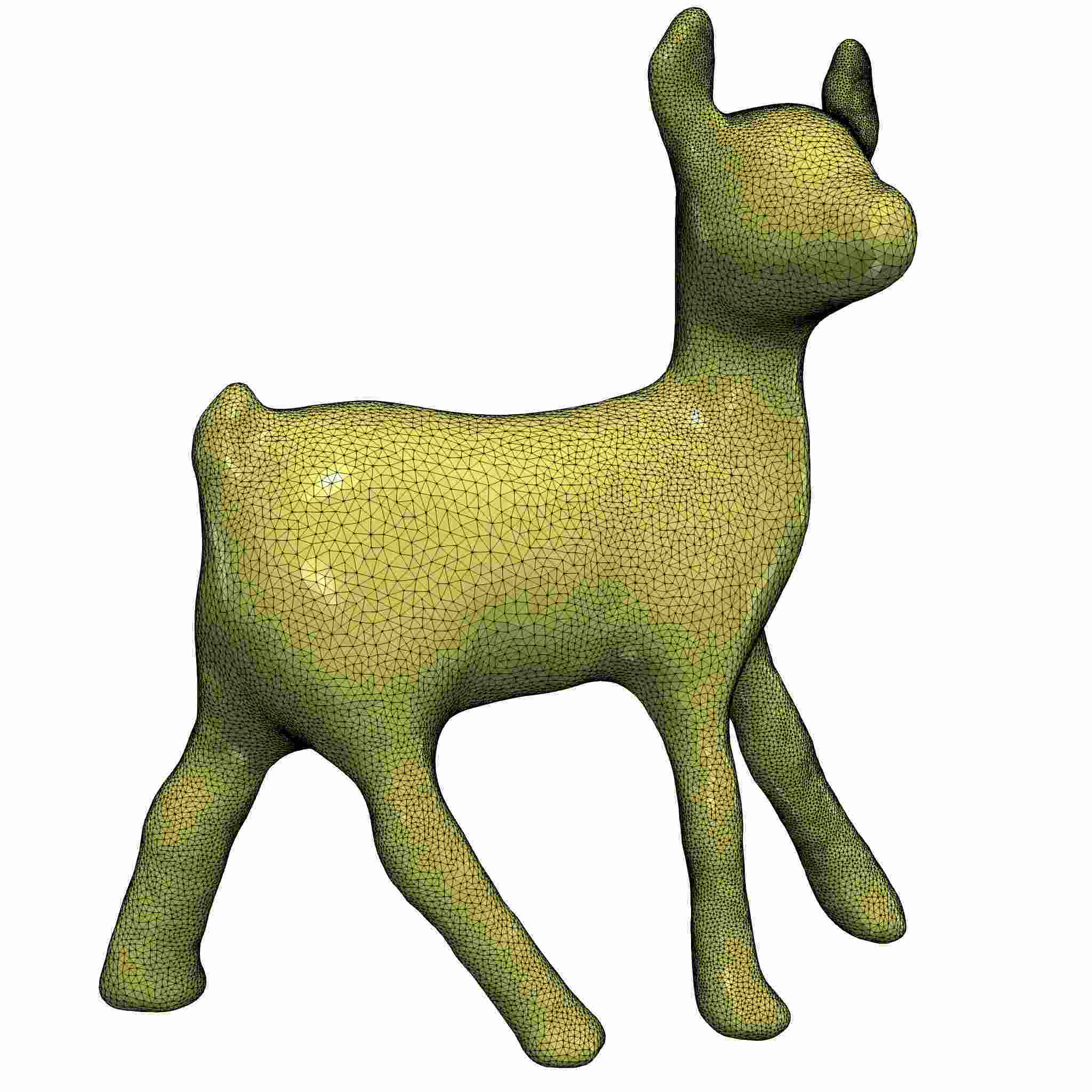}
			\\  		
			&$\quad\:\:\theta_{min}, Q_{min}, d_{H}$
			&$\quad\:\:\theta_{min}, Q_{min}, d_{H}$
			&$\quad\:\:\theta_{min}, Q_{min}, d_{H}$
			&$\quad\:\:\theta_{min}, Q_{min}, d_{H}$
			&$\quad\:\:\theta_{min}, Q_{min}, d_{H}$	
			\\
			&$\quad\:\:30^{\circ}, 0.47, 0.374 $
			&$\quad\:\:28^{\circ}, 0.45, 0.396 $
			&$\quad\:\:29^{\circ}, 0.47, 0.371 $
			&$\quad\:\:29^{\circ}, 0.45, 0.383 $
			&$\quad\:\:30^{\circ}, 0.47, 0.403 $	
			\\
			\rotatebox{90}{\shortstack[c]{$L = 0.1$}}
			& \includegraphics[width=1.0\linewidth]{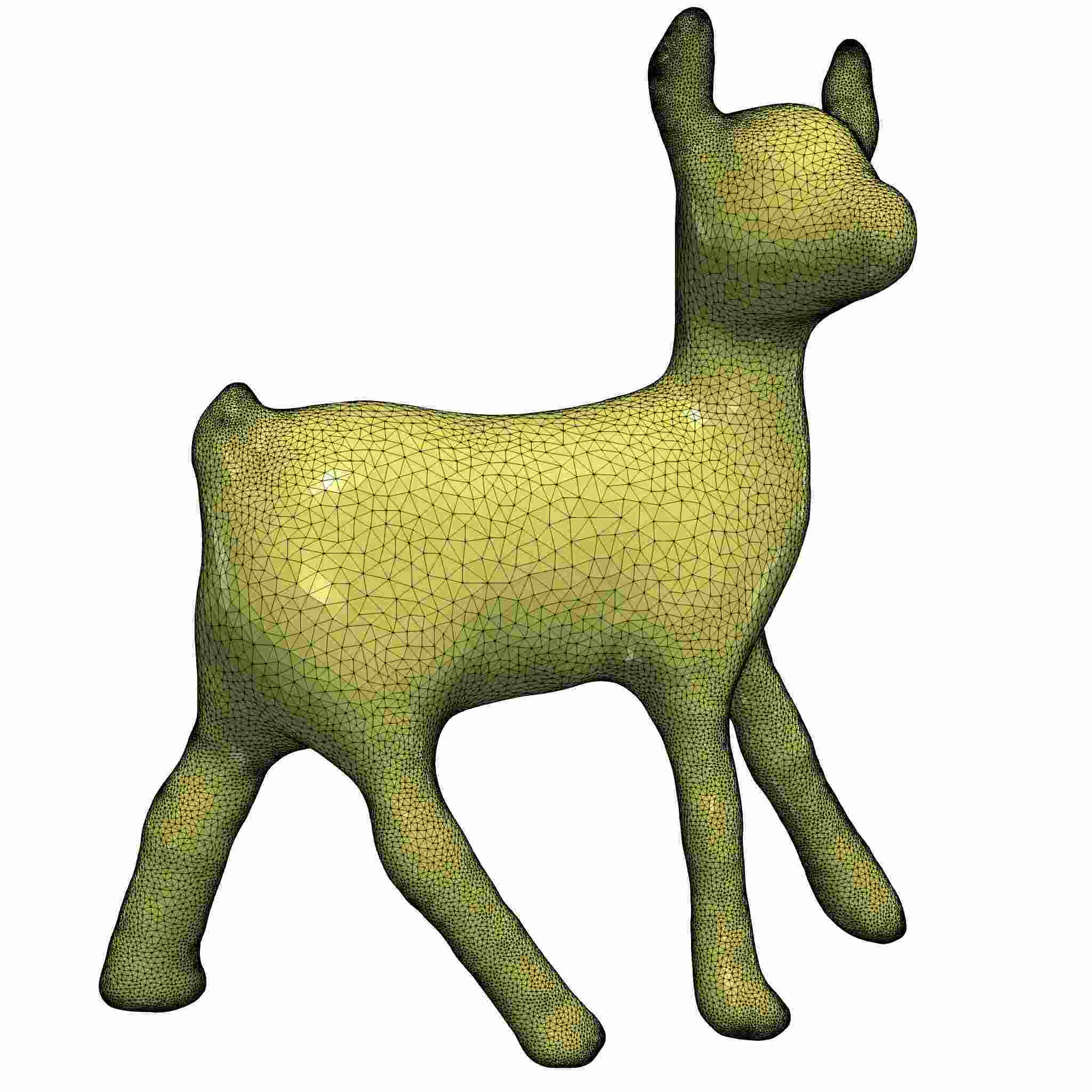}
			& \includegraphics[width=1.0\linewidth]{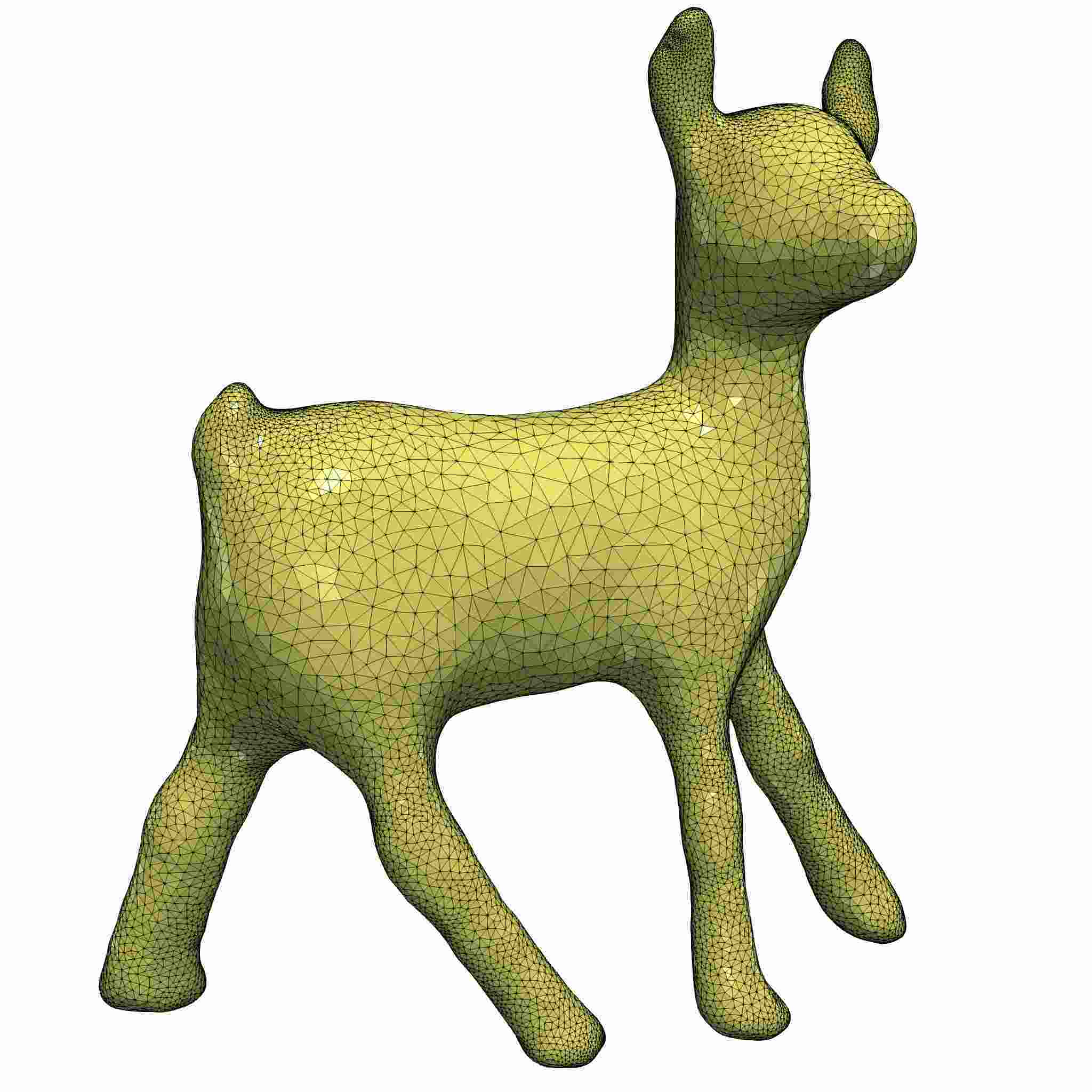}
			& \includegraphics[width=1.0\linewidth]{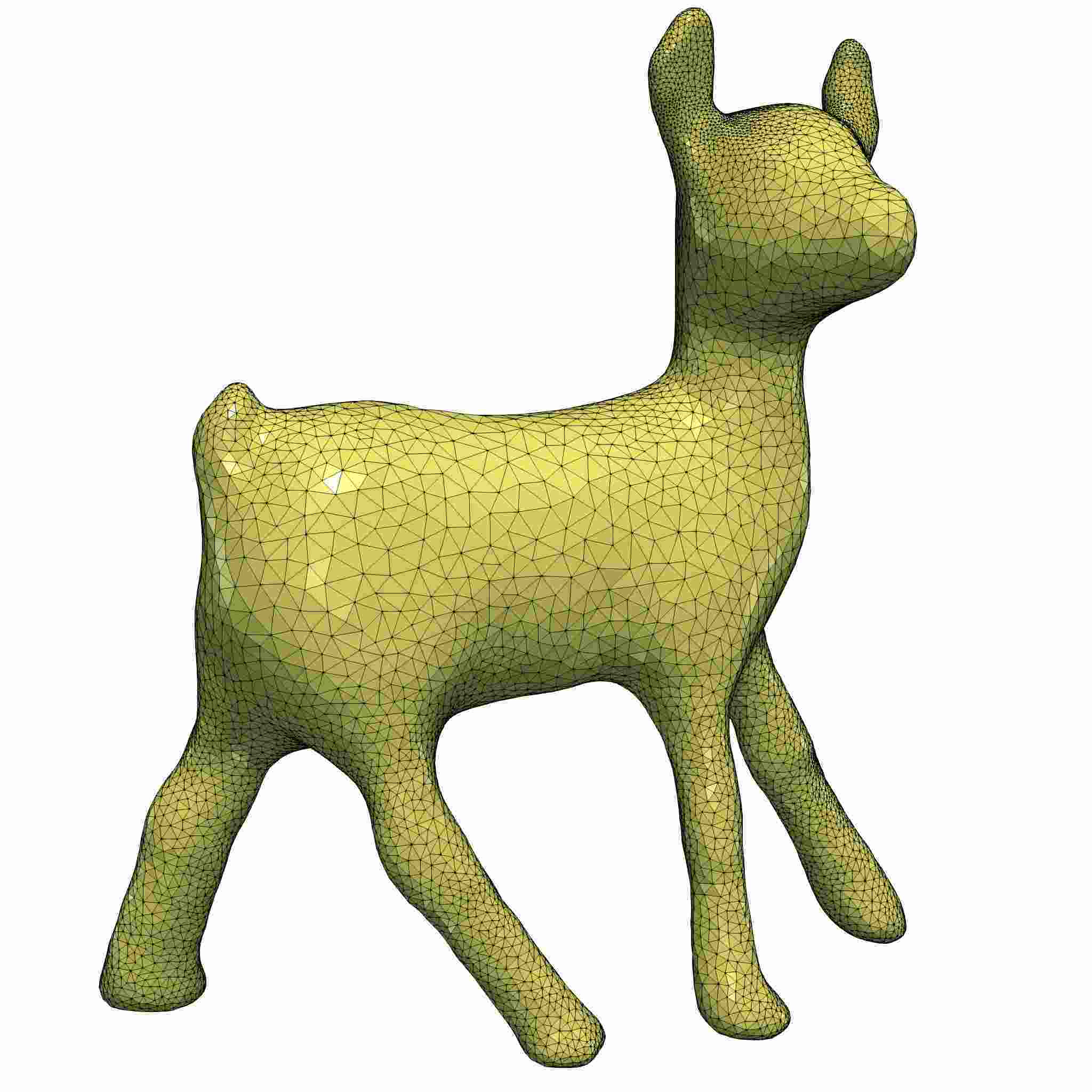}
			& \includegraphics[width=1.0\linewidth]{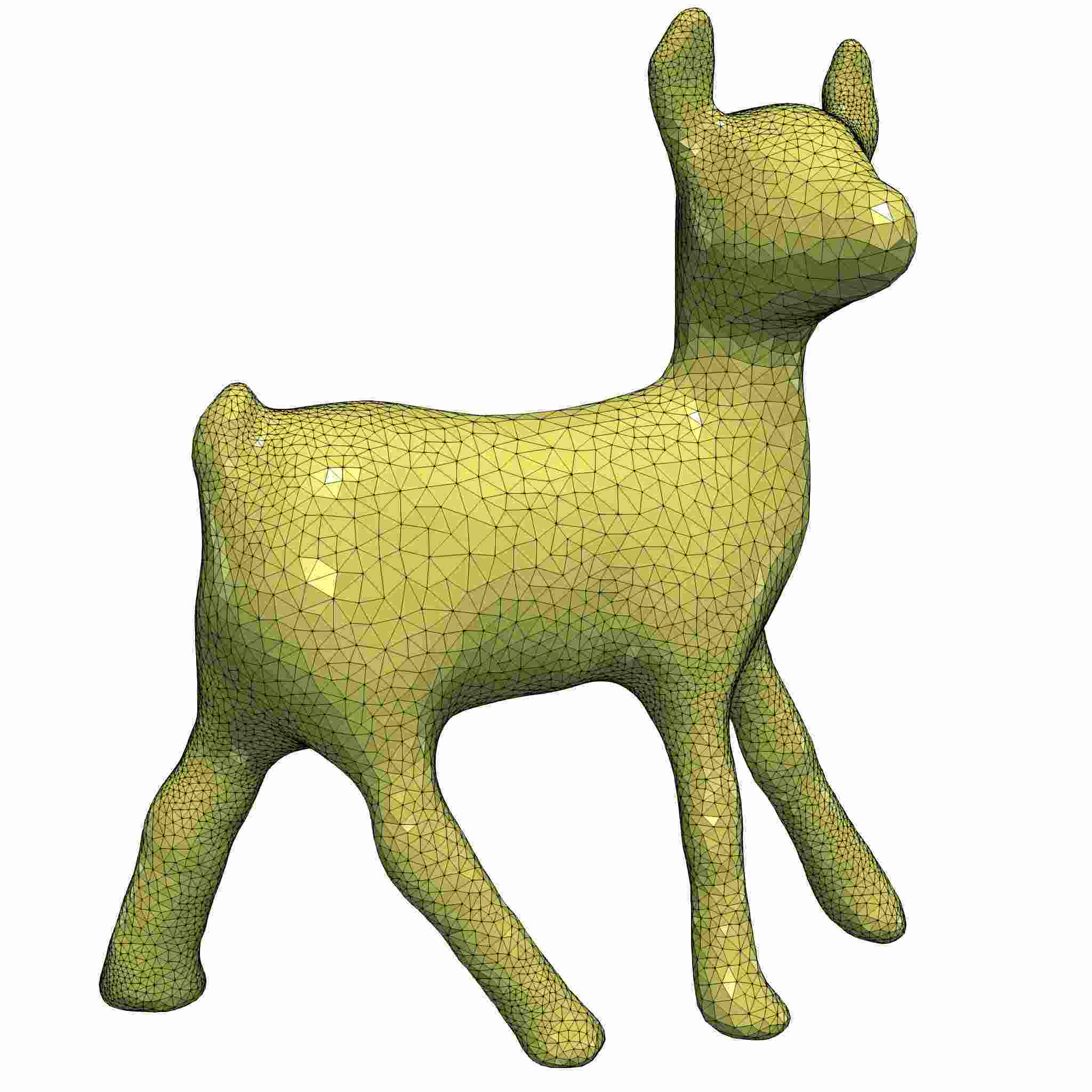}
			& \includegraphics[width=1.0\linewidth]{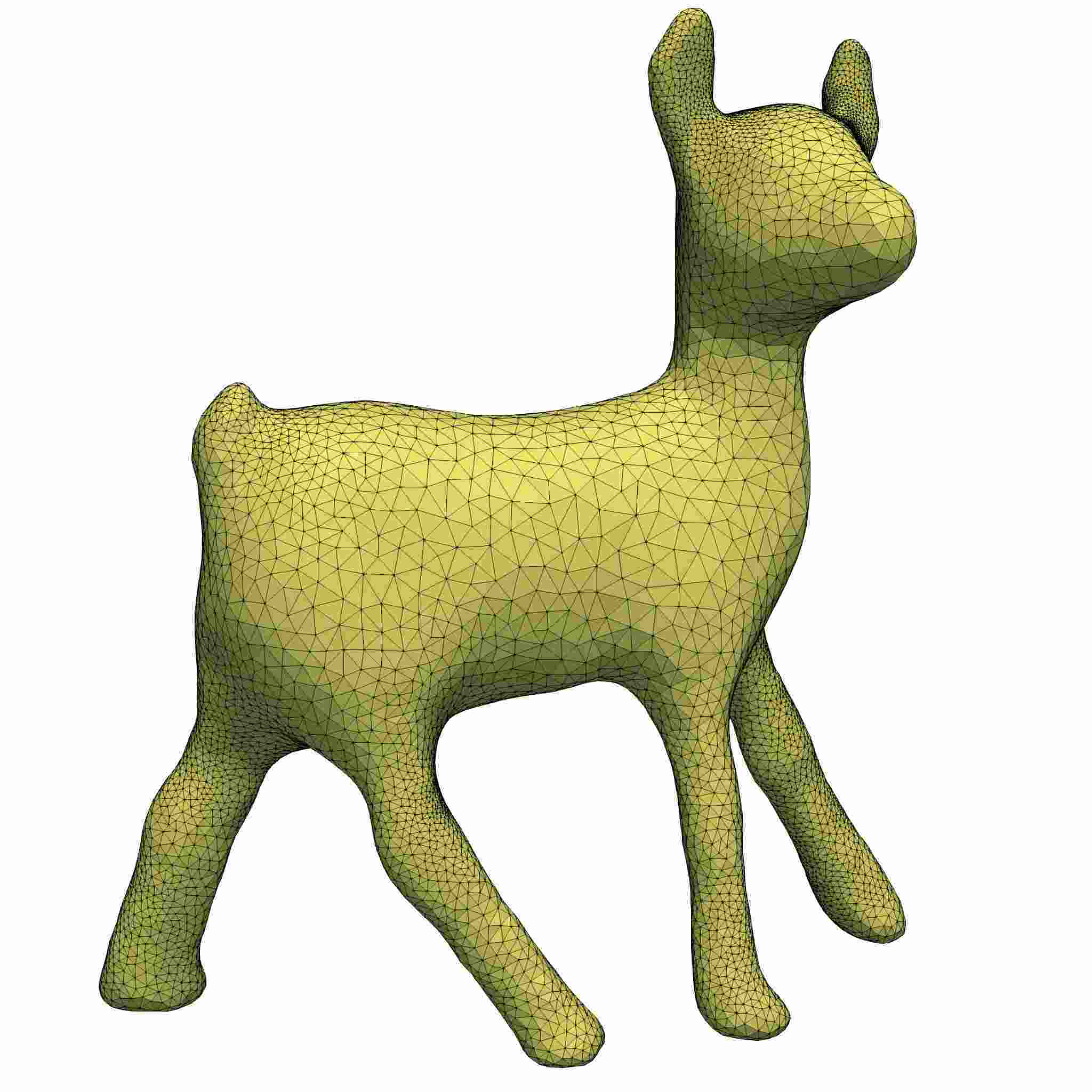}			
			\\			
			&$\quad\:\:\theta_{min}, Q_{min}, d_{H}$
			&$\quad\:\:\theta_{min}, Q_{min}, d_{H}$
			&$\quad\:\:\theta_{min}, Q_{min}, d_{H}$
			&$\quad\:\:\theta_{min}, Q_{min}, d_{H}$
			&$\quad\:\:\theta_{min}, Q_{min}, d_{H}$						
			\\
			&$\quad\:\:27^{\circ}, 0.45, 0.368 $
			&$\quad\:\:28^{\circ}, 0.44, 0.451 $
			&$\quad\:\:28^{\circ}, 0.46, 0.44 $
			&$\quad\:\:28^{\circ}, 0.45, 0.622 $
			&$\quad\:\:28^{\circ}, 0.46, 0.514 $		
			\\	
			\rotatebox{90}{\shortstack[c]{$L = 0.3$}}
			& \includegraphics[width=1.0\linewidth]{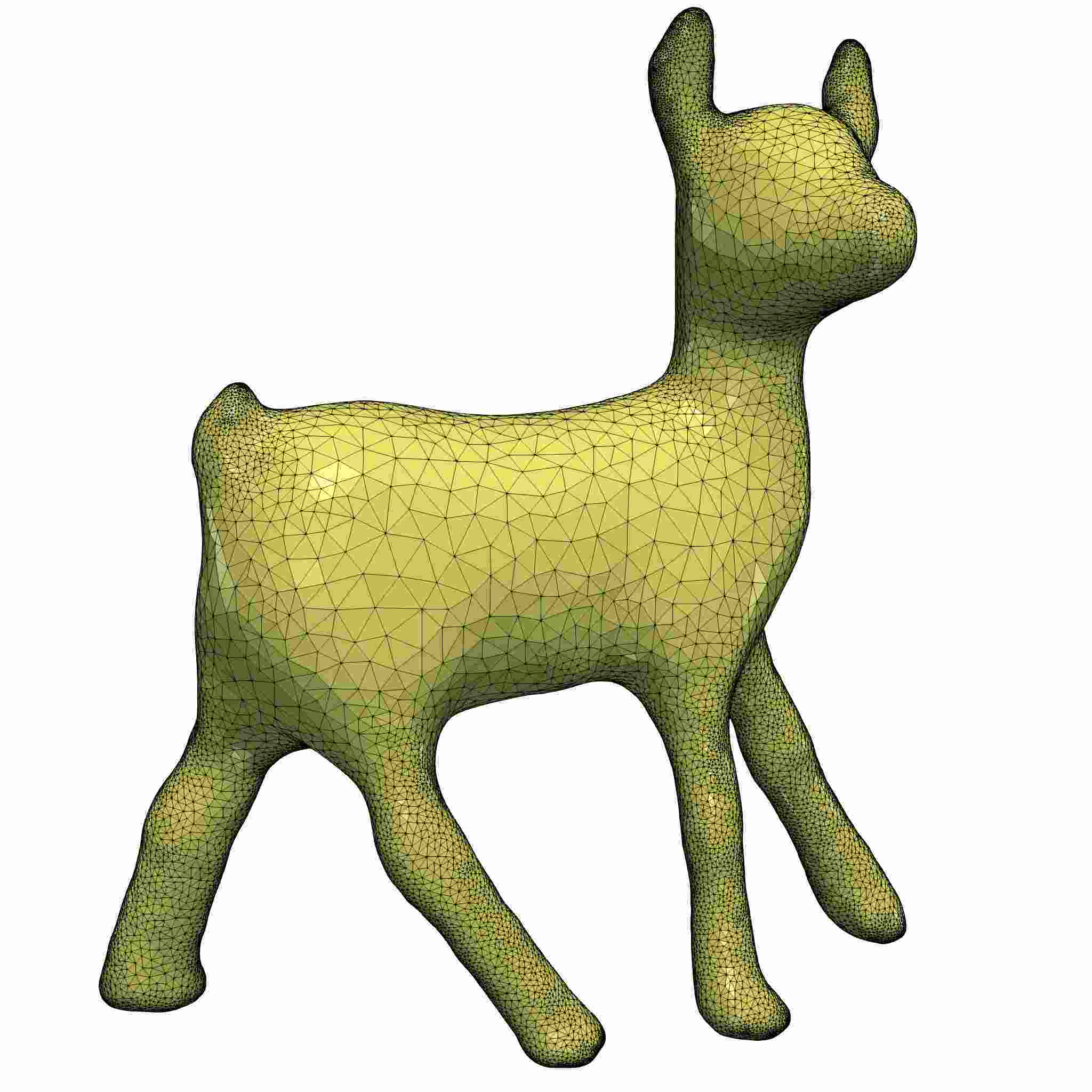}
			& \includegraphics[width=1.0\linewidth]{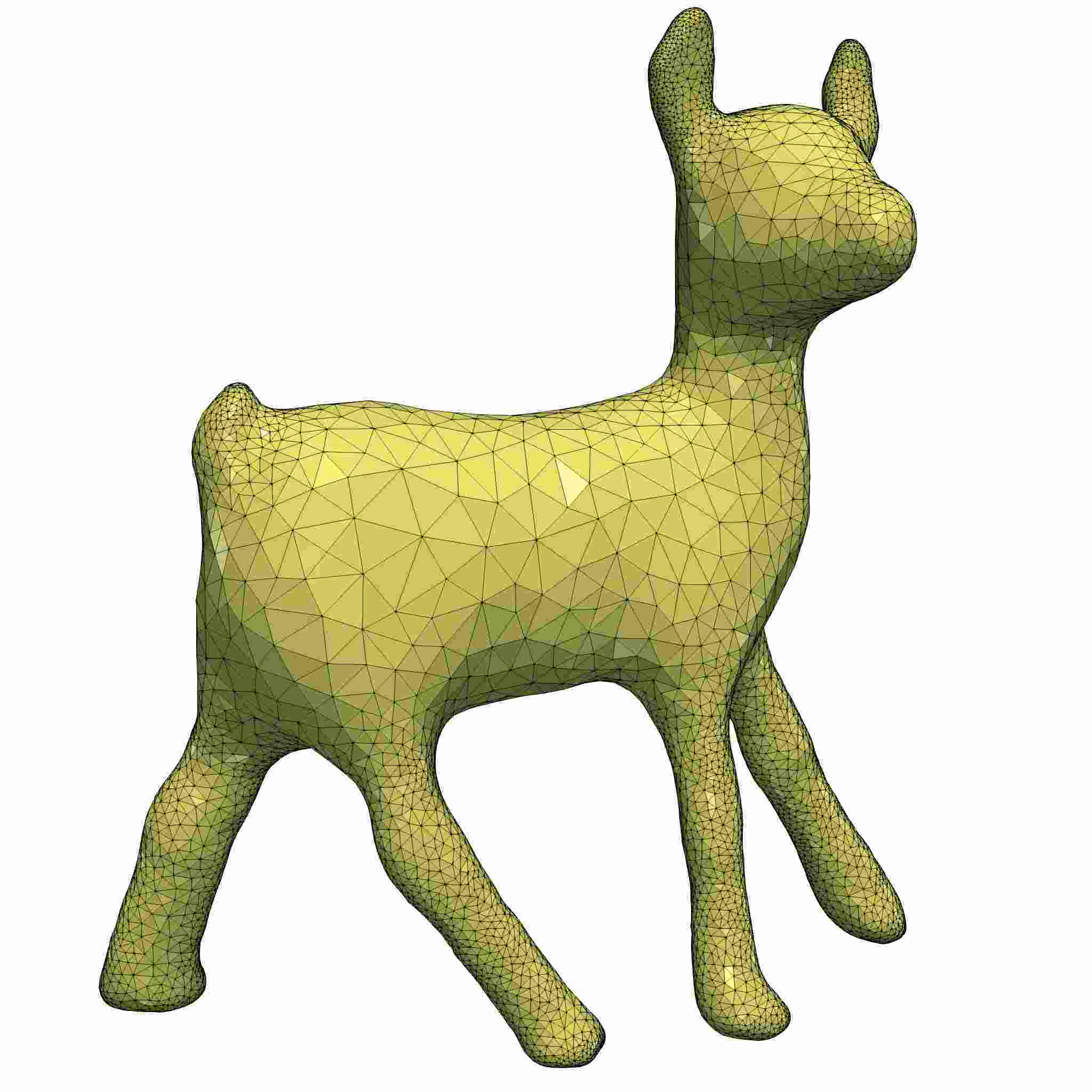}
			& \includegraphics[width=1.0\linewidth]{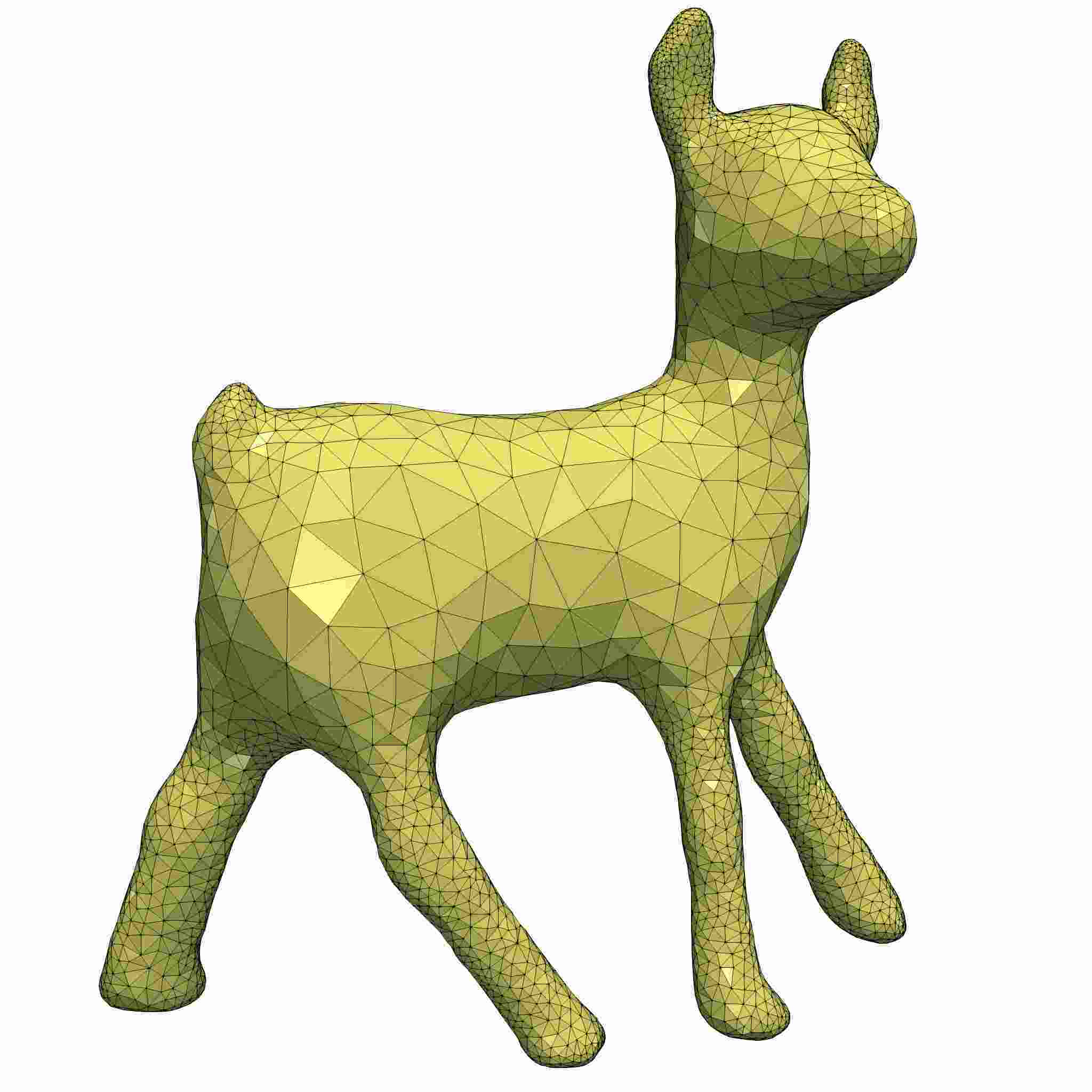}
			& \includegraphics[width=1.0\linewidth]{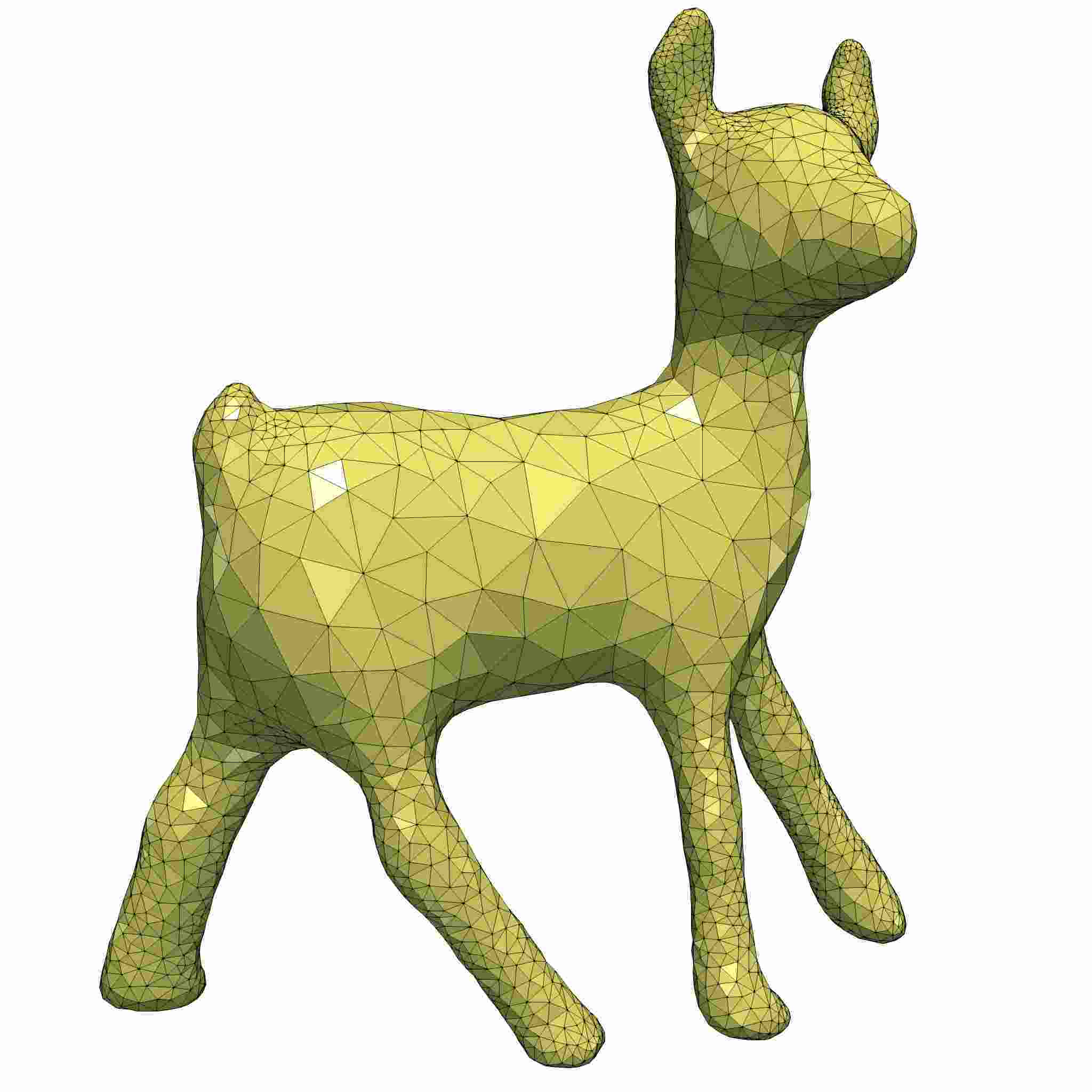}
			& \includegraphics[width=1.0\linewidth]{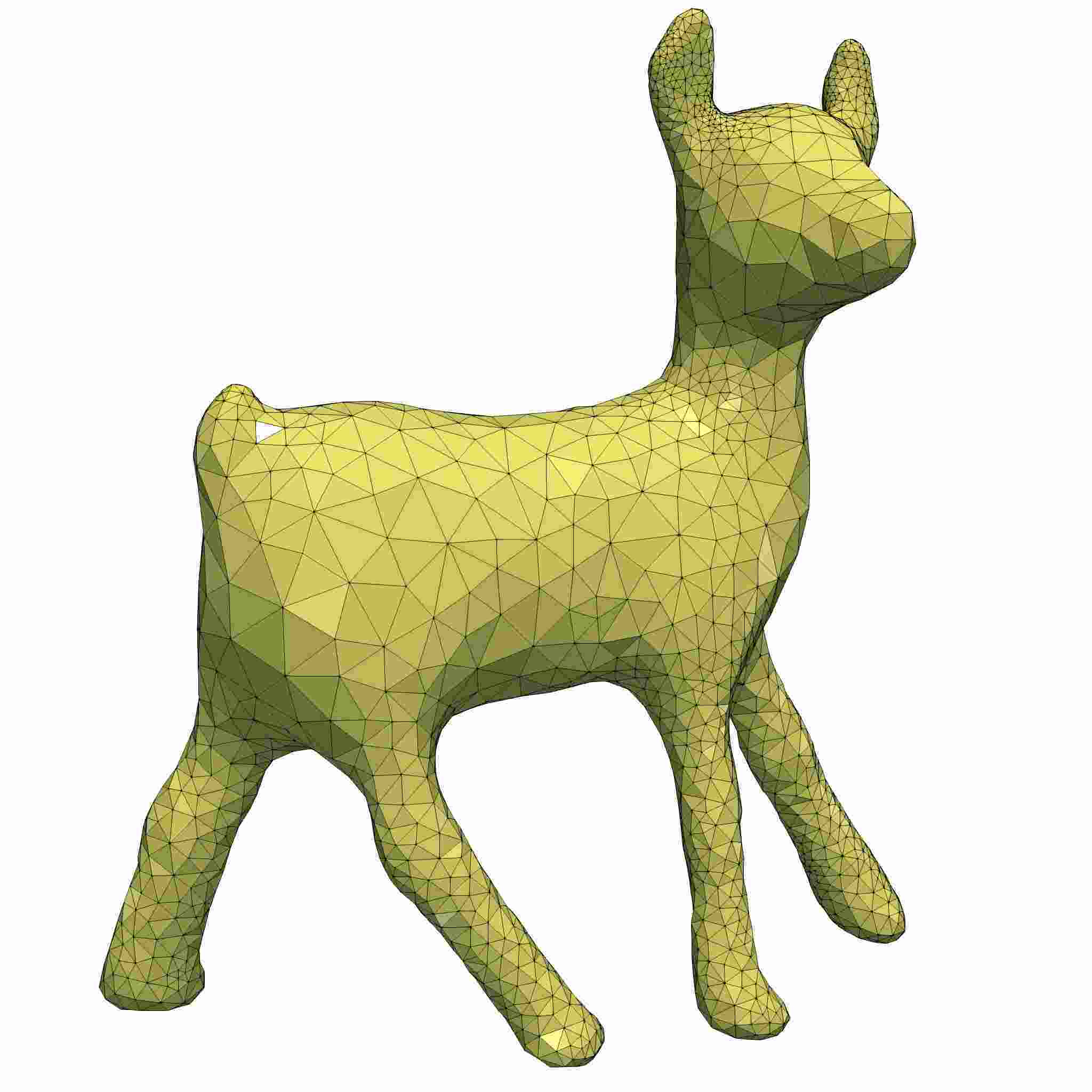}
			\\			
			\\			
			&$\quad\:\:\theta_{min}, Q_{min}, d_{H}$
			&$\quad\:\:\theta_{min}, Q_{min}, d_{H}$
			&$\quad\:\:\theta_{min}, Q_{min}, d_{H}$
			&$\quad\:\:\theta_{min}, Q_{min}, d_{H}$
			&$\quad\:\:\theta_{min}, Q_{min}, d_{H}$	
			\\
			&$\quad\:\:21^{\circ}, 0.35, 0.373 $
			&$\quad\:\:22^{\circ}, 0.38, 0.565 $
			&$\quad\:\:23^{\circ}, 0.39, 0.566 $
			&$\quad\:\:24^{\circ}, 0.4,  0.668 $
			&$\quad\:\:23^{\circ}, 0.41, 0.79 $	
			\\
			\rotatebox{90}{\shortstack[c]{$L = 0.5$}}
			& \includegraphics[width=1.0\linewidth]{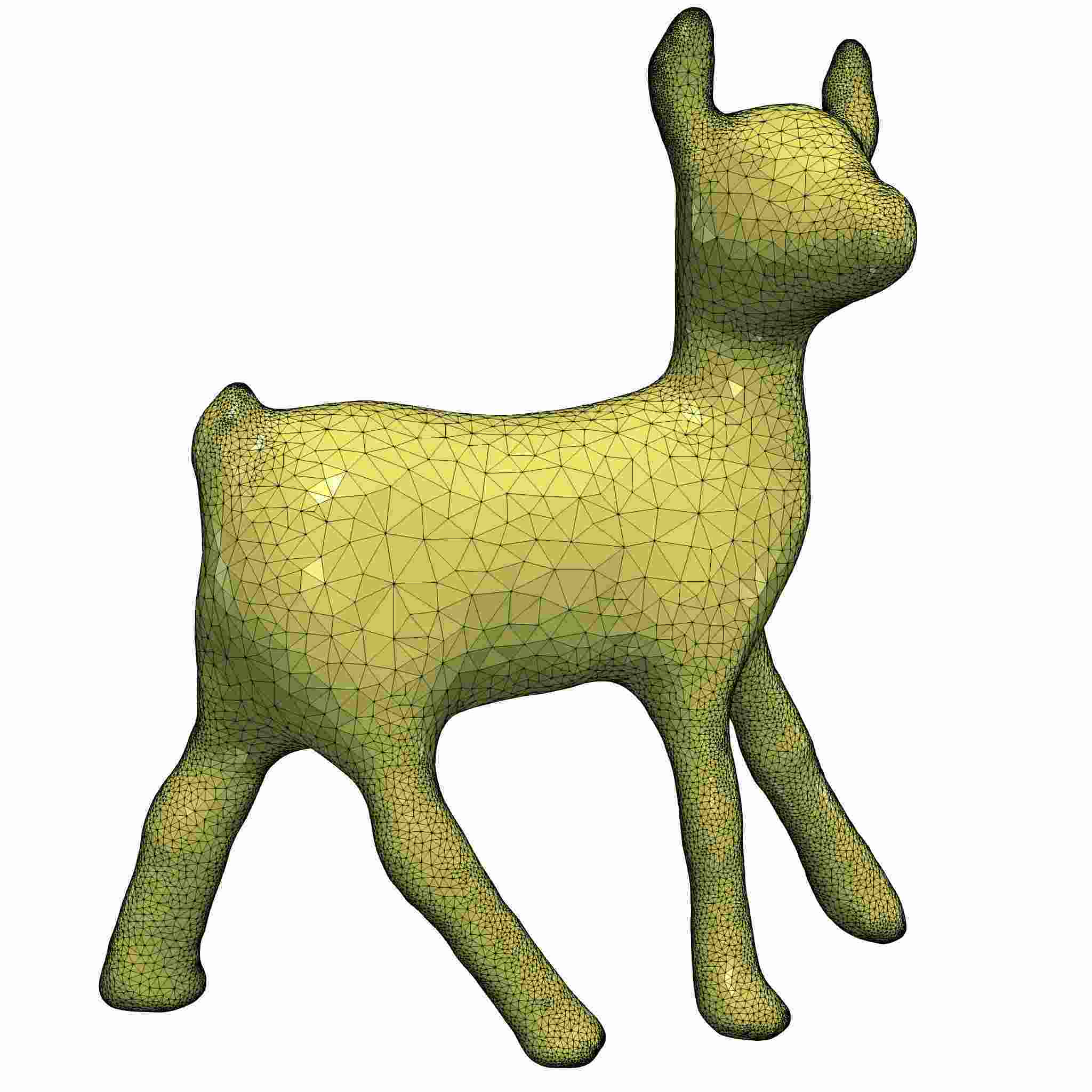}
			& \includegraphics[width=1.0\linewidth]{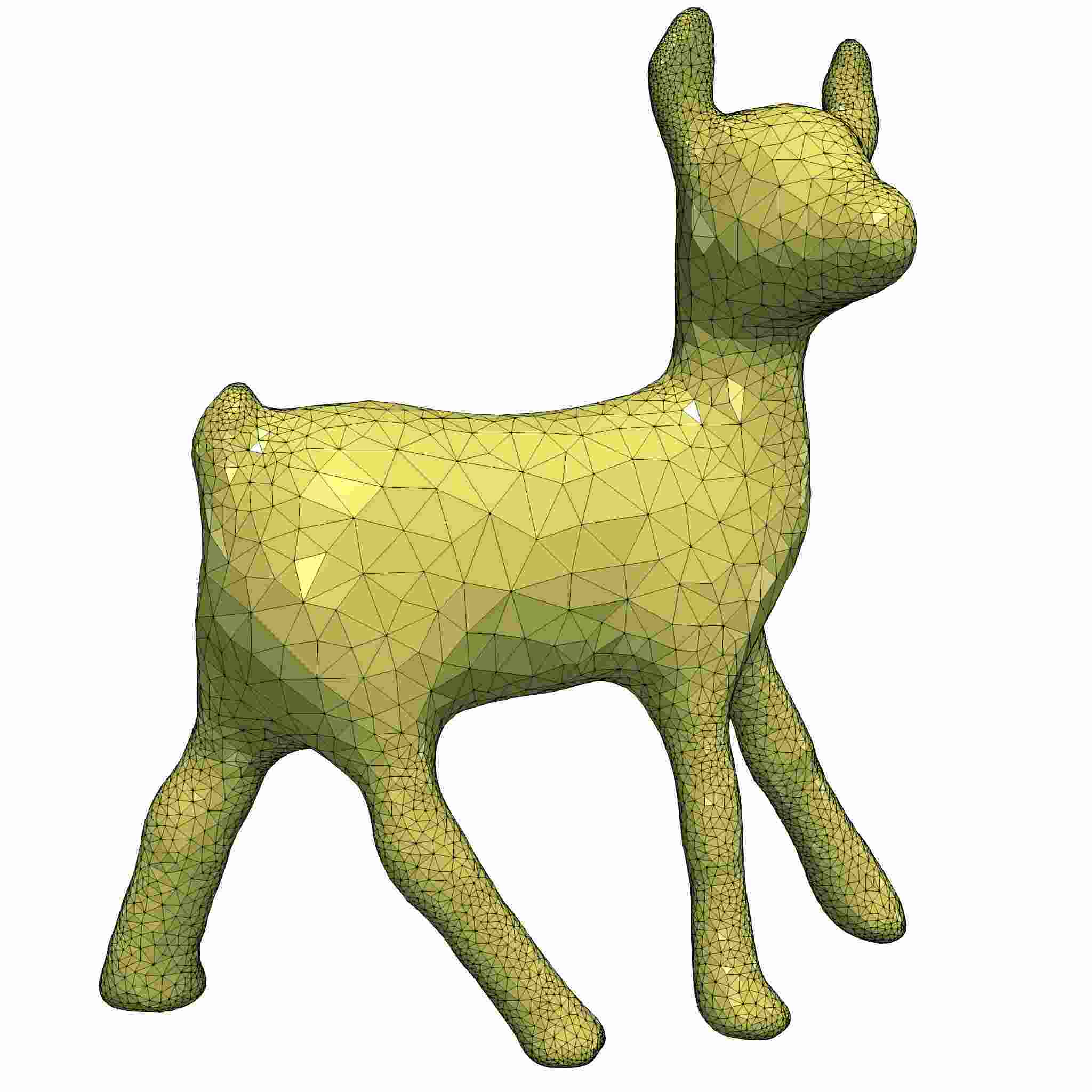}
			& \includegraphics[width=1.0\linewidth]{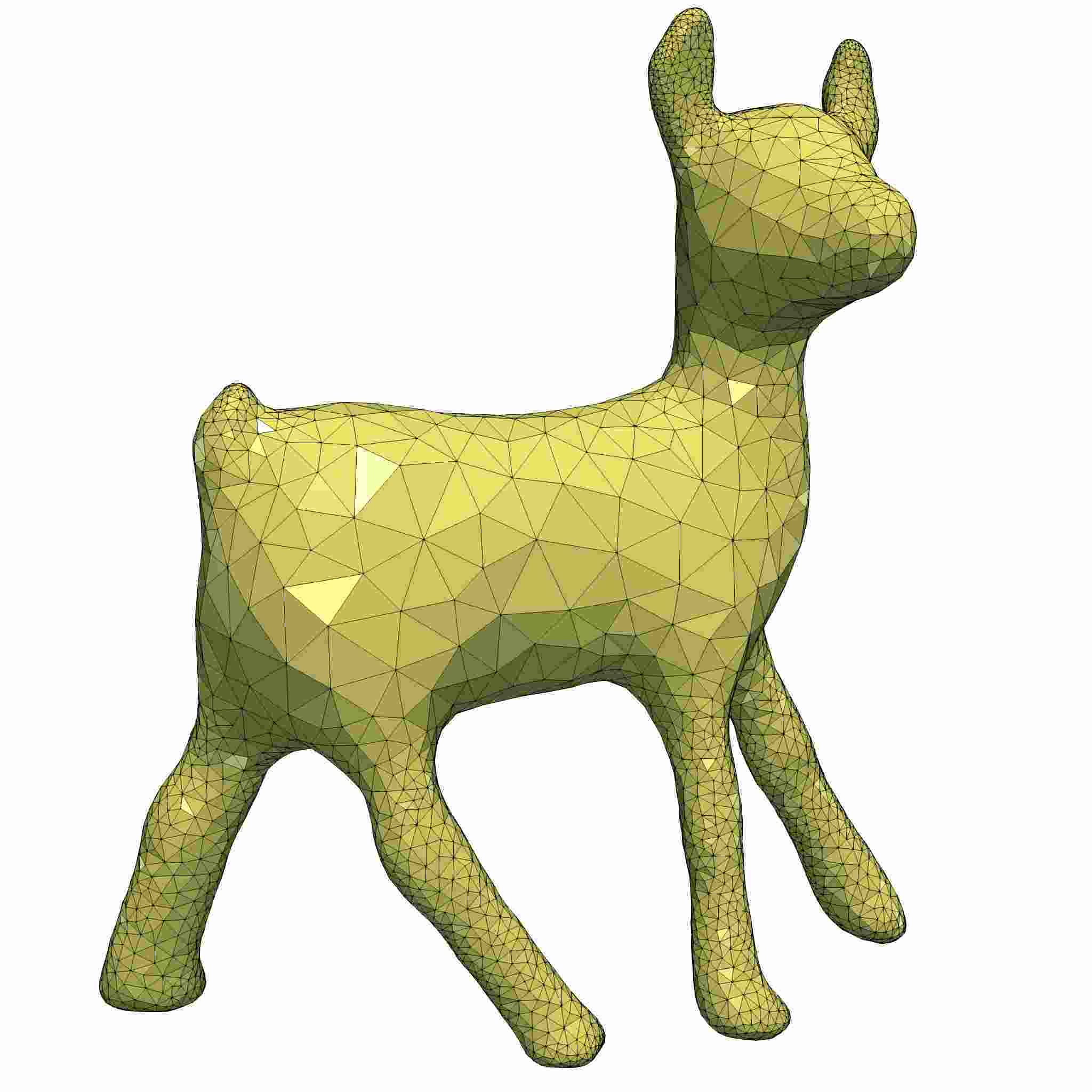}
			& \includegraphics[width=1.0\linewidth]{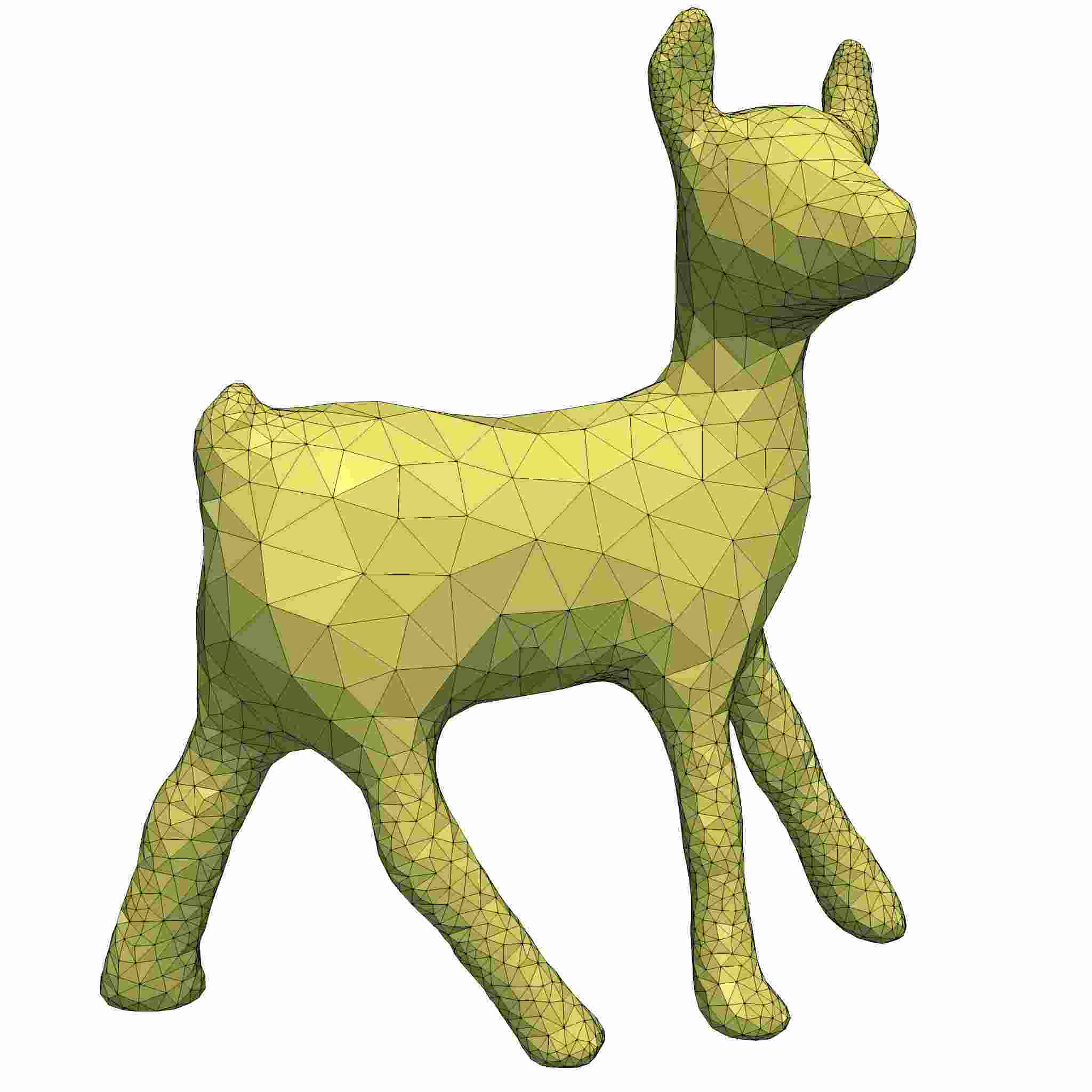}
			& \includegraphics[width=1.0\linewidth]{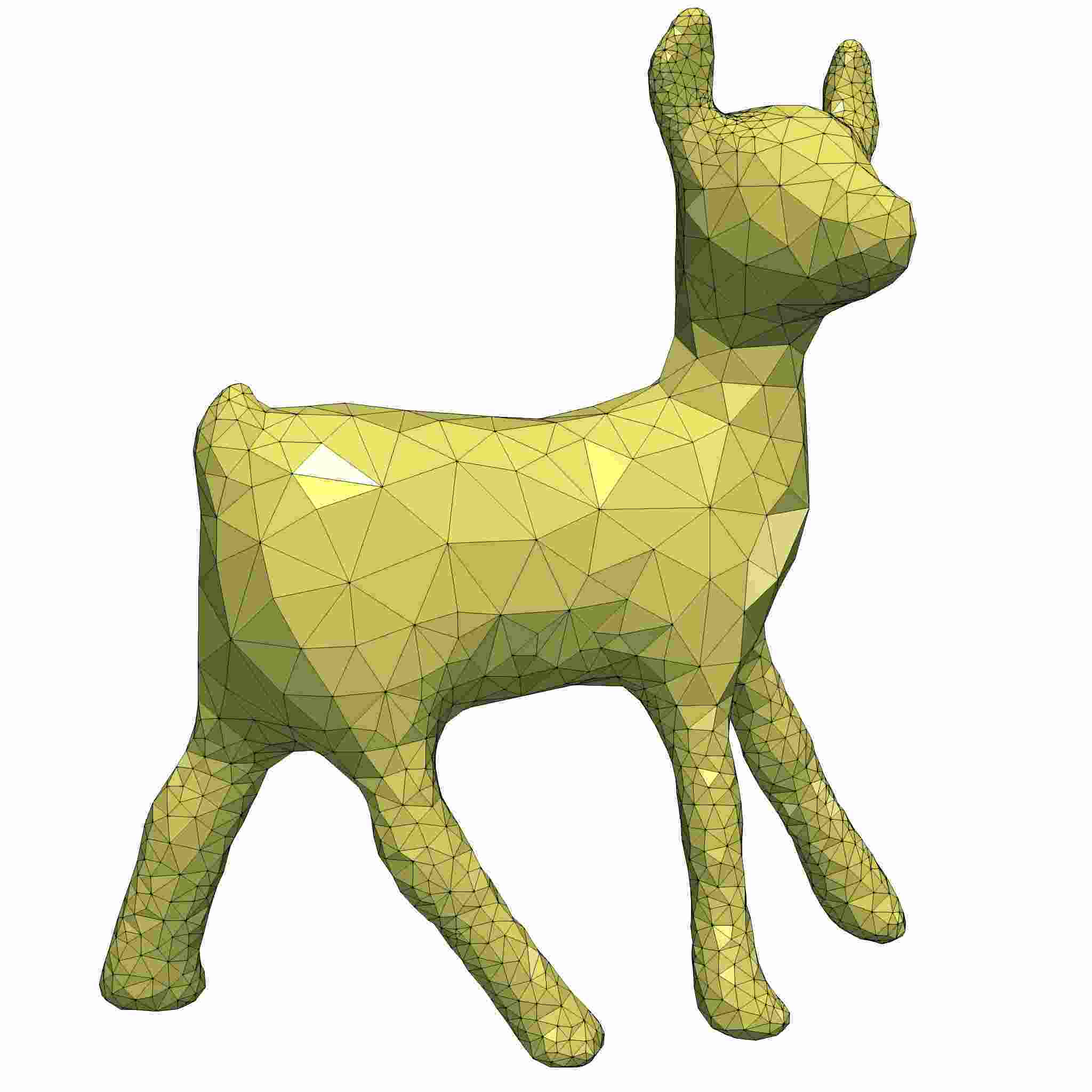}
			\\			
			&$\quad\:\:\theta_{min}, Q_{min}, d_{H}$
			&$\quad\:\:\theta_{min}, Q_{min}, d_{H}$
			&$\quad\:\:\theta_{min}, Q_{min}, d_{H}$
			&$\quad\:\:\theta_{min}, Q_{min}, d_{H}$
			&$\quad\:\:\theta_{min}, Q_{min}, d_{H}$	
			\\
			&$\quad\:\:16^{\circ}, 0.25, 0.433 $
			&$\quad\:\:15^{\circ}, 0.28, 0.508 $
			&$\quad\:\:17^{\circ}, 0.29, 0.606 $
			&$\quad\:\:17^{\circ}, 0.29, 0.822 $
			&$\quad\:\:17^{\circ}, 0.34, 1.0 $	
			\\			
			\rotatebox{90}{\shortstack[c]{$L = 0.9$}}
			& \includegraphics[width=1.0\linewidth]{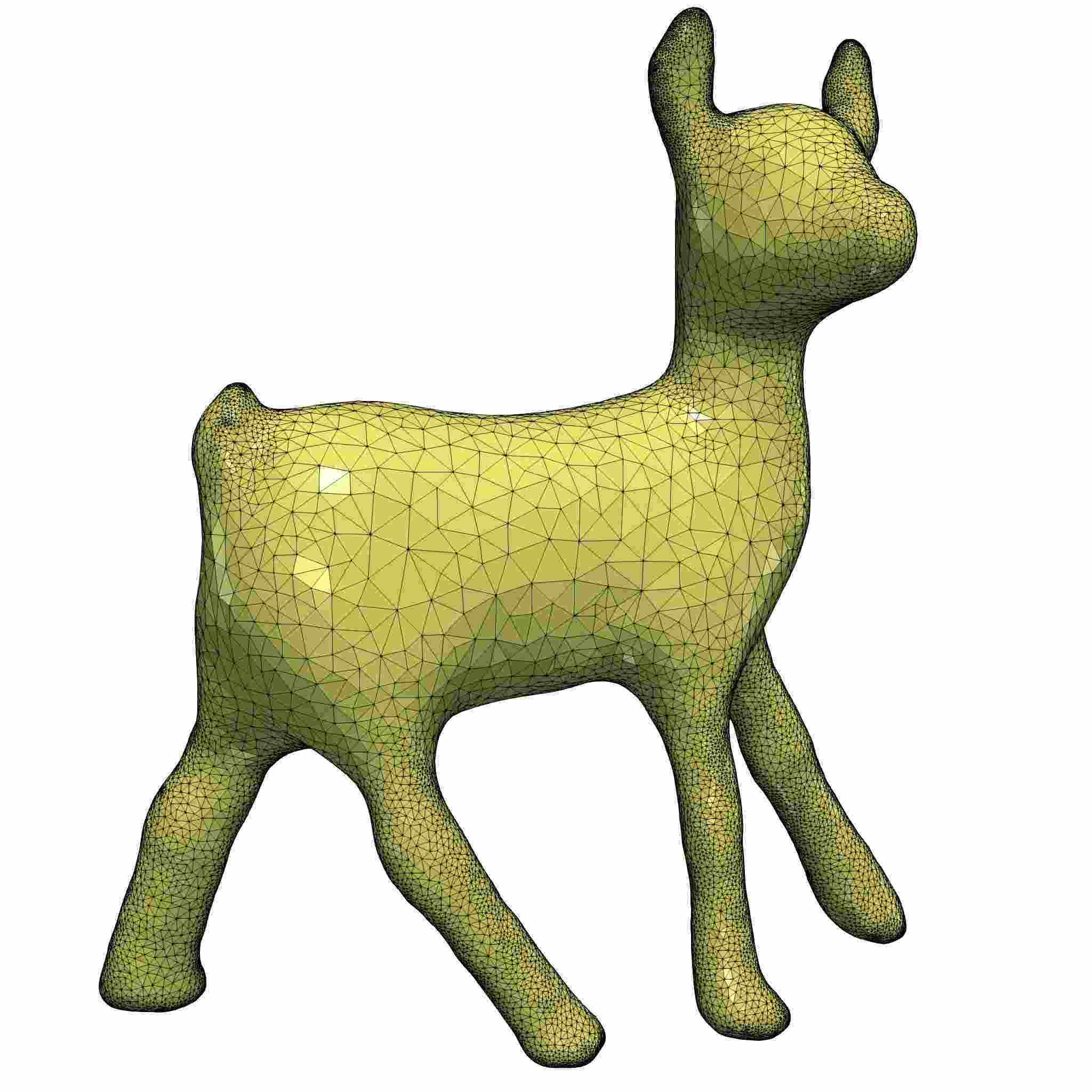}
			& \includegraphics[width=1.0\linewidth]{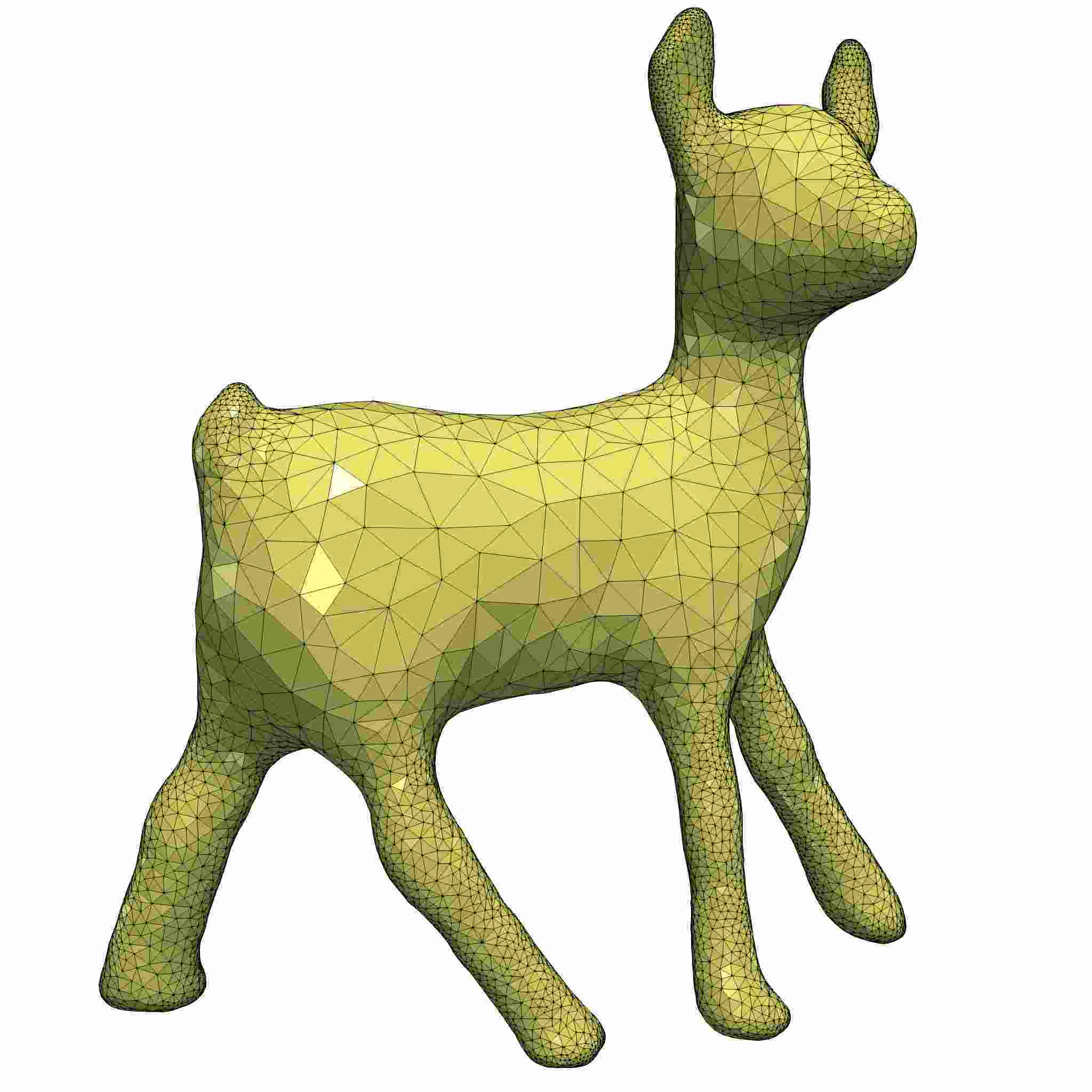}
			& \includegraphics[width=1.0\linewidth]{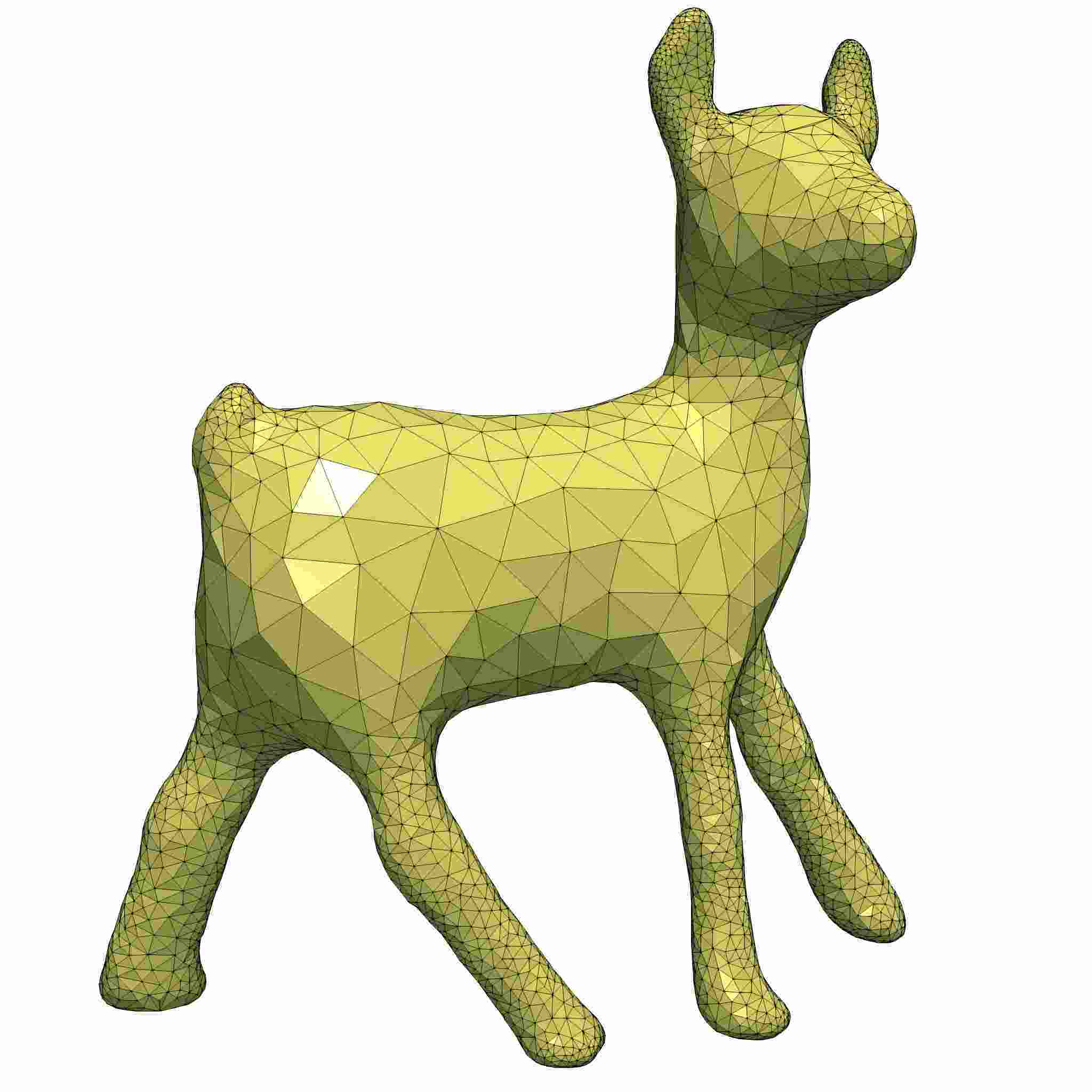}
			& \includegraphics[width=1.0\linewidth]{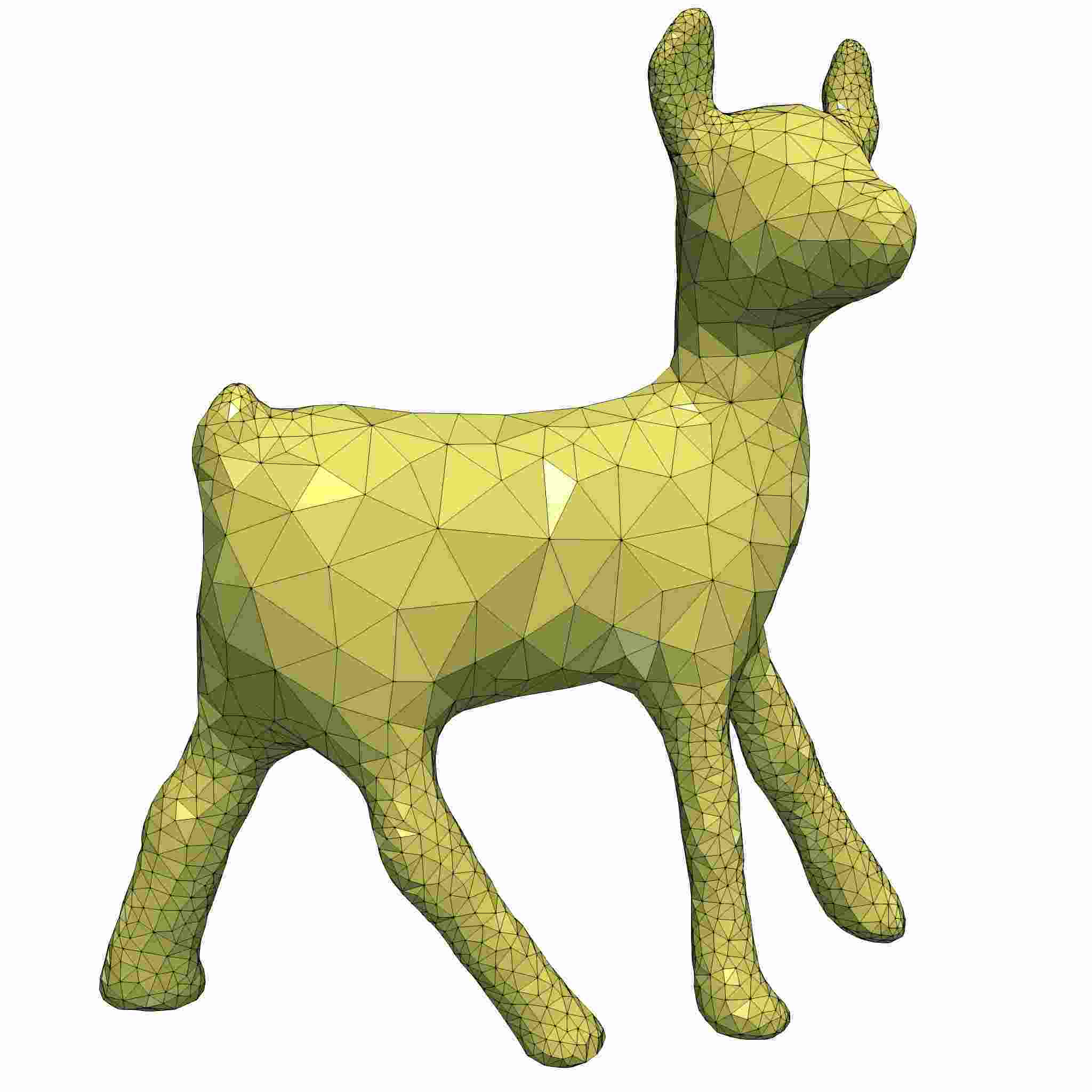}
			& \includegraphics[width=1.0\linewidth]{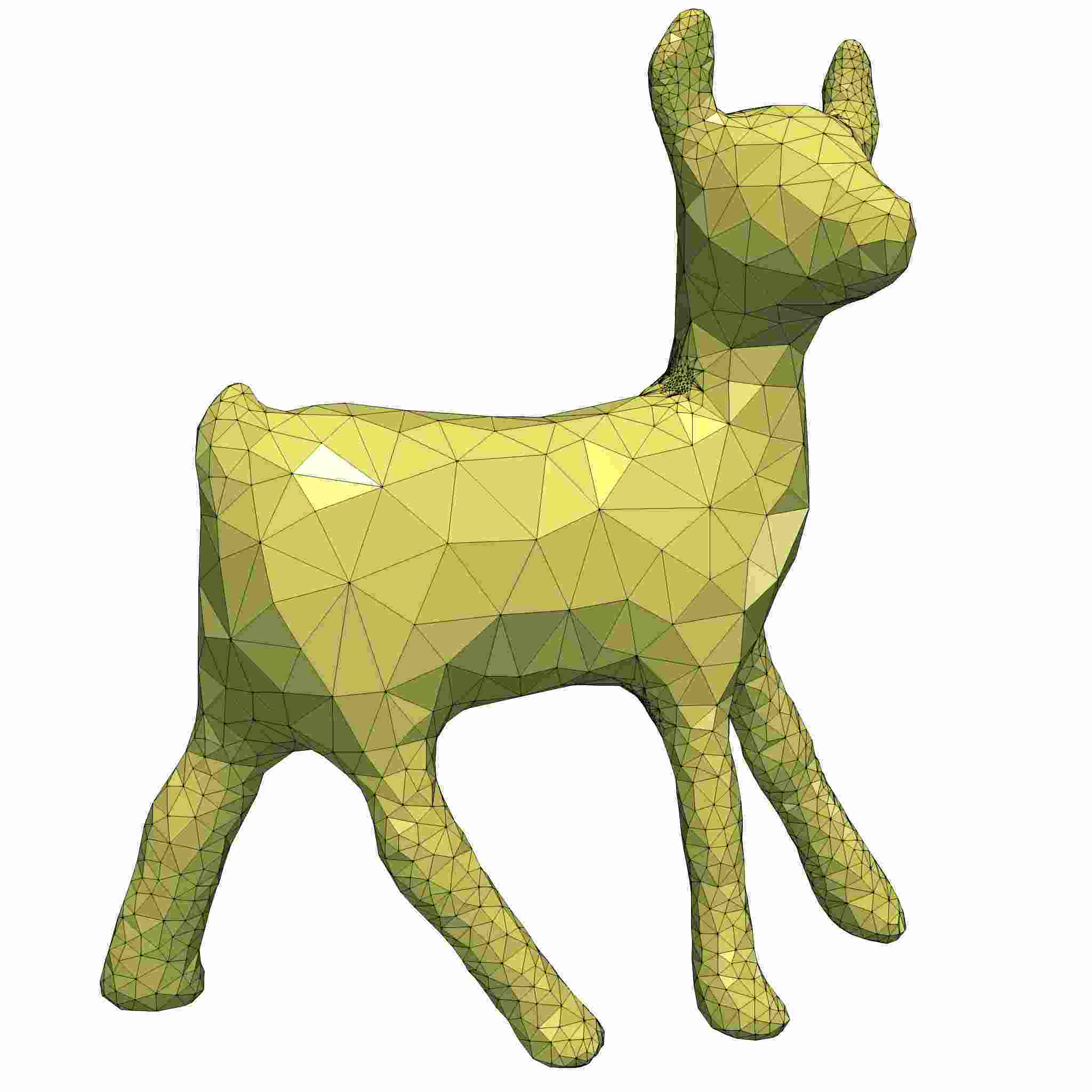}			
			\\						
			&$\quad\:\:\theta_{min}, Q_{min}, d_{H}$
			&$\quad\:\:\theta_{min}, Q_{min}, d_{H}$
			&$\quad\:\:\theta_{min}, Q_{min}, d_{H}$
			&$\quad\:\:\theta_{min}, Q_{min}, d_{H}$
			&$\quad\:\:\theta_{min}, Q_{min}, d_{H}$	
			\\
			&$\quad\:\:12^{\circ}, 0.2, 0.385 $
			&$\quad\:\:12^{\circ}, 0.25, 0.666 $
			&$\quad\:\:12^{\circ}, 0.19, 0.643 $
			&$\quad\:\:4^{\circ}, 0.1, 0.860 $
			&$\quad\:\:3^{\circ}, 0.08, 0.791 $	
			\\
			\bottomrule
			\rule{0pt}{10pt}
			&\qquad $\vcTheta = 25^{\circ}$
			&\qquad $\vcTheta = 40^{\circ}$
			&\qquad $\vcTheta = 55^{\circ}$
			&\qquad $\vcTheta = 70^{\circ}$
			&\qquad $\vcTheta = 85^{\circ}$						
		\end{tabular}
	}
	\vspace{-6pt}
	\caption{The Impact of input parameters $L$ and $\vcTheta$ on the surface quality and approximation error, demonstrated on the smooth Goat model. Besides the visual comparison, we report the minimum angle in the surface mesh $\theta_{min}$, the minimum triangle quality $Q_{min}$, and the Hausdorff error $d_{H}$ ($\times 10^{-2}$) normalized by the diagonal of the bounding box.
	}
	\label{fig:parameters}
\end{figure*}

Next, we study the impact of varying both $L$ and $\vcTheta$. We chose a relatively simple smooth model to better assess the degradation in surface approximation. Figure~\ref{fig:parameters} illustrates VoroCrust output on the Goat model for $5 \times 5$ combinations of parameter settings. As shown earlier, smaller values of $L$ result in more regular meshes with superior element quality per the minimum triangle angle. On the other hand, the parameter $\vcTheta$ controls the surface approximation. Namely, higher values of $\vcTheta$ result in higher Hausdorff errors.

\begin{figure}[htb]
	\centering
	\setlength{\tabcolsep}{4pt}
	\resizebox{1.0\linewidth}{!}{
		\begin{tabular}{c*{3}{m{0.34\linewidth}}}
			& \includegraphics[width=1.0\linewidth]{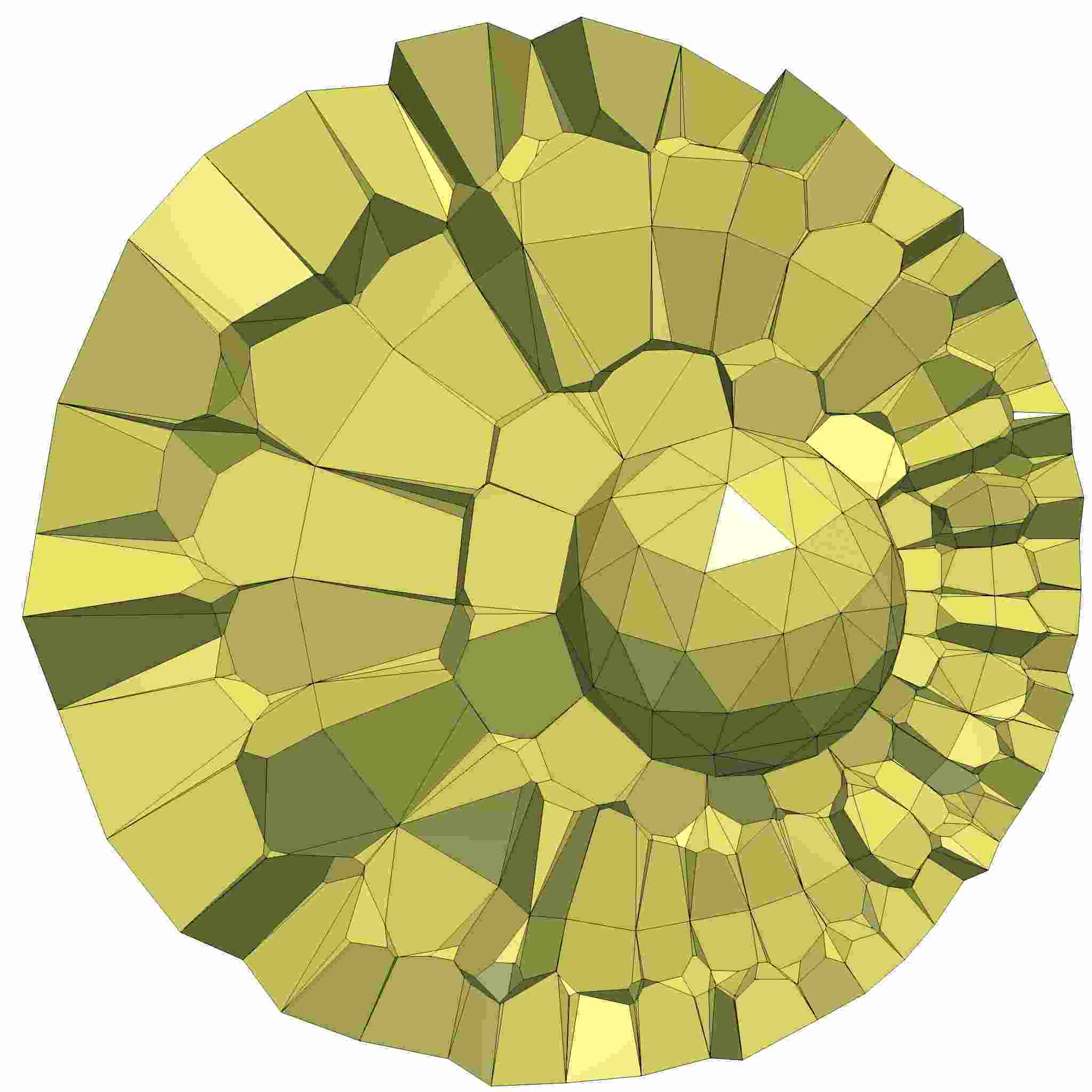}
			& \includegraphics[width=1.0\linewidth]{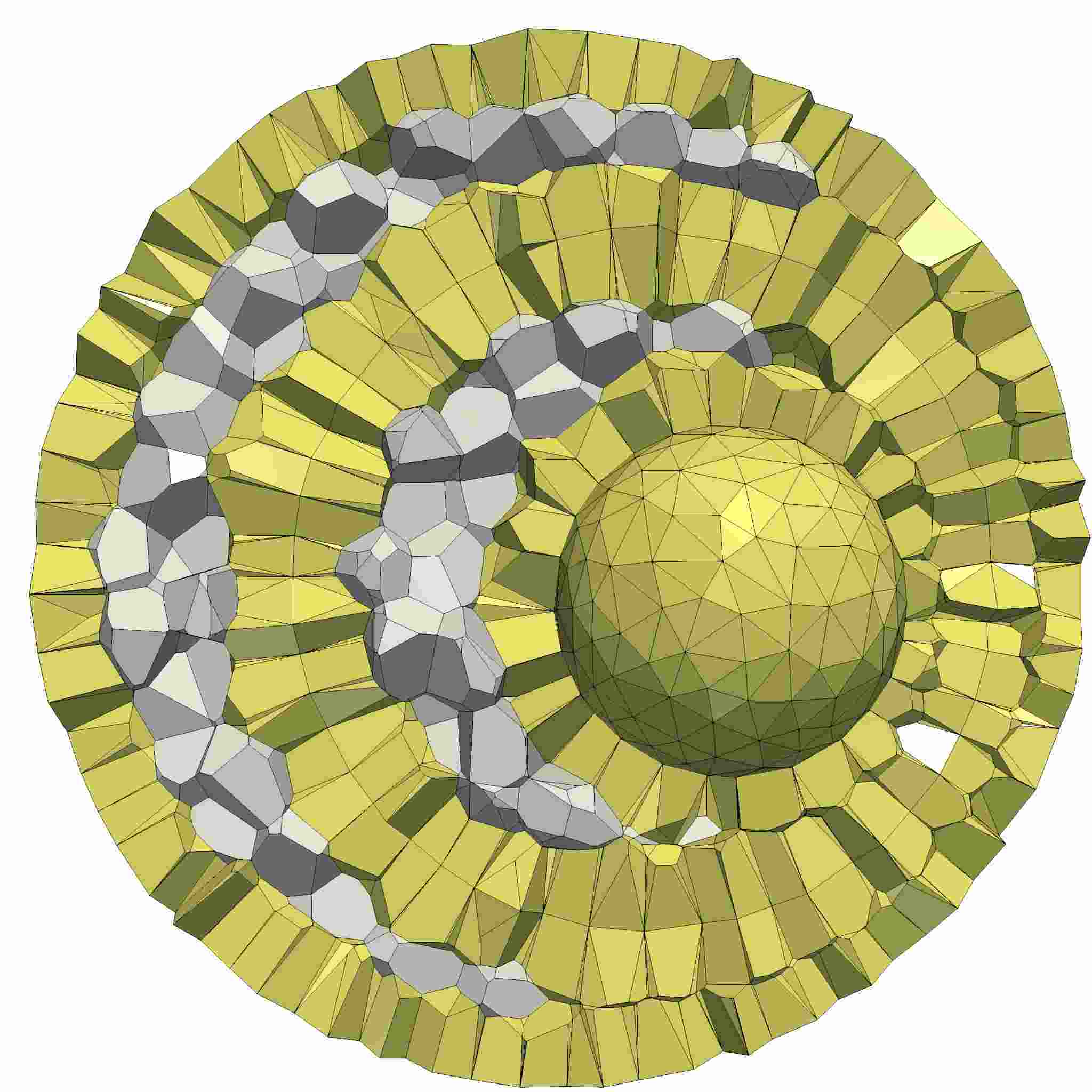}
			& \includegraphics[width=1.0\linewidth]{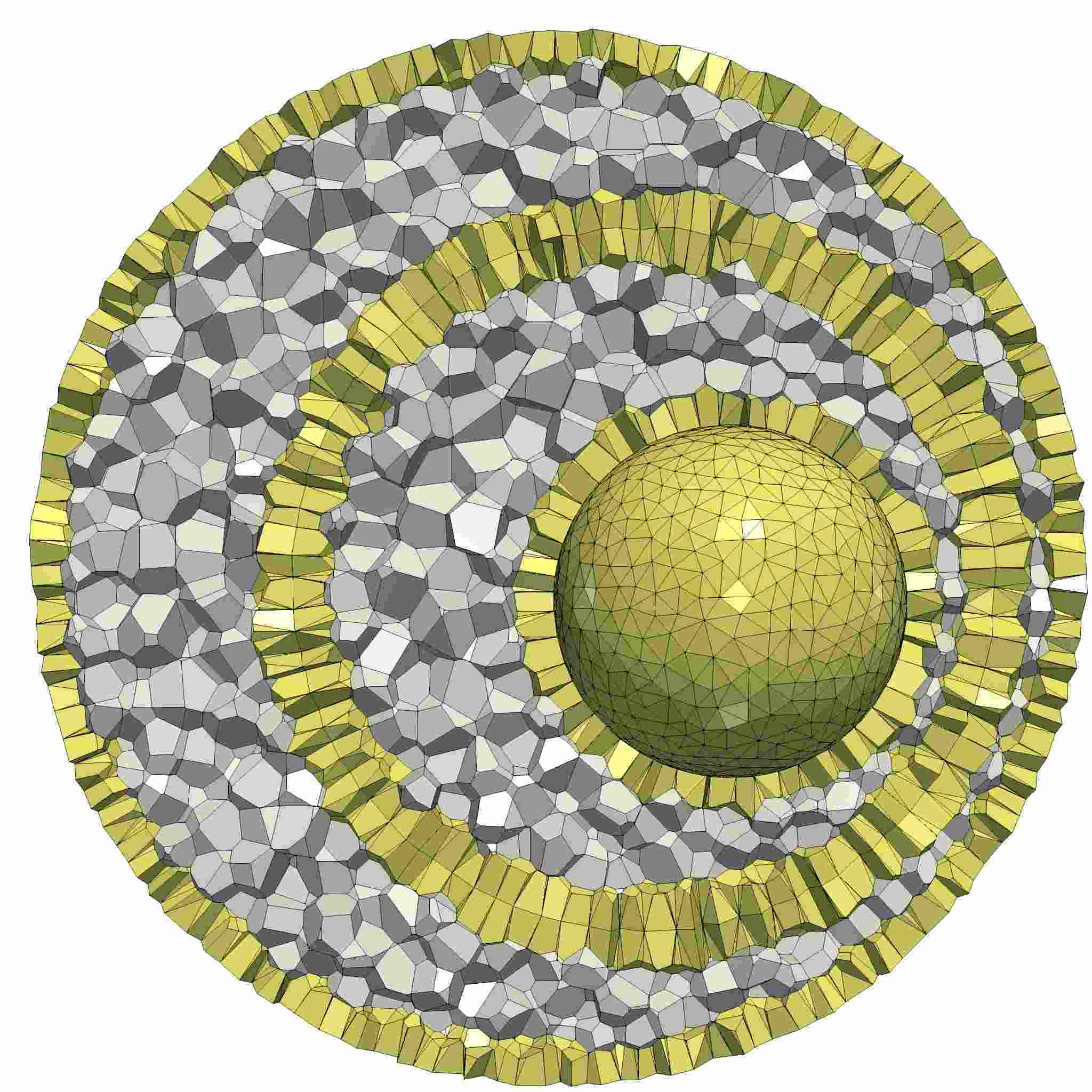}
			\\			
		\end{tabular}
	}
	\caption{Impact of the sizing field parameter $\rmax$ on the nested spheres model. From left to right: $\rmax = \infty$ (default), $\rmax = 1$, and $\rmax = 0.5$.}
	\label{fig:r_max_fig}
\end{figure}

Finally, we study the impact of the input sizing field $\rmax$ on the multi-layered nested spheres models. Figure~\ref{fig:r_max_fig} shows how $\rmax$ can be used to directly control ball radii to enforce further refinement. The default setting of $\rmax=\infty$ incurs the minimum level of refinement required by the geometry of the domain according to the quality requirements indicated by the parameters $L$ and $\vcTheta$. We note that $\rmax$ can be specified as a spatially varying sizing field.

\noindent \textbf{Non-manifold models.} These are particularly important in physical simulations with multiple materials of different properties. VoroCrust detects non-manifold features in the input mesh, as described in Section~\ref{sec:preprocessing}, and the ball conditions described in Section~\ref{sec:refinement} guide the refinement to protect those features, ensuring their correct recovery in the output mesh. Figure~\ref{fig:nonmanifold} shows VoroCrust output for a collection of non-manifold models. In addition, Figure~\ref{fig:turbine} shows VoroCrust output for a complex mechanical model. \newline

\usetikzlibrary{spy,shapes.misc}

\usetikzlibrary{spy,shapes.misc}

\begin{figure}[htb]
\centering
\setlength{\tabcolsep}{4pt}
\resizebox{\linewidth}{!}{
	\begin{tabular}{cc}	 
 	\begin{tikzpicture}
	[spy using outlines={rounded rectangle, magnification=3,width=2.5cm, height=1.5cm, connect spies}]
	\node {\includegraphics[width=0.7\linewidth]{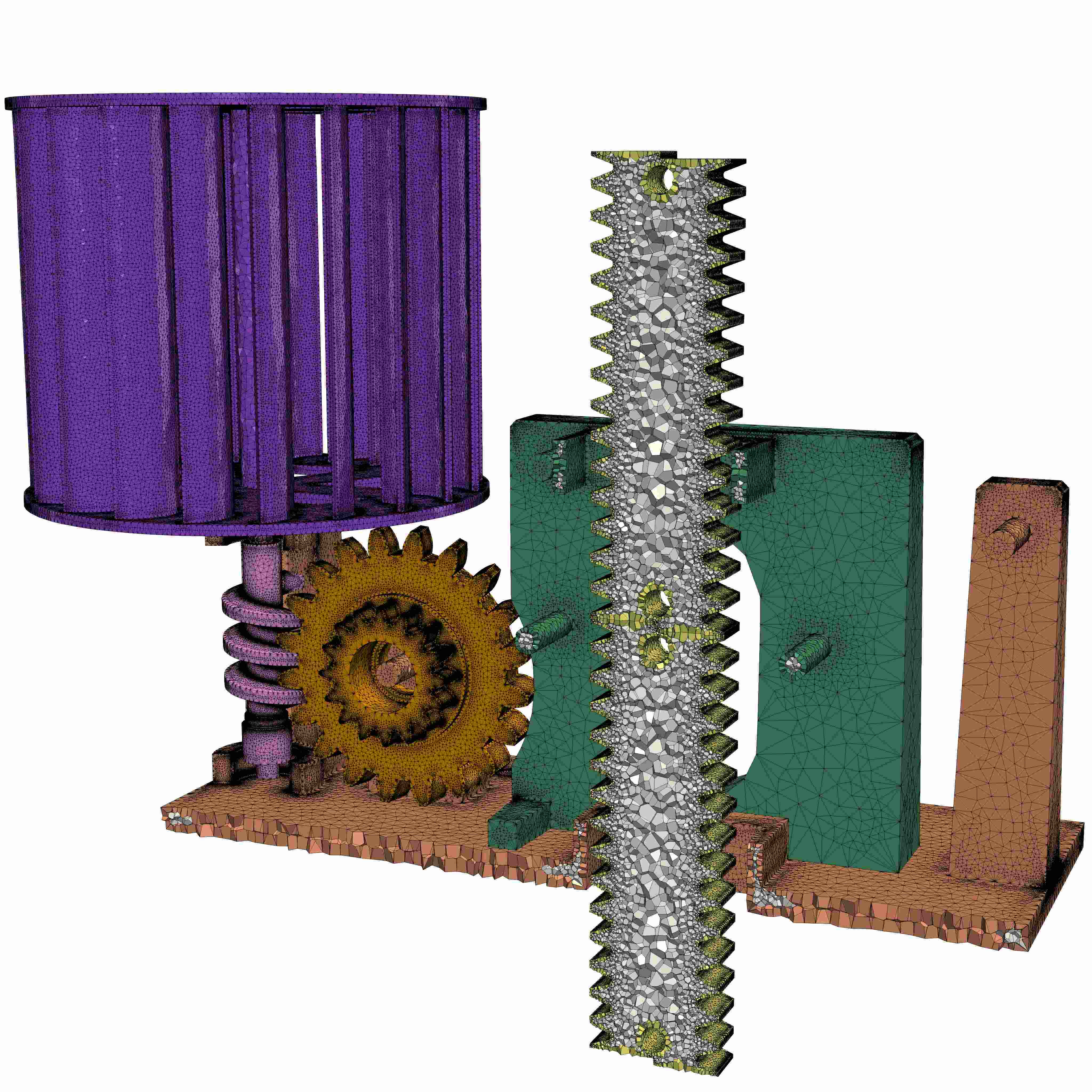}};	
	\spy on (-1.5,-0.6) in node at (-1.2,-2.3);
	\spy on (0.6,-0.4) in node at (2.2,2.0);
	
	\end{tikzpicture} 		
	\end{tabular}}
\caption{VoroCrust output for complex mechanical parts sharing non-manifold contact interfaces with detailed sharp features.}
\label{fig:turbine}
\end{figure}


In summary, this study demonstrates the flexibility of the VoroCrust algorithm to accommodate a wide range of parameter settings that cater to the requirements of different applications. In particular, the set of parameters provided allows the user to trade-off the quality of the surface mesh, approximation error, output complexity, and running time. \newline

\noindent \textbf{Comparison.} We compare against the restricted Voronoi diagram (RVD) of Yan et al.~\shortcite{yan2010efficient} as a representative of state-of-the-art polyhedral meshing algorithms based on clipped Voronoi cells. While RVD is typically used within CVT-based algorithms to speed up energy calculations, we are only interested in its robust clipping capabilities which provide a suitable baseline for comparison. For all models, we use the interior VoroCrust seeds $\interiorSeeds$ as input to RVD clipping. As shown in Figure~\ref{fig:teaser}, VoroCrust achieves superior quality in terms of the surface mesh, where RVD clipping produces an imprint of the input mesh with many small facets. In particular, by examining the ratio of the shortest to longest edge length per surface facet, it is clear that RVD clipping results in many skinny facets which can be problematic for many applications. Moreover, RVD clipping possibly results in non-convex cells for non-convex models, e.g., Figure~\ref{fig:RVD_nonconvex}. In our experiments, the ratio of non-convex cells in RVD output varies between $3\%$ and $96\%$, depending on the curvature of the input surface and the chosen set of Voronoi seeds. In contrast, VoroCrust output conforms to the boundary with true Voronoi cells, which are guaranteed to be convex, while achieving much better quality of surface elements. We note that clipping the Voronoi cells of a given set of seeds can be performed much faster, as in the parallel RVD implementation of Yan et al.~\shortcite{yan2010efficient}, compared to the multiple iterations and non-trivial steps of VoroCrust refinement; see Section~\ref{sec:algorithm}. Additional comparisons against RVD are provided in the supplemental materials.


\section{Limitations}
\label{sec:limitations}
The main limitation of the presented algorithm is the possible presence of short Voronoi edges in the interior of the output mesh, which can lead to small time steps in numerical simulations significantly increasing their cost. To eliminate such short edges, mesh improvement techniques may be applied as postprocessing~\cite{cgf.13256,sieger2010optimizing}.

\begin{figure}[h]
\centering
\setlength{\tabcolsep}{3pt}
\resizebox{\linewidth}{!}{
        \begin{tabular}{cc}
        \rotatebox{90}{\qquad \qquad \qquad \qquad \begin{Huge}Open Hand\end{Huge}}
     \includegraphics[width=1.0\linewidth]{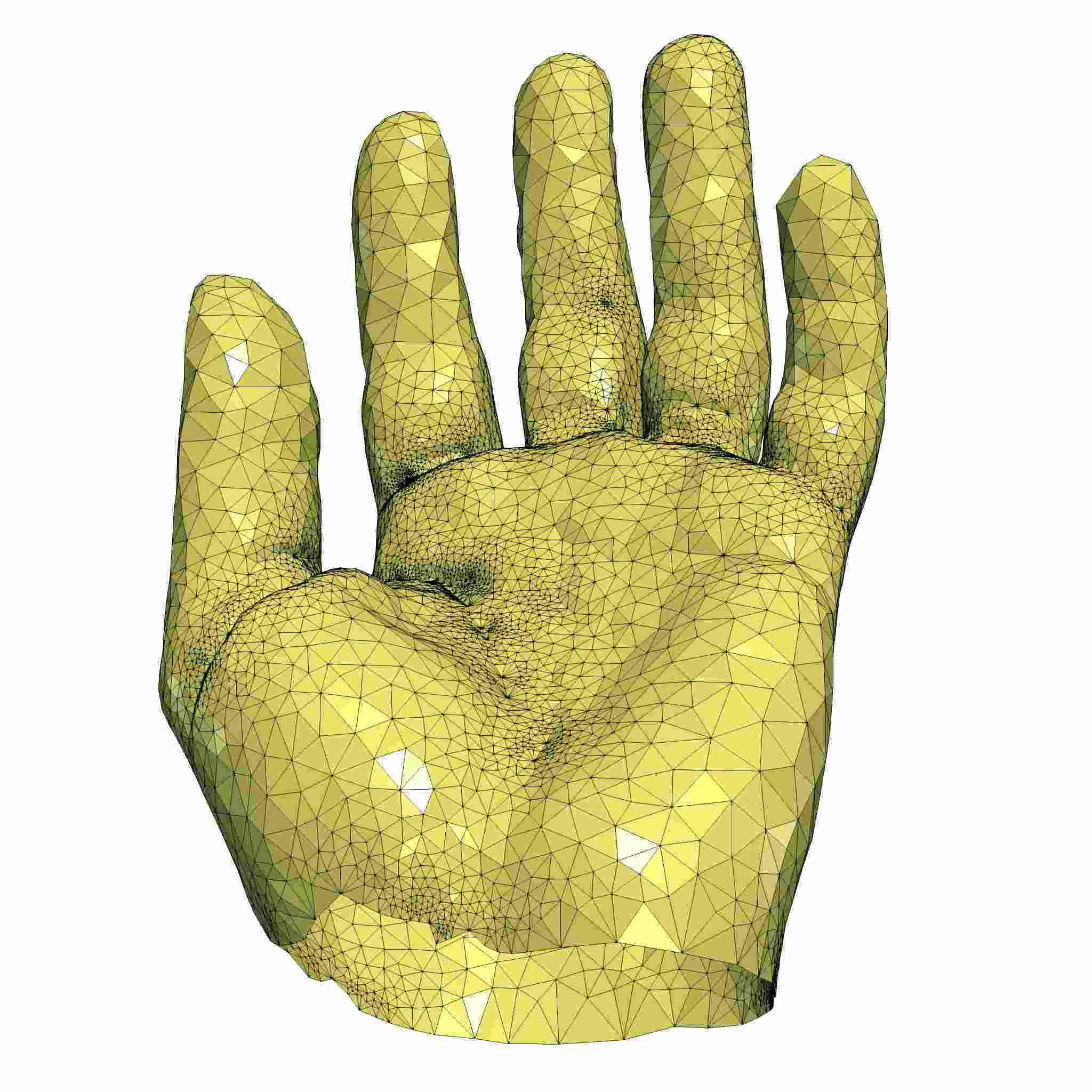} &
     \includegraphics[width=1.0\linewidth]{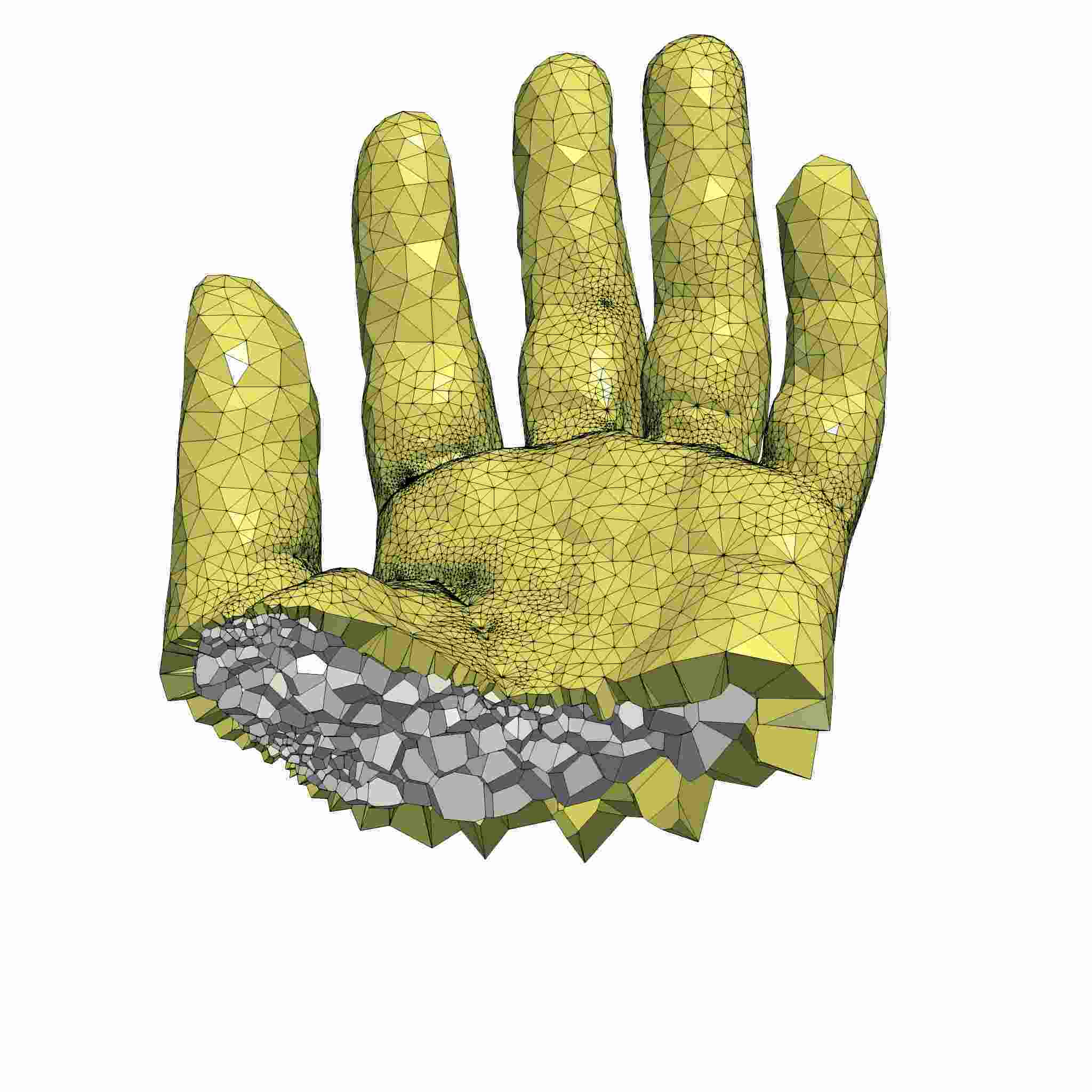}\\
        \end{tabular}}
        \caption{VoroCrust can handle surfaces with boundary. Volume samples within a suitable bounding box can be filtered, e.g., manually, as shown.}
\label{fig:open_hand}
\end{figure}

Another limitation is the requirement that the input triangulation is a faithful approximation of the domain. This inhibits the application of this approach to implicit forms~\cite{10.1007/978-3-319-46487-9_11}, noisy inputs~\cite{Mederos:2005:SRN}, or unclean geometries~\cite{Attene:2013:PMR}. In particular, the algorithm does not fill holes or undesirable cracks in non-watertight inputs~\cite{hole_filling}. Nonetheless, VoroCrust readily handles surfaces with boundary as shown in Figure~\ref{fig:open_hand}.

Finally, the isotropic nature of the proposed sampling process may result in an unnecessarily large number of cells in narrow regions. For such geometries, boundary layers of elongated cells enable higher fidelity near the boundary~\cite{doi:10.1002/1097-0207,10.1007/978-3-642-33573-0_29}. In cases of strong anisotropy, aligning the cells, e.g., to the eigenvectors of a Hessian~\cite{Fu:2014:ASM,Budninskiy:2016:OVT}, better captures the variation of physical quantities.

\section{Conclusion and Future Work}
\label{sec:conclusion}
Voronoi cells provide a competitive alternative to traditional mesh elements with many desirable features that follow naturally from their definition, e.g., convexity, convex planar facets, and orthogonal duals provided by the corresponding Delaunay elements. As general polyhedral elements, Voronoi cells enjoy greater degrees of freedom that enable better mesh connectivity and the ability to conform to complex geometries undergoing large deformations. What hindered their wide-scale adoption is the lack of a robust Voronoi meshing algorithm that can handle broad classes of domains exhibiting arbitrary curved boundaries and sharp features.

We developed the VoroCrust algorithm to fill this gap. VoroCrust is based on a well-principled mirroring approach combined with state-of-the-art techniques for automatic sizing estimation to mesh piecewise-smoothed complexes. We proved strong theoretical guarantees on the correctness of the VoroCrust algorithm and the quality of its output, as demonstrated through a variety of models. By conducting an extensive comparison against state-of-the-art polyhedral meshing methods based on clipped Voronoi cells, we established the advantage of VoroCrust output.

For future work, we consider speed-ups by parallelization as well as anisotropic meshing and boundary layers. We believe that VoroCrust refinement can be extended to accommodate additional requirements catering to the quality of the cells while preserving the surface approximation. To match the quality of the output surface mesh, improving the quality of the volume mesh by eliminating short Voronoi edges possibly present in the interior is particularly important. Such short edges arise when the distance from a Voronoi vertex $v$ to its $d+1$ nearest Voronoi seeds is only slightly less than its distance to the $(d+2)^{th}$ nearest Voronoi seed $s$. In other words, the seed $s$ lies close to the \emph{Delaunay sphere} centered at $v$. A potential approach to avoid such configurations is to define a \emph{buffer zone} to penalize the placement of Voronoi seeds that result in short Voronoi edges. The buffer zone can be defined as a thickening of the Delaunay sphere into a spherical shell whose thickness is a small fraction of the radius. We are currently exploring a restricted sampling technique based on this idea. Finally, the increased flexibility enabled by weighted vertices~\cite{Goes:2014:WTG} for a weighted-version of conforming Voronoi meshing would be an interesting extension especially in the context of primal-dual meshing.

\section*{Acknowledgements}
Meshes are courtesy of Carl Gable (geological models), and Carlo H. S\'equin (Heptoroid), \emph{Thingi10k}~\cite{Thingi10K}, the Stanford 3D Scanning Repository~\cite{Stanford}, and the Aim@Shape Shape Repository~\cite{aimshape}.

\bibliographystyle{ACM-Reference-Format}
\bibliography{siggraph19}

\appendix
\section{Implementation Details}
\label{sec:implementation}
We start by describing the speed-ups for proximity queries against the input PLC $\dsurf$ and the set of balls $\ballset$. Then, we describe the generation of interior samples.


\subsection{Supersampling the boundary}\label{subsec:supersampling}
The algorithm constructs one \kdtree{} for each type of strata to speed up proximity queries against $\dsurf$. The \kdtree{} indexing the sharp corners is simply populated using the set of sharp corners. In order to populate the \kdtree{} indexing the creases, the algorithm generates a set of $10^5$ points sampled uniformly at random from all sharp edges. Similarly, the \kdtree{} indexing the surface patches is populated using a set of $10^6$ points sampled uniformly from all facets. Each generated sample $q$ stores a vector $v_{\sigma,q}$ for each edge or facet $\sigma \ni q$.

\subsection{Querying the Boundary \kdtrees{}}\label{subsec:noncosmooth_query}
Given a point $p$ on a face $\sigma$, the algorithm estimates the distance to the nearest non-co-smooth point on the input mesh $\dsurf$ by querying the three boundary \kdtrees{} indexing the sharp corners, creases and surface patches. Let $K$ denote any of the boundary \kdtrees{}. As the query aims to determine the nearest non-co-smooth point, the co-smoothness test described in Section 2.3 can be used to filter the set of points indexed by $K$. We implemented a custom \kdtree{} that performs this filtration on-the-fly. As in the standard \kdtree{}, the query maintains an estimate of the distance to the nearest point which can be initialized to any sufficiently large value, e.g., the diameter of $\dsurf$ or $\infty$. By comparing the current estimate against the distance from $p$ to the splitting plane associated with the current node, the query discards an entire subtree if it cannot improve the estimate. The only difference is that due to the filtration defined by the co-smoothness test, a node associated with a point which is co-smooth with $p$ does not provide a distance to update the estimate.

\subsection{Ball Neighborhood}\label{subsec:ball-neighborhood}

To find the set of balls overlapping a given ball $b_p$, a naive search would be costly. Instead, we find an upper bound on the distance between $p$ and any sample $q$ such that $b_q$ may overlap $b_p$. Then, we use this bound to query the \kdtrees.

Consider two overlapping balls $b_p$ and $b_q$ generated by the MPS procedure, with radii $r_p$ and $r_q$. W.l.o.g., assume $r_q \geq r_p > 0$. The $L$-Lipschitzness condition implies that $r_q \leq r_p + L \cdot \|p - q\|$. Since the two ball overlap: $\|p - q\| < r_p + r_q$. Combining the two inequalities, it follows that: $\|p - q\| < r_p + r_q + L \cdot \|p - q\|$. We conclude that $\|p - q\| \leq \frac{2}{1-L}\cdot r_p$. Hence, we query the \kdtrees{} for all balls whose centers are within that distance from $p$ and check if they overlap $b_p$.

\subsection{Point Neighborhood}\label{subsec:point-neighborhood}
The deep coverage condition is checked for each new sample $p$. To speed up this check, we derive an upper bound on the distance between $p$ and the center of any ball that may cover it, and use this to query the \kdtrees.

Let $q$ denote the center of the closest ball to $p$, which we find by a standard nearest-neighbor query to the \kdtree{} in question. The radius of a ball placed at $p$ respecting $L$-Lipschitzness can be estimated as $r_p \leq r_q + L \cdot \|p - q\|$.

Consider a ball $b_s$ that barely covers $p$. It follows that $r_s \leq r_p + L \cdot \|p - s\|$, where $\|p - s\| \leq r_s$. Combining the two inequalities, it follows that $r_s \leq r_q + L \cdot \|p - q\| + L \cdot r_s$, implying $r_s \leq \frac{r_q + L \cdot \|p - q\|}{1-L}$. Hence, we query the \kdtree{} for all balls whose centers are within that distance from $p$ and check if they contain $p$.

\subsection{Sampling the interior}
\label{sec:imp_interior}

The algorithm starts by computing a bounding box $\textrm{bb}$ enclosing the input mesh $\dsurf$; we expand $\textrm{bb}$ to the box $3\times$ larger with the same center. This box is used to initialize the set of interior seeds $\interiorSeeds$ using a lightweight dart-throwing phase. Additional samples are added as needed using the more efficient spoke-darts algorithm~\cite{spokedarts18}. To guide interior sampling, and ensure a sufficient distance between interior seeds and surface seeds, each surface seed $s \in \surfaceSeeds$ is assigned a radius $r_s$ by averaging the radii of the three balls in $\ballset$ defining it. As was done for the set of surface balls $\ballset$, we maintain two \kdtrees{} $K^\updownarrow$ and $K^\ddarrow$ for all balls centered at seeds in $\surfaceSeeds$ or $\interiorSeeds$, respectively.

To initialize $\interiorSeeds$, a new sample $z$ is generated uniformly at random from $\textrm{bb}$. Then, the closest seed $s \in \surfaceSeeds$ to $z$ is found by a nearest-neighbor query to $K^\updownarrow$. If $\|z - s\| < r_s$, z is rejected. Otherwise, $z$ gets the label of $s$ and a radius $r_z = r_s + L \cdot \|z - s\|$, which extends the estimated sizing function to the interior of the domain~\cite{LipschitzExtension}. Similarly, the closest interior seed $z^\ast \in \interiorSeeds$ to $z$ is found by querying $K^\ddarrow$ and $z$ is rejected if $\|z - z^\ast\| < r_{z^\ast}$. Whenever a new sample is rejected, we increment a \emph{miss counter} and otherwise reset it back to 0 if the sample was successfully added into $\interiorSeeds$. Initialization terminates when the miss counter reaches 100.

Then, we continue to add seeds into $\interiorSeeds$ using the spoke-darts algorithm~\cite{spokedarts18} as follows. We populate a queue $Q$ with all seeds generated by dart-throwing. While the queue is not empty, we pop the next sample $z$ and do the following. Letting $b_z$ be the ball centered at $z$ with radius $r_z$, we choose a random direction $\delta$ and shoot a \emph{spoke} (ray) starting at $z$ in that direction to obtain a new point $z_\delta$ at distance $2\cdot r_z$ from $z$. Then, we query the \kdtrees{} to find all balls potentially containing $z_\delta$. For each such ball, we trim the line segment $\ell_\delta$ between $z$ and $z_\delta$ by pushing $z_\delta$ to lie on the boundary of that ball. Once we are done, if $z_\delta$ was pushed all the way into the ball $b_z$, we increment the miss counter. Otherwise, we sample a point $z^+$ uniformly at random on $\ell_\delta$, add it as a seed, and reset the miss counter to 0. As before, $z^+$ is assigned a label and a radius before pushing it into $Q$. When the miss counter reaches 100, we discard the current point and pop a new point from $Q$. This process terminates when $Q$ is empty. Finally, we enforce $L$-Lipschitzness on all interior samples, shrinking balls as necessary, before repopulating $Q$ with all seeds and repeating until no ball gets shrunk.

\subsection{Code Profiling and Bottlenecks}
\label{sec:profiling}
We instrument our code to collect more detailed timing statistics for the main procedures of the algorithm; see Section~\ref{sec:algorithm}. As would be expected, the most time consuming component of the algorithm is surface coverage, with related MPS iterations as described under ``Protection and Coverage,'' and to a lesser extent volume sampling per Section ~\ref{sec:imp_interior}; other procedures including preprocessing, sharp feature protection, and sliver elimination are not as demanding. In particular, each surface sample requires a sizing estimate by querying the boundary \kdtrees{} which store a dense sampling of surface elements; see Section ~\ref{subsec:supersampling}. In addition, whenever we shrink a surface ball, checking for uncovered surface patches requires restarting the surface MPS procedure. For example, Table~\ref{tab:timing} summarizes the running time on two sample models.

\begin{table}[h]
  \centering
\begin{tabular}{| c | c | c |}
\hline
  Procedure & Smooth & Sharp Features \\
\hline
  Corner protection & 0 & 0.213\\
\hline
  Edge protection & 0 & 4.157  \\
\hline
  Surface coverage & 671.165 & 180.986  \\
\hline
  Fixing \Lipschitz{} violations & 17.255 & 2.962  \\
\hline
  Sliver elimination & 14.127 & 3.216  \\
\hline
  Interior sampling & 13.981 & 36.395  \\
\hline
\end{tabular}
  \caption{Timing breakdown for the smooth model shown in Figure~\ref{fig:teaser} and the model with sharp features shown in Figure~\ref{fig:RVD_nonconvex}.}
  \label{tab:timing}
\end{table}

Per the table above, \Lipschitz{} violations and sliver elimination incur higher overhead for the smooth model with higher surface curvature compared to the model with sharp features and otherwise flat regions.


\end{document}